\documentclass[onecolumn]{emulateapj}

\usepackage{natbib}
\usepackage[usenames]{color}

\newcommand{\HI}{\ion{H}{1}}

\newcommand{\kms}{\mbox{km~s$^{-1}$}}
\newcommand{\Msol}{\mbox{M$_\odot$}}
\newcommand{\surm}{\mbox{M$_\odot$ pc$^{-2}$}}

\newcommand{\siggas}{\mbox{$\Sigma_{\rm gas}$}}
\newcommand{\sigsfr}{\mbox{$\Sigma_{\rm SFR}$}}
\newcommand{\sigstar}{\mbox{$\Sigma_{*}$}}
\newcommand{\sighi}{\mbox{$\Sigma_{\rm HI}$}}
\newcommand{\sightwo}{\mbox{$\Sigma_{\rm H_2}$}}

\newcommand{\coj}{\mbox{$^{12}$CO ($J=1\rightarrow0$)}}
\newcommand{\um}{\mbox{$\micron$}}
\newcommand{\ac}{\mbox{$\arcsec$}}
\newcommand{\Htwo}{\mbox{H$_2$}}
\newcommand{\rhohtwo}{\mbox{$\rho_{\rm H_2}$}}
\newcommand{\rhohi}{\mbox{$\rho_{\rm HI}$}}

\slugcomment{Accepted for publication in \aj}

\begin{document}

\title{The Interstellar Medium and Star Formation in Edge-On Galaxies. II. NGC  4157, 4565, and 5907}

\author{Kijeong Yim\altaffilmark{1,2}, Tony Wong\altaffilmark{1}, Rui Xue\altaffilmark{1}, Richard J. Rand\altaffilmark{3}, 
Erik Rosolowsky\altaffilmark{4}, \\
J. M. van der Hulst\altaffilmark{2}, Robert Benjamin\altaffilmark{5}, Eric J. Murphy\altaffilmark{6,7}}
\altaffiltext{1}{Department of Astronomy, University of Illinois, 
1002 West Green Street, Urbana, IL 61801, USA;  k.yim@astro.rug.nl}
\altaffiltext{2}{Kapteyn Astronomical Institute, University of Groningen, P.O. Box 800, 9700 AV Groningen, The Netherlands}
\altaffiltext{3}{Department of Physics and Astronomy, University of New Mexico, 1919 Lomas Blvd NE, Albuquerque, NM 87131-1156, USA}
\altaffiltext{4}{University of British Columbia Okanagan, 3333 University Way, Kelowna, BC V1V 1V7, Canada}
\altaffiltext{5}{Department of Physics, University of Wisconsin-Whitewater, 800 West Main Street, Whitewater, WI 53190, USA}
\altaffiltext{6}{Observatories of the Carnegie Institution for Science, 813 Santa Barbara Street, Pasadena, CA 91101, USA}
\altaffiltext{7}{Infrared Processing and Analysis Center, California Institute of Technology, MC 220-6, Pasadena CA, 91125, USA}

\begin{abstract}
We present a study of the vertical structure of the gaseous and stellar disks in a sample of edge-on galaxies (NGC 4157, 4565, and 5907) using BIMA/CARMA \coj, VLA \HI, and {\it Spitzer} 3.6 \um\ data. In order to take into account projection effects when we measure the disk thickness as a function of radius, we first obtain the inclination by modeling the radio data. Using the measurement of the disk thicknesses and the derived radial profiles of gas and stars, we estimate the corresponding volume densities and vertical velocity dispersions. Both stellar and gas disks have smoothly varying scale heights and velocity dispersions, contrary to assumptions of previous studies. Using the velocity dispersions, we find that the gravitational instability parameter $Q$ follows a fairly uniform profile with radius and is $\geq$ 1 across the star forming disk. The star formation law has a slope that is significantly different from those found in more face-on galaxy studies, both in deprojected and pixel-by-pixel plots.  
Midplane gas pressure based on the varying scale heights and velocity dispersions appears to roughly hold a power-law correlation with the midplane volume density ratio.

\end{abstract}


\section{Introduction}
\label{intro}

Our understanding of the relationship between star formation and properties of the interstellar medium (ISM), such as gas content and interstellar pressure, is still rudimentary. 
Observational studies (e.g., \citealt{1998ApJ...498..541K}; \citealt{2002ApJ...569..157W}; \citealt{2008AJ....136.2846B}) have derived a power-law relationship between the star formation rate and gas surface density (\sigsfr\ $\propto$ \siggas$^{n}$), a relationship referred to as the star formation law or Kennicutt-Schmidt law. 
Although the relationship is usually expressed in terms of surface density, fundamentally it is the volume density of the gas that should be related to star formation (\citealt{1959ApJ...129..243S}; \citealt{1998ApJ...506L..19F}; \citealt{2012ApJ...745...69K}).

In spite of the importance of the volume density, however, most star formation studies have used surface density in combination with an assumed constant value of gas scale height since the disk thickness is usually difficult to measure.
The disk thickness is, however, expected to increase with radius for a nearly  isothermal disk and such flaring of the gas disk has been measured in  recent observations of edge-on galaxies (e.g., \citealt{2010A&A...515A..62O};  \citealt{2011AJ....141...48Y}).
An edge-on galaxy provides the most direct measurement of disk thickness, which for the gas disk enables us to derive the volume density, a quantity that may be better correlated with star formation rate (SFR) than surface density. 

\citet{1993ApJ...411..170E} derived a theoretical relationship between the molecular gas fraction and the hydrostatic midplane pressure. 
In this model a high intercloud pressure enhances the formation of \Htwo. 
Therefore, the pressure is perhaps a critical ingredient for describing star formation, which has been demonstrated to proceed primarily from molecular gas.
More recent studies (e.g., \citealt{2002ApJ...569..157W}; \citealt{2004ApJ...612L..29B}) have established a strong empirical correlation between the molecular to atomic gas ratio and the hydrostatic midplane pressure. 
This pressure is inferred by assuming a constant gas velocity dispersion and stellar scale height. However, this assumption may not be valid in general. For example, recent studies (e.g., \citealt{2009AJ....137.4424T}; \citealt{2011AJ....141...48Y}) have found that the gas velocity dispersion changes with radius. Therefore, it is important to examine the role of the gas pressure in controlling the molecular to atomic gas ratio when radial variations of the velocity dispersions and the scale heights are taken into account. 
Again, edge-on galaxies provide an important laboratory for this study.

In a previous paper (\citealt{2011AJ....141...48Y}, hereafter Paper I), we studied the ISM and star formation in the prototypical edge-on galaxy NGC 891. In this second paper, we study the vertical structure of three additional edge-on galaxies (NGC 4157, 4565, and 5907) to investigate how the ISM properties are related to molecular cloud and star formation. We carry out this study using CO imaging from the Berkeley-Illinois-Maryland Association (BIMA) and the Combined Array for Research in Millimeter-wave Astronomy (CARMA), \HI\ imaging from the Very Large Array (VLA), and IR imaging from the {\it Spitzer Space Telescope}. We have selected these galaxies primarily based on (1) availability of multi-wavelength data including CO, HI, and IR (3.6 and 24 \um), (2) high inclination ($i > 80 \degr$), (3) presence of strong CO emission, and (4) active star formation as measured at 60 \um\ by IRAS. The properties of the three galaxies are shown in Table \ref{galprop}. Note that we do not use the inclinations shown in Table \ref{galprop}, but obtain inclinations by modeling in Section \ref{inc}.

\begin{table}[!bt]
\begin{center}
\caption{Galaxy Properties\label{galprop}}
\begin{tabular}{ccccccccc}
\tableline\tableline
Galaxy &R.A.& Decl. &$D_{25}$  &PA& Distance& Inclination &$V_{\rm sys}$&K\\
&\multicolumn{2}{c}{(J2000)}&($\arcmin$)&(\degr)&(Mpc)& (\degr)&(\kms)&mag\\
(1)&(2)&(3)&(4)&(5)&(6)&(7)&(8)&(9)\\
\tableline
NGC 4157&12 11 04.34&50 29 06.32&6.73&63&12.9&82&770&7.363 \\
NGC 4565&12 36 29.18&25 55 27.23&16.72&135&9.7&87.5&1230&6.060 \\
NGC 5907&15 15 53.55&56 19 43.00&12.60&115&11.0&86.5&670&6.757\\
\tableline
\end{tabular}
\tablecomments{Column (2) and (3) are the galaxy centers used in this study. Column (4) is the optical diameter in arcminutes and the references for NGC 4157, 4565, and 5907 are \cite{2001A&A...370..765V}, \cite{1991AJ....102...48R}, and \cite{2010AJ....140..753L}, respectively. Column (5) is the position angle obtained from the 3.6 \um\ image using the MIRIAD task IMFIT. Column (6) is the distance from the literature: \cite{1999AJ....117.2102I} for NGC 4157, \cite{2005A&A...432..475D} for NGC 4565, and  \cite{2006A&A...459..703J} for NGC 5907. 
Column (7) is the inclination from the literature: \cite{2001A&A...370..765V} for NGC 4157, \cite{2012ApJ...760...37Z} for NGC 4565, and \citealt{1997A&A...319..450G} for NGC 5907. 
Column (8) is the heliocentric systemic velocity adopted in this paper.  Column (9) is the K$_s$-band magnitude from the SIMBAD database.}
\end{center}
\end{table}

The organization of this paper is as follows.  We summarize the observations and the reduction of CO, \HI, and IR data in Section \ref{obs} and show the radial distribution of \sightwo, \sighi, \sigstar, and \sigsfr\ in Section \ref{radial}. We measure the inclinations of the less edge-on galaxies (NGC 4157, 4565, and 5907) and scale heights of the gas and stars and derive the vertical velocity dispersions in Section \ref{vertical}. We examine the Kennicutt-Schmidt law and the role of the instability parameter $Q$ using our sample of galaxies in Section \ref{sflaw}.  We investigate and discuss the relationship between the molecular to atomic gas ratio and the interstellar gas pressure when the scale heights and the velocity dispersions vary with radius in Section \ref{rmol}. Finally, our results are summarized in Section \ref{sum}.

\section{Observations and Data Reduction}
\label{obs}

\subsection{CO Observations}
The BIMA \coj\ observations were taken in 2004 toward the edge-on galaxies NGC 4157 (in B, C, and D conﬁgurations), NGC 4565 (C and D), and NGC 5907 (C and D) with a 5 field mosaic oriented along the major axis with a 50\ac\ separation between the primary beam centers. In 2010 and 2011, we observed NGC 4565 and 5907 using CARMA (in C and D configurations) in order to obtain higher angular resolution. The CARMA observations are mosaics of 11 pointings toward NGC 4565 and 13 pointings toward NGC 5907 with a half beam spacing of 30\ac.
The median baseline lengths are roughly 278 m (B), 93 m (C), and 40 m (D) for BIMA and 278 m (C) and 111 m (D) for CARMA.   
The instruments and configurations used are summarized in Table \ref{obstable}. 
All the CO data were calibrated and imaged using the MIRIAD package. 
The calibration for each track involved use of MFCAL for bandpass and gain calibration and the BOOTFLUX task for flux calibration.
For NGC 4565 and 5907, we combined the separately calibrated BIMA and CARMA data  using the MIRIAD task INVERT. All the CO images were produced using natural weighting to achieve the highest possible sensitivity. 

\begin{figure}[!tbp]
\begin{center}
\includegraphics[width=0.6\textwidth,angle=270]{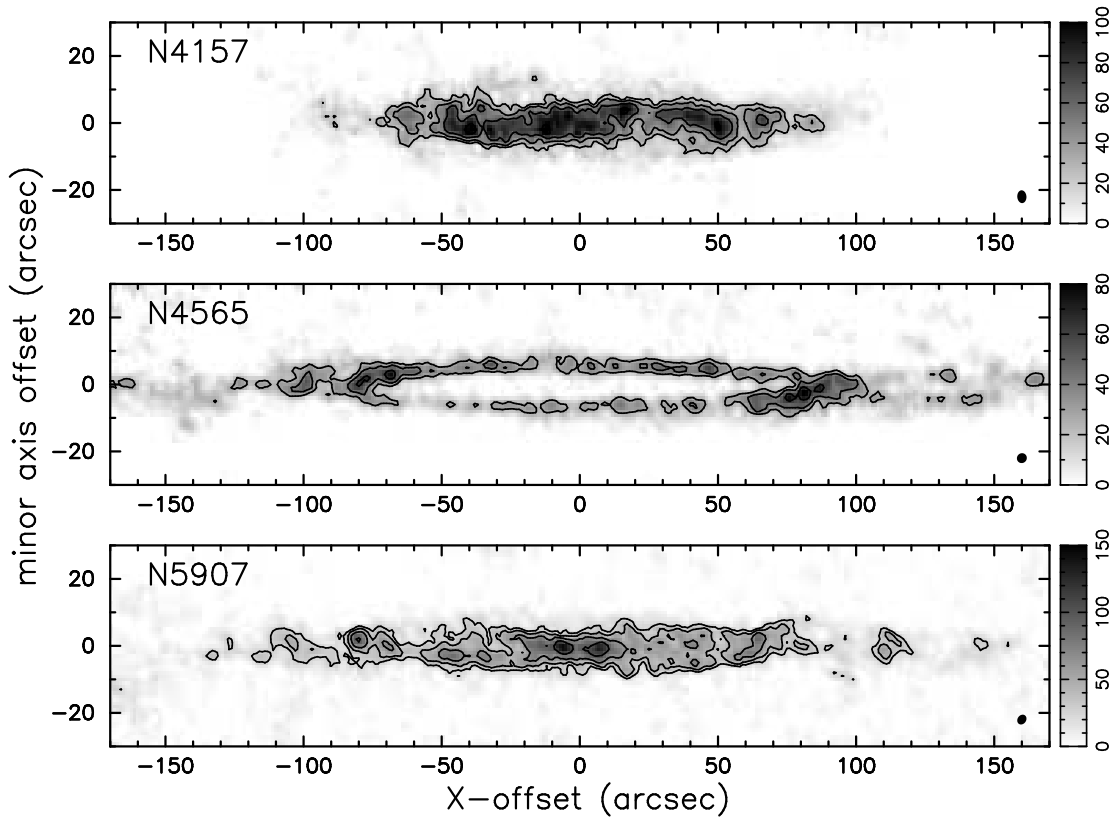}
\caption[CO integrated intensity maps]{CO integrated intensity maps of NGC 4157 ($top$), NGC 4565 ($middle$), and NGC 5907 ($bottom$). Contour levels are $26 \times 1.4^n$ K \kms, with n=0, 1, 2, 3. The synthesized beam is shown in the lower right corner of each box.
\label{comaps}}
\end{center}
\end{figure}

The CO integrated intensity maps of NGC 4157, 4565, and 5907 are shown in Figure \ref{comaps}. 
Throughout this paper, we use $x$ for the coordinate along the major axis (line of nodes), $y$ for the coordinate perpendicular to $x$ in the galaxy plane, and $z$ for the coordinate perpendicular to the galaxy plane. The coordinate along the apparent minor axis is actually a combination of $y$ and $z$ (if not exactly edge-on), so we refer to it as the ``minor axis offset". 
The angular and velocity resolutions, total CO flux in Jy \kms, and the RMS noise per channel in K are presented in Table \ref{obstable}. For the total flux, we  used a masked integrated intensity map derived from a gain corrected image cube, limited to the region where the map noise is less than twice its minimum (central) value.  The masked map is obtained by blanking regions below a 3$\sigma$ threshold in a smoothed cube.
We used a noise-flattened map convolved to 10\ac\ for the smoothed cube, where the noise-flattened map is the original cube divided by the normalized sensitivity  image. The average RMS noise is obtained from line-free edge channels of the gain corrected cube. 
Each map is rotated to place the blue-shifted side of the disk (lower velocities than $V_{\rm sys}$) on the negative $x$-axis. 

\begin{table}[!bt]\tiny
\begin{center}
\caption{CO and \HI\ Observing Parameters\label{obstable}}
\begin{tabular}{ccccccc}
\tableline\tableline
Galaxy&\multicolumn{2}{c}{NGC 4157}&\multicolumn{2}{c}{    NGC 4565}&\multicolumn{2}{c}{  NGC 5907}\\
& CO&HI&CO&HI&CO&HI\\
\tableline
Telescope&BIMA&VLA&BIMA \& CARMA&VLA&BIMA \& CARMA&VLA\\
Array&BCD&CD&CD \& CD&BCD&CD \& CD&CD\\
$\theta$ (\ac)&3.84$\times$3.30&15.69$\times$14.87&3.6$\times$2.95&6.26$\times$5.59&3.48$\times$2.76&15.43$\times$13.84\\
$\Delta v$ (\kms)&10&20&10&20&10&20\\
Total Flux (Jy \kms)&1280&110&1030&274&1530&259\\
Channel Noise (K)&0.23&0.67&0.16&2.73&0.15&0.81\\
Velocity Range (\kms)&550--990&550--990&960--1490&970--1490&420--920&410--910\\
Circular Beam&4\ac\ (250 pc)&15.7\ac\ (981 pc)&3.7\ac\ (174 pc)& 6.3\ac\ (296 pc)&3.6\ac\ (190 pc)&15.5\ac\ (821 pc)\\
\tableline
\end{tabular}
\end{center}
\tablecomments{$\theta$ is the angular resolution. $\Delta v$ is the velocity resolution. The original beam has been convolved to a circular beam for the analysis in Section \ref{inc}. }
\end{table}


\subsection{\HI\ Observations}
\label{h1obs}
We have obtained \HI\ data of the galaxies from the NRAO VLA archive. 
The array configurations are C and D for NGC 4157 and 5907 and B, C, and D for NGC 4565. 
A data reduction pipeline implemented in the CASA (Common Astronomy Software Applications) software package (\citealt{2007ASPC..376..127M}) was built for reprocessing the whole dataset in a uniform procedure, which we describe below.

For the data taken from each individual track, flagging of shadowed data was automatically performed first. Edge channels in each spectral window were also automatically flagged, if a test bandpass calibration showed their solution amplitudes falling below 0.75. Any additional bad visibilities from either interference or instrumental problems were located and examined using tasks PLOTMS and VIEWER, and flagged using FLAGCMD with mode = `manualflag' or `quack'. Then the flux calibrator visibility model was set using SETJY, and
we computed the bandpass and flux-scaled gain tables from the primary and secondary calibrator data, using tasks BANDPASS, GAINCAL, and FLUXSCALE. Those solution tables were later applied to the data with the task APPLYCAL.
 
For the calibrated visibilities of each track, we usually performed a continuum subtraction in  uv-space using task UVCONTSUB. A first order polynomial fit was used to estimate the continuum emission from line-free channels. Then the data from different tracks were combined using task CONCAT, and we imaged and cleaned the spectral data cubes using the CASA task CLEAN, with the ROBUST weighting algorithm (the ROBUST parameter was set to R = 0.5). The clean depth went to a brightness level of 2.5 times the RMS derived from the line free channels of a dirty cube.

\begin{figure}[!tbp]
\begin{center}
\includegraphics[width=0.6\textwidth,angle=270]{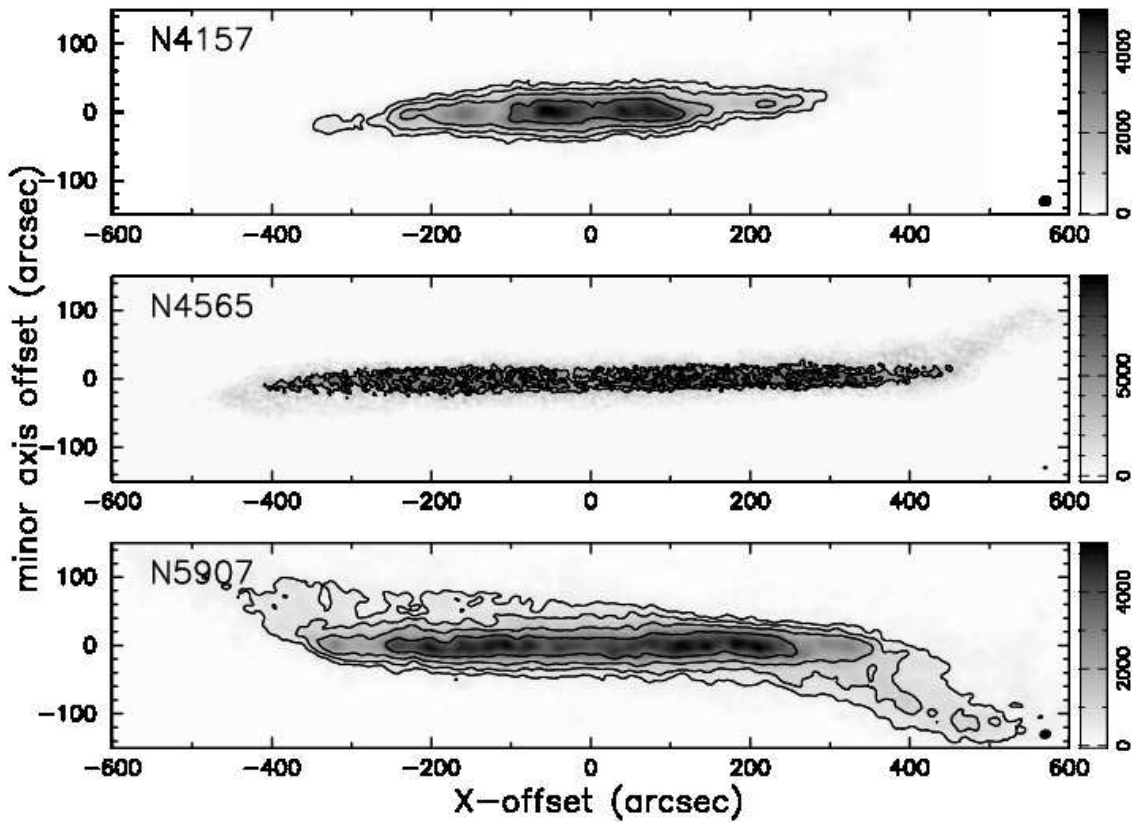}
\caption[\HI\ integrated intensity maps]{\HI\ integrated intensity maps of NGC 4157 ($top$), NGC 4565 ($middle$), and NGC 5907 ($bottom$). Contour levels are $460 \times 1.9^n$ K \kms, with n=0, 1, 2, 3 for NGC 4157 and NGC 5907 and $3400 \times 1.4^n$ K \kms, with n=0, 1, 2, 3 for NGC 4565.  The synthesized beam is shown in the lower right corner of each box. 
\label{h1maps}}
\end{center}
\end{figure}

Figure \ref{h1maps} shows the \HI\ integrated intensity maps of NGC 4157, 4565, and 5907. The observing parameters including the angular and velocity resolutions are shown in Table \ref{obstable}. 
The total flux in units of Jy \kms\ has been obtained from integration of  the masked cube, where the mask was constructed from the 3$\sigma$ contour of a cube smoothed to 20\ac. The average RMS noise is measured in the line-free edge channels of the original cube. 
The strong warp of NGC 5907 noted in previous studies (e.g., \citealt{1976A&A....53..159S}) is evident.  
In addition, slight HI warps are shown in the other two galaxies (NGC 4157 and 4565).  
However, we do not see any evidence for warps in non-\HI\ data for all three galaxies.  

\subsection{IR Observations}
The IR data (IRAC 3.6 \um\ and MIPS 24 \um) were obtained from the {\it Spitzer} archive: Program ID 69 (PI: G. Fazio) for both 3.6 \um\ and 24 \um\ images of NGC 4157,  Program ID 3 (PI: G. Fazio) for 3.6\um\ images of NGC 4565 and 5907, and Program ID 20268 (PI: R. de Jong) for 24 \um\ images of NGC 4565 and 5907. We have downloaded the basic calibrated data (BCD) and used MOPEX (Mosaicking and Point Source Extraction) to remove background variations in the BCD images and mosaic the images. In the case of 24 \um, we have removed a residual flat field in the BCD data before background matching and mosaicking. 
In addition, we have downloaded 3.6 \um\ images along with the supplied masks from the S$^4$G archive \citep{2010PASP..122.1397S} and masked out the stars on the images for stellar modeling in Section \ref{stmod}.

\begin{figure}[!tbp]
\begin{center}
\includegraphics[width=0.6\textwidth,angle=270]{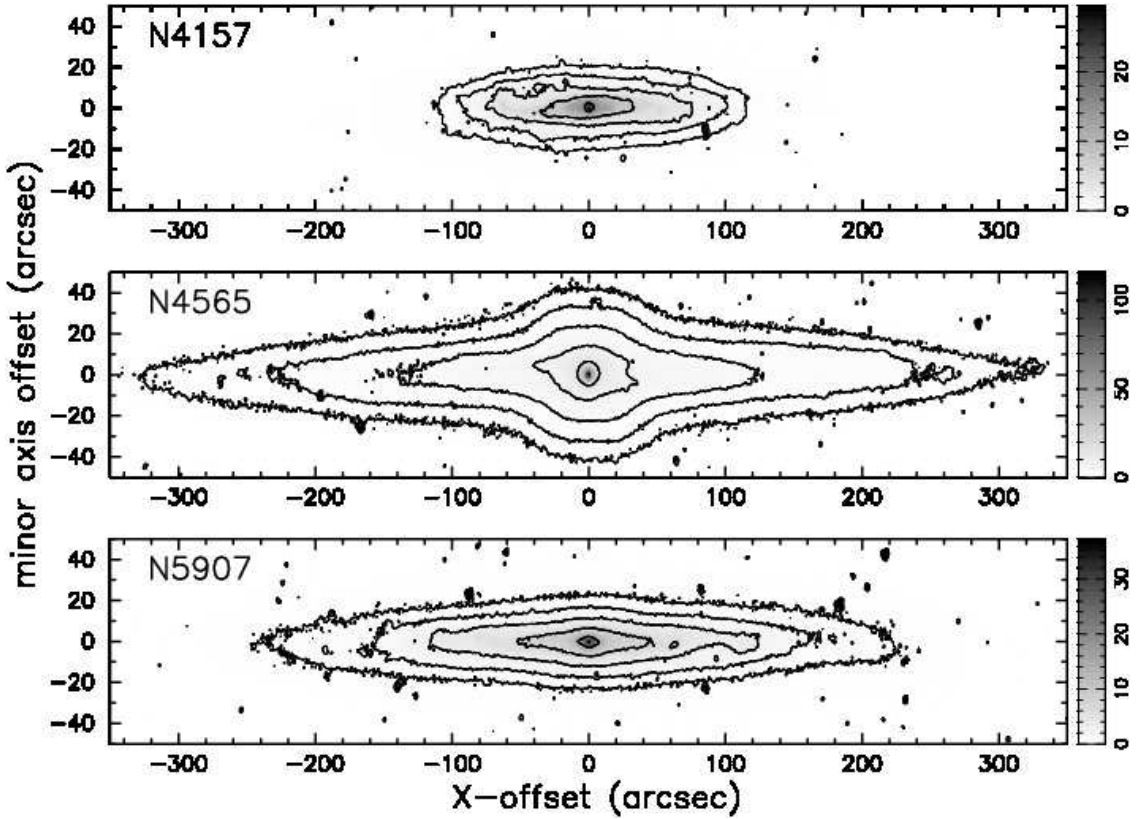}
\caption[{\it Spitzer} 3.6 \um\ Images]{{\it Spitzer} 3.6 \um\ images of NGC 4157 ($top$), NGC 4565 ($middle$), and NGC 5907 ($bottom$). Contour levels are $0.4 \times 2.6^n$ MJy sr$^{-1}$, with n=0, 1, 2, 3, 4.
\label{3p6maps}}
\end{center}
\end{figure}

\begin{figure}[!tbp]
\begin{center}
\includegraphics[width=0.6\textwidth,angle=270]{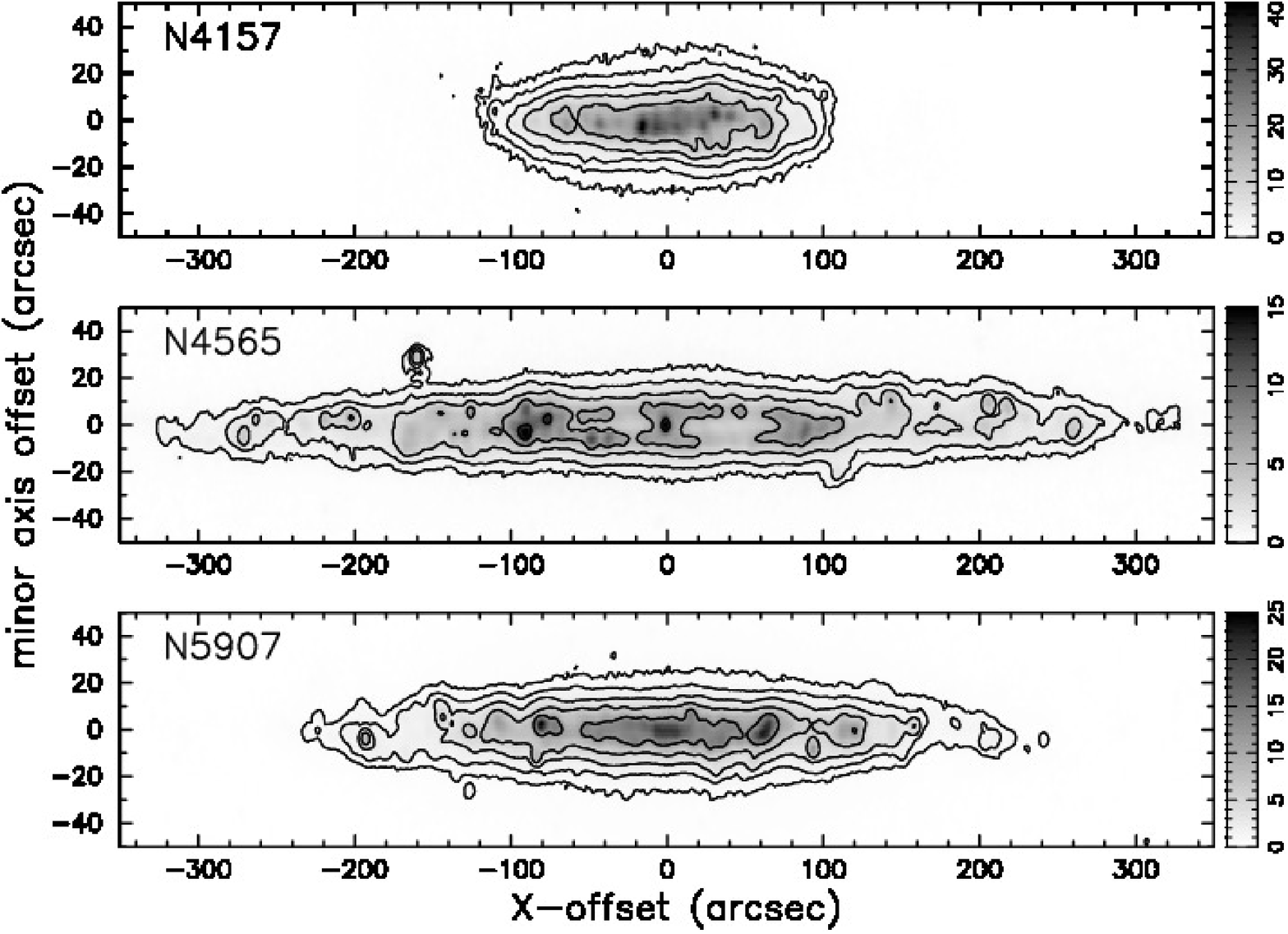}
\caption[{\it Spitzer} 24 \um\ Images]{{\it Spitzer} 24 \um\ images of NGC 4157 ($top$), NGC 4565 ($middle$), and NGC 5907 ($bottom$). Contour levels are $0.6 \times 2.1^n$ MJy sr$^{-1}$, with n=0, 1, 2, 3, 4.
\label{24maps}}
\end{center}
\end{figure}

The reduced 3.6 \um\ and 24 \um\ images of NGC 4157, 4565, and 5907 are shown in Figure \ref{3p6maps} and Figure \ref{24maps}, respectively. The ring seen in the CO map of NGC 4565 is clearly shown in the 24 \um\ image.  
In the 3.6 \um\ images, bright stars located nearby or projected against the galaxies  are blanked and replaced by values from neighboring pixels using the Groningen Image Processing System (GIPSY; \citealt{1992ASPC...25..131V}) tasks BLOT and PATCH before deriving their radial distributions. 



\section{Radial Surface Density Distributions}
\label{radial}

\subsection{Molecular and Atomic Gas}
\label{gasprofile}

\begin{figure}[!tbp]
\begin{center}
\includegraphics[width=0.7\textwidth,angle=270]{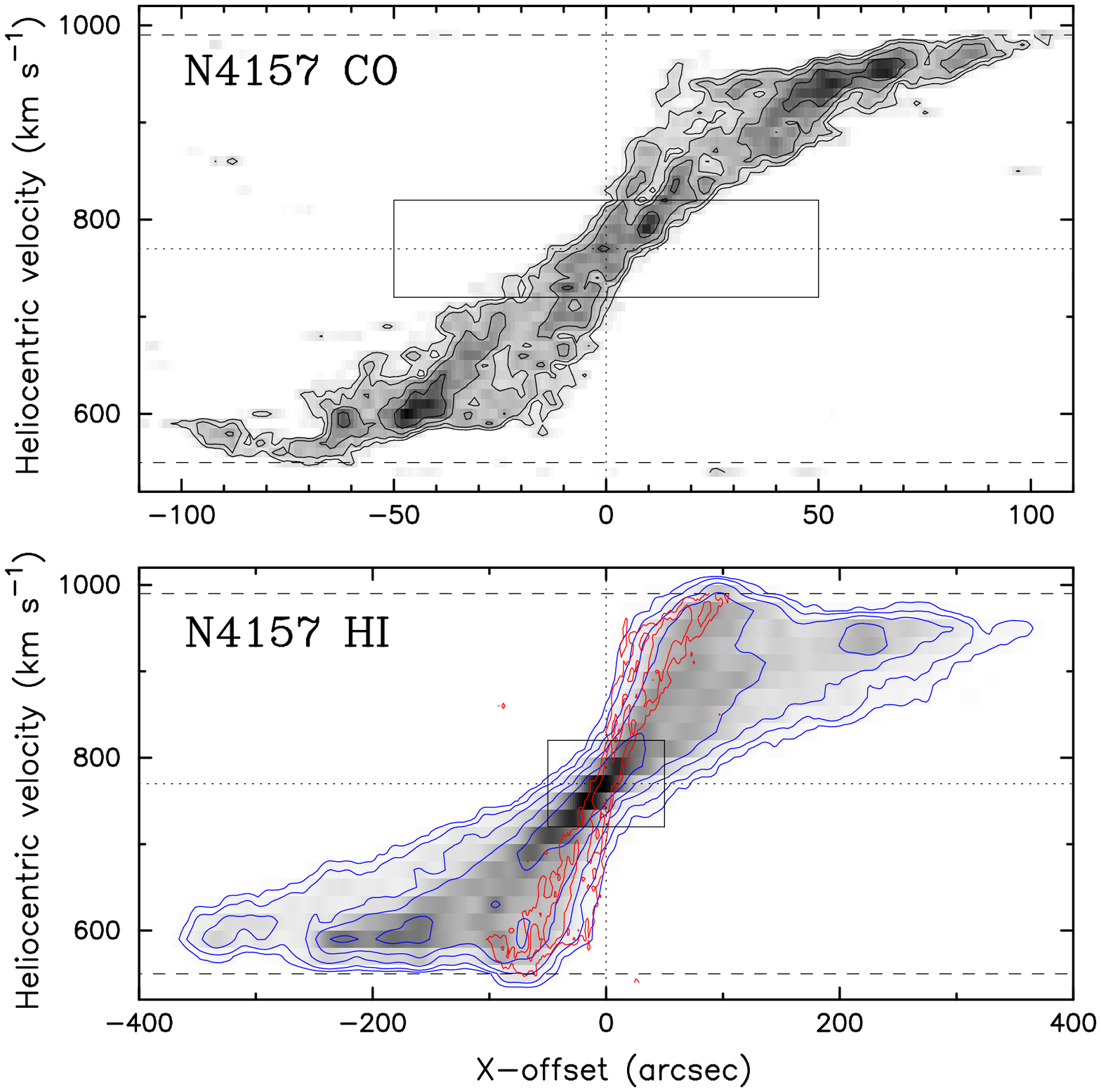}
\caption[Vertically integrated position-velocity diagram of NGC 4157]{Vertically integrated position-velocity diagrams of NGC 4157 CO ($top$) and \HI\ ($bottom$). CO contours are $3.0 \times 1.7^n$ K arcsec, with n=0, 1, 2, 3. 
\HI\ contour levels are $93.0 \times 2.1^n$ K arcsec, with n=0, 1, 2, 3, 4. 
CO contours (red) are overlaid on \HI\ contours (blue). 
The dashed lines show the assumed circular speed (220 \kms) used to derive the radial profile using the PVD method. The box represents the excluded region in the PVD method. 
\label{pv1}}
\end{center}
\end{figure}

\begin{figure}[!tbp]
\begin{center}
\includegraphics[width=0.7\textwidth,angle=270]{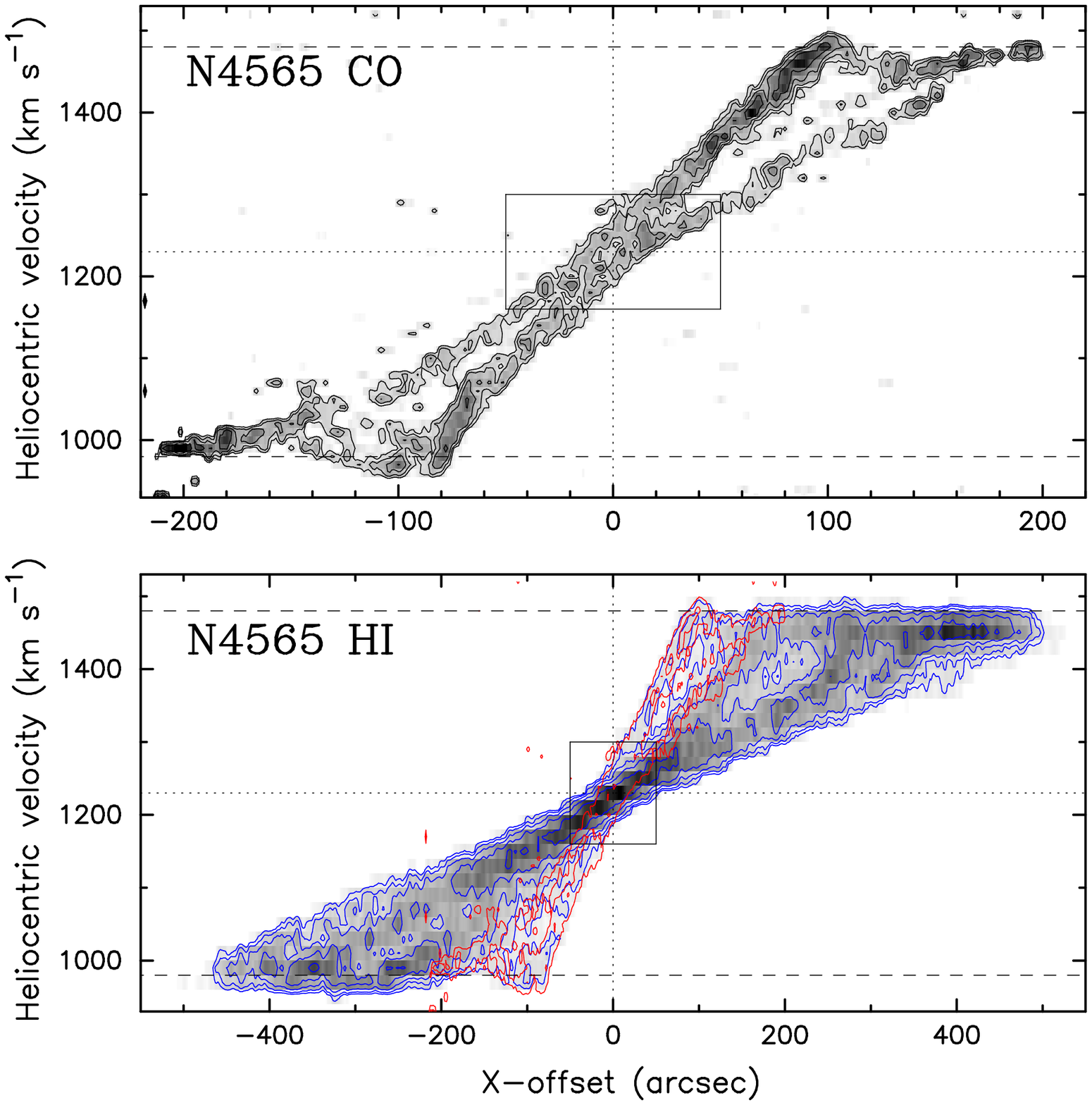}
\caption[Vertically integrated position-velocity diagram of NGC 4565]{Vertically integrated position-velocity diagrams of NGC 4565 CO ($top$) and \HI\ ($bottom$). 
CO contours are $2.2 \times 1.7^n$ K arcsec, with n=0, 1, 2, 3. 
\HI\ contour levels are $450.0 \times 1.6^n$ K arcsec, with n=0, 1, 2, 3, 4. 
CO contours (red) are overlaid on \HI\ contours (blue). 
The dashed lines show the assumed circular speed (250 \kms) used to derive the radial profile using the PVD method. The box represents the excluded region in the PVD method. 
\label{pv2}}
\end{center}
\end{figure}

\begin{figure}[!tbp]
\begin{center}
\includegraphics[width=0.7\textwidth,angle=270]{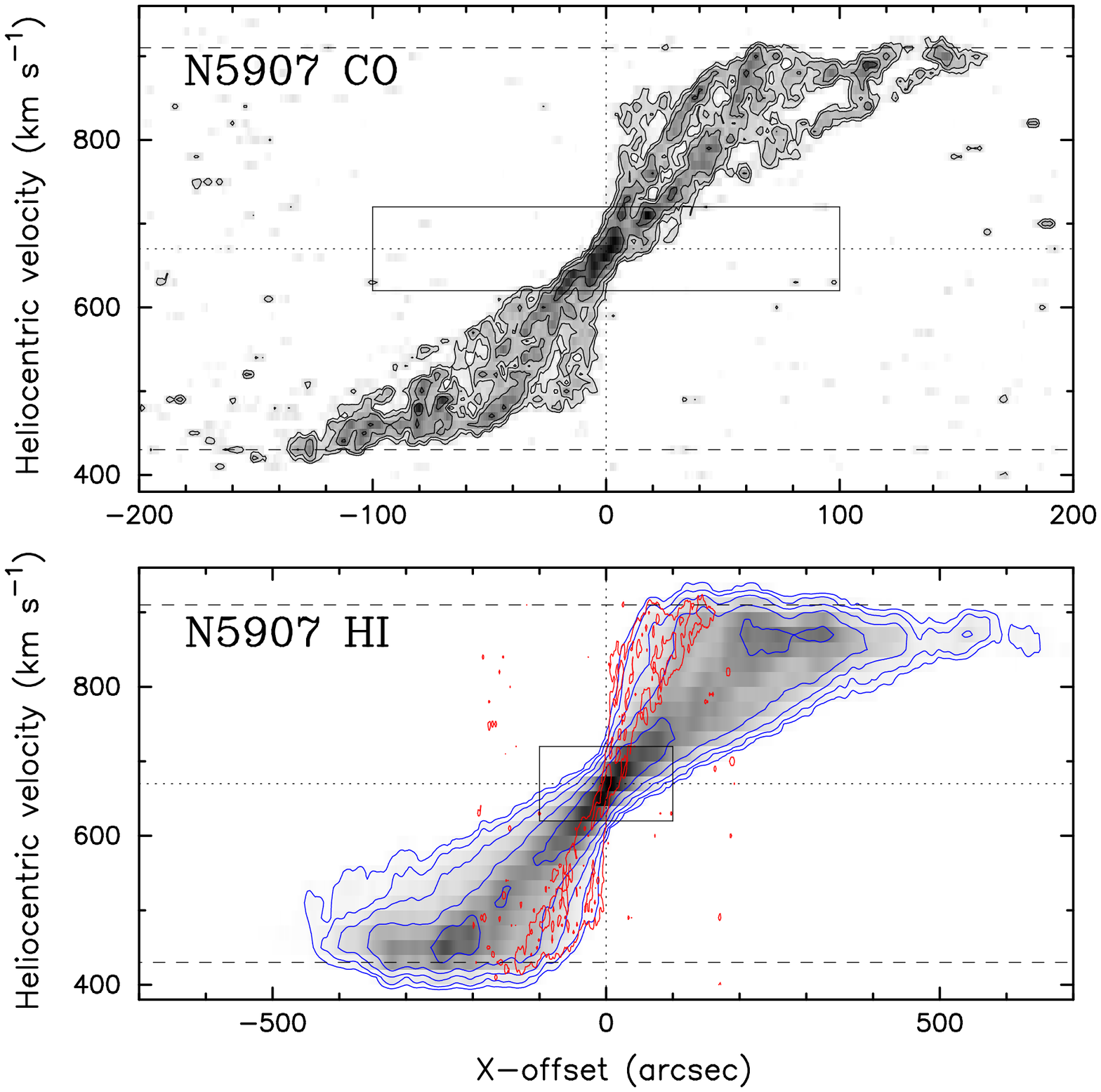}
\caption[Vertically integrated position-velocity diagram of NGC 5907]{Vertically integrated position-velocity diagrams of NGC 5907 CO ($top$) and \HI\ ($bottom$). CO contours are $2.2 \times 1.7^n$ K arcsec, with n=0, 1, 2, 3. 
\HI\ contour levels are $96.0 \times 2.1^n$ K arcsec, with n=0, 1, 2, 3, 4. 
CO contours (red) are overlaid on \HI\ contours (blue). 
The dashed lines show the assumed circular speed (240 \kms) used to derive the radial profile using the PVD method. The box represents the excluded region in the PVD method. 
\label{pv3}}
\end{center}
\end{figure}

\begin{figure}[!tbp]
\begin{center}
\includegraphics[width=0.5\textwidth]{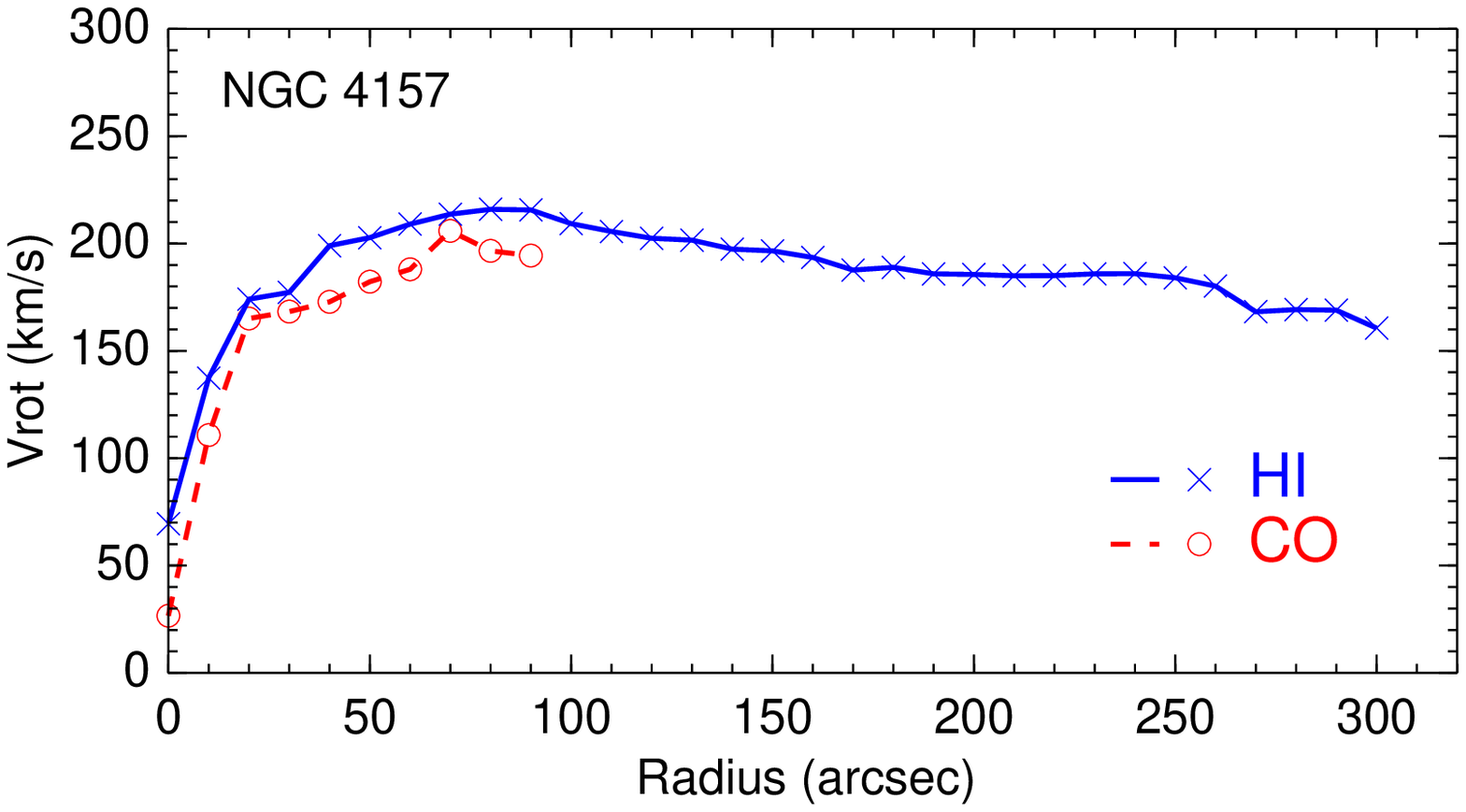}\\
\includegraphics[width=0.5\textwidth]{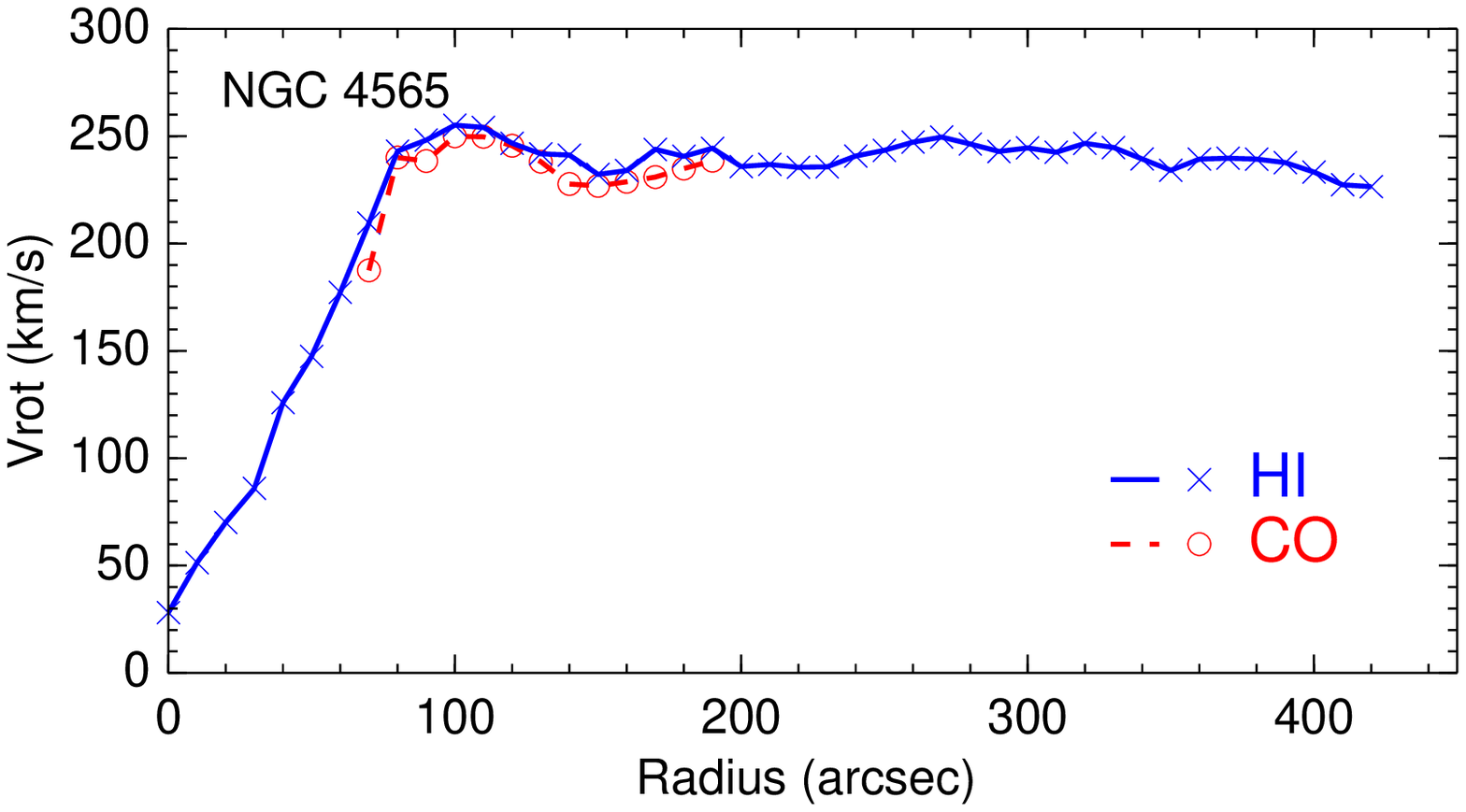}\\
\includegraphics[width=0.5\textwidth]{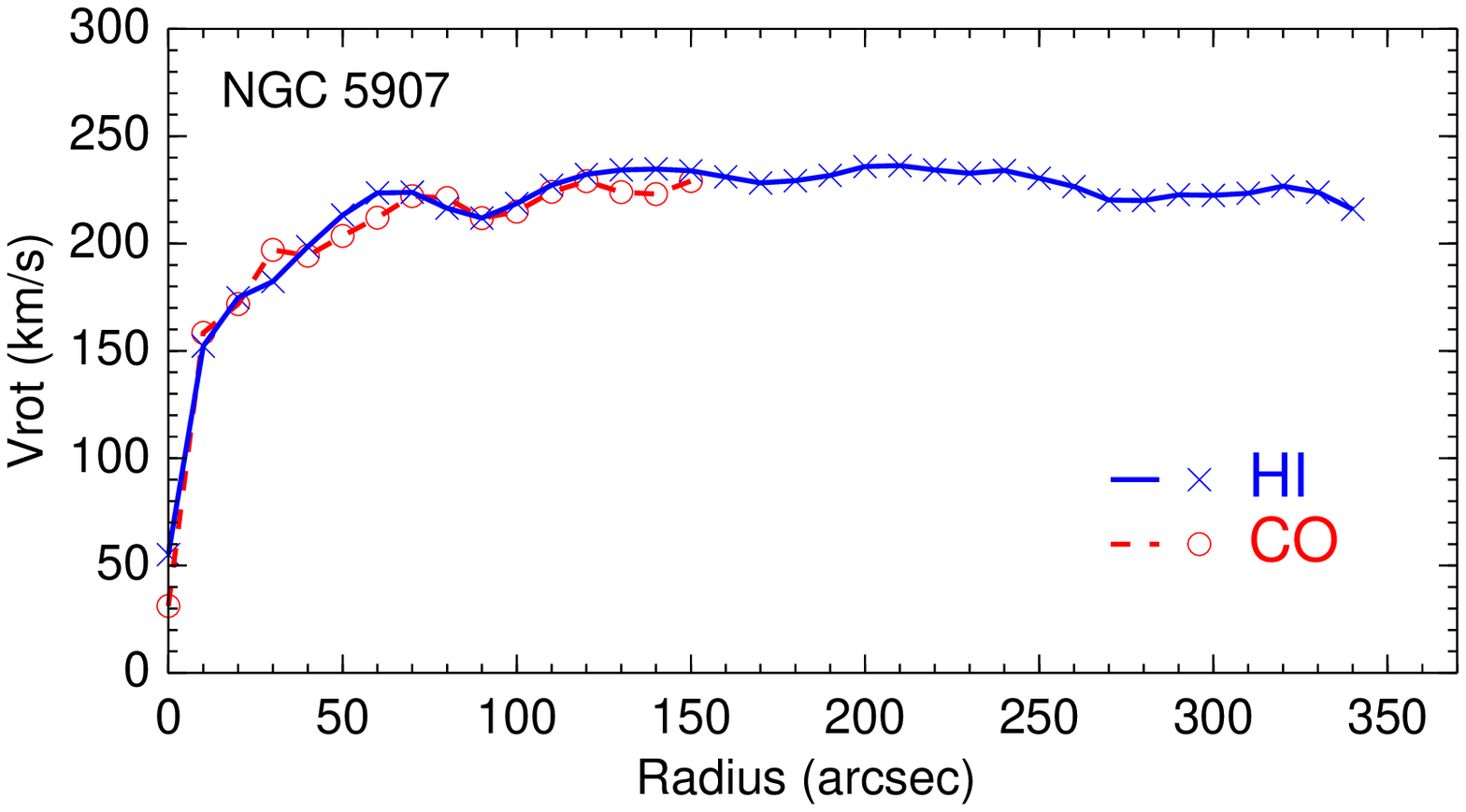}
\caption[Rotation curve]{Rotation curves obtained from position-velocity diagrams along the midplane. The red circles show the CO rotation curve  and blue crosses represent the  \HI\ rotation velocity. 
\label{rot}}
\end{center}
\end{figure}

\begin{figure}[!tbp]
\begin{center}
\begin{tabular}{c@{\hspace{0.1in}}c@{\hspace{0.1in}}c}
\includegraphics[width=0.32\textwidth]{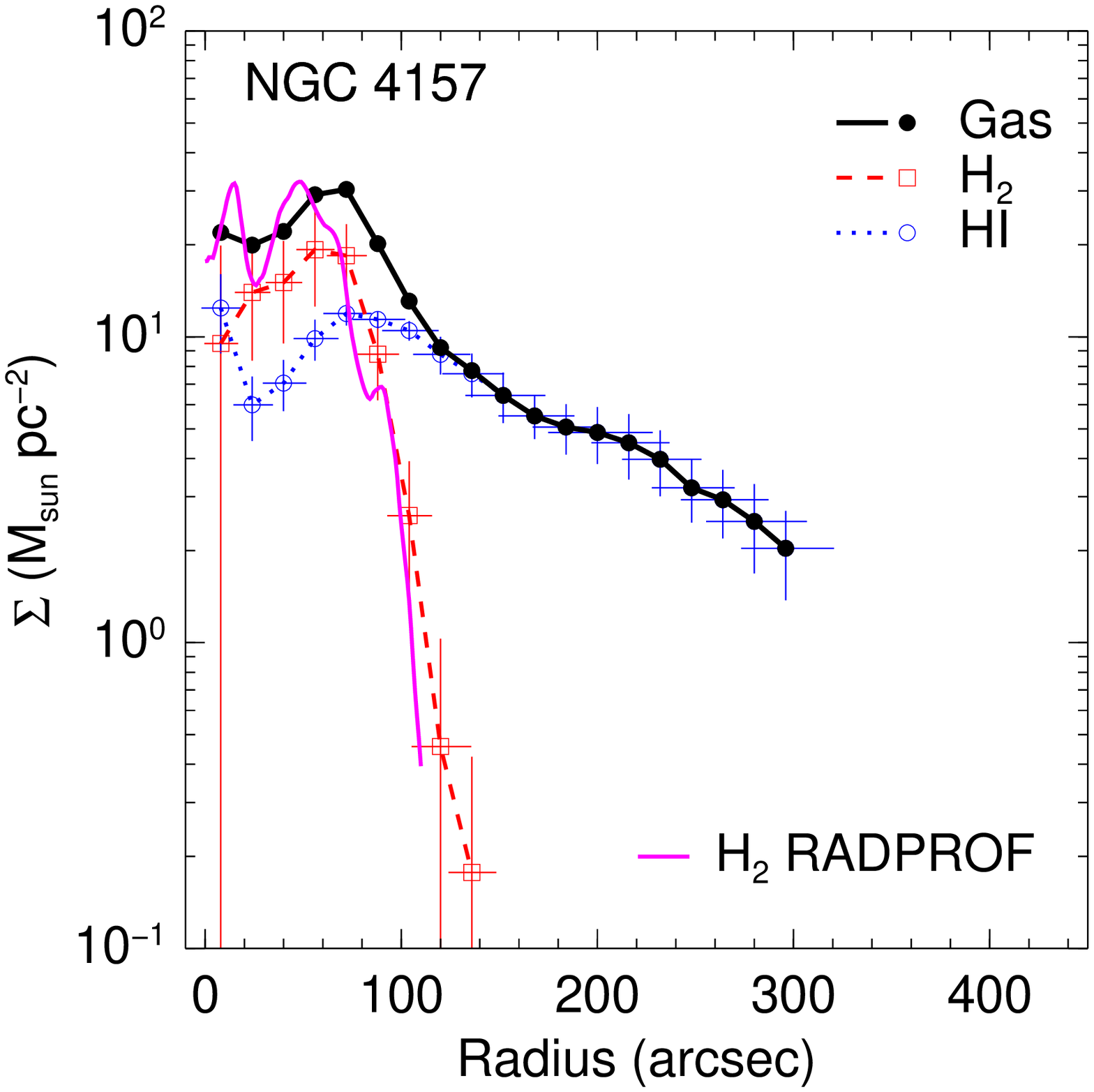}&
\includegraphics[width=0.32\textwidth]{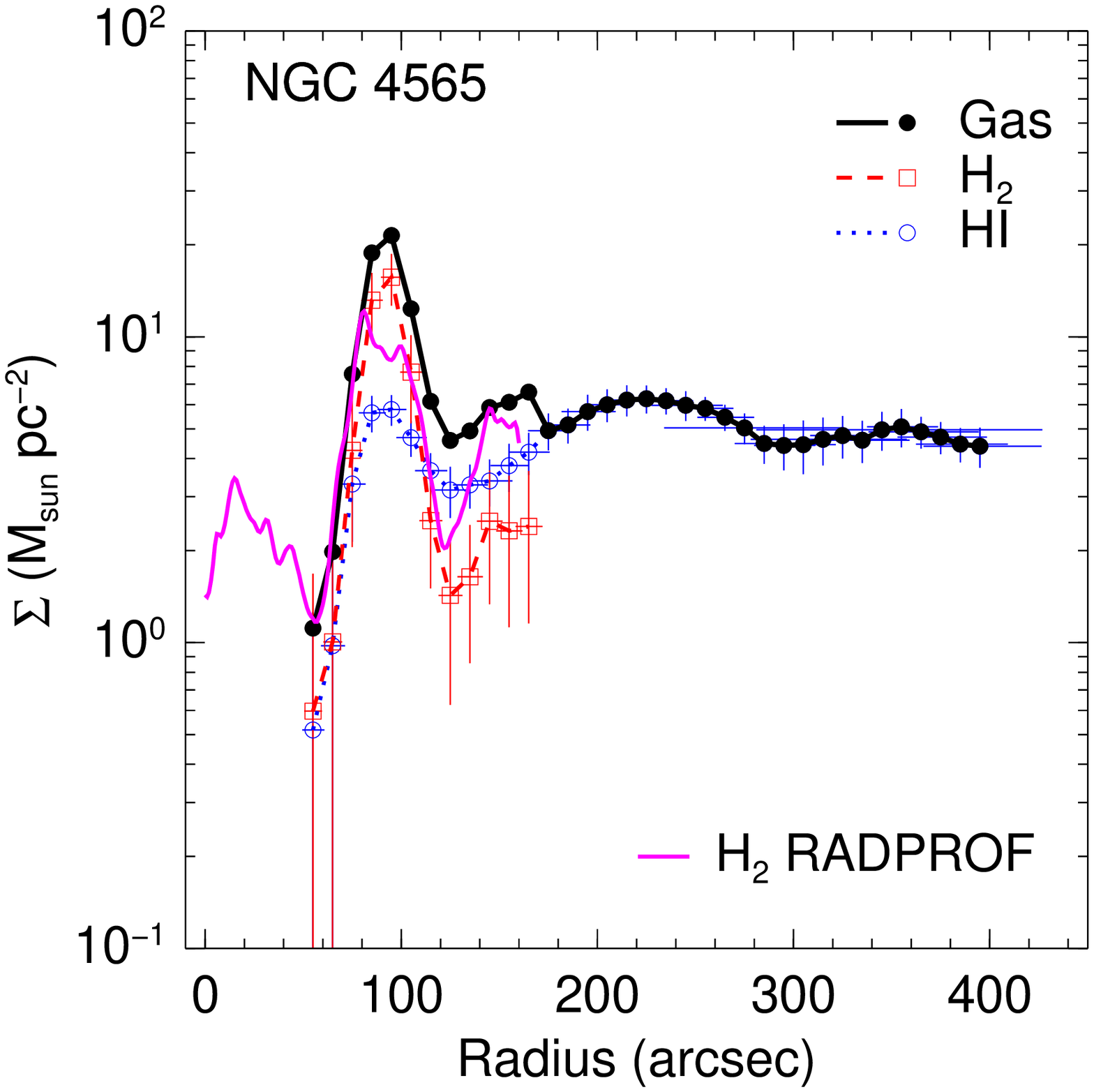}&
\includegraphics[width=0.32\textwidth]{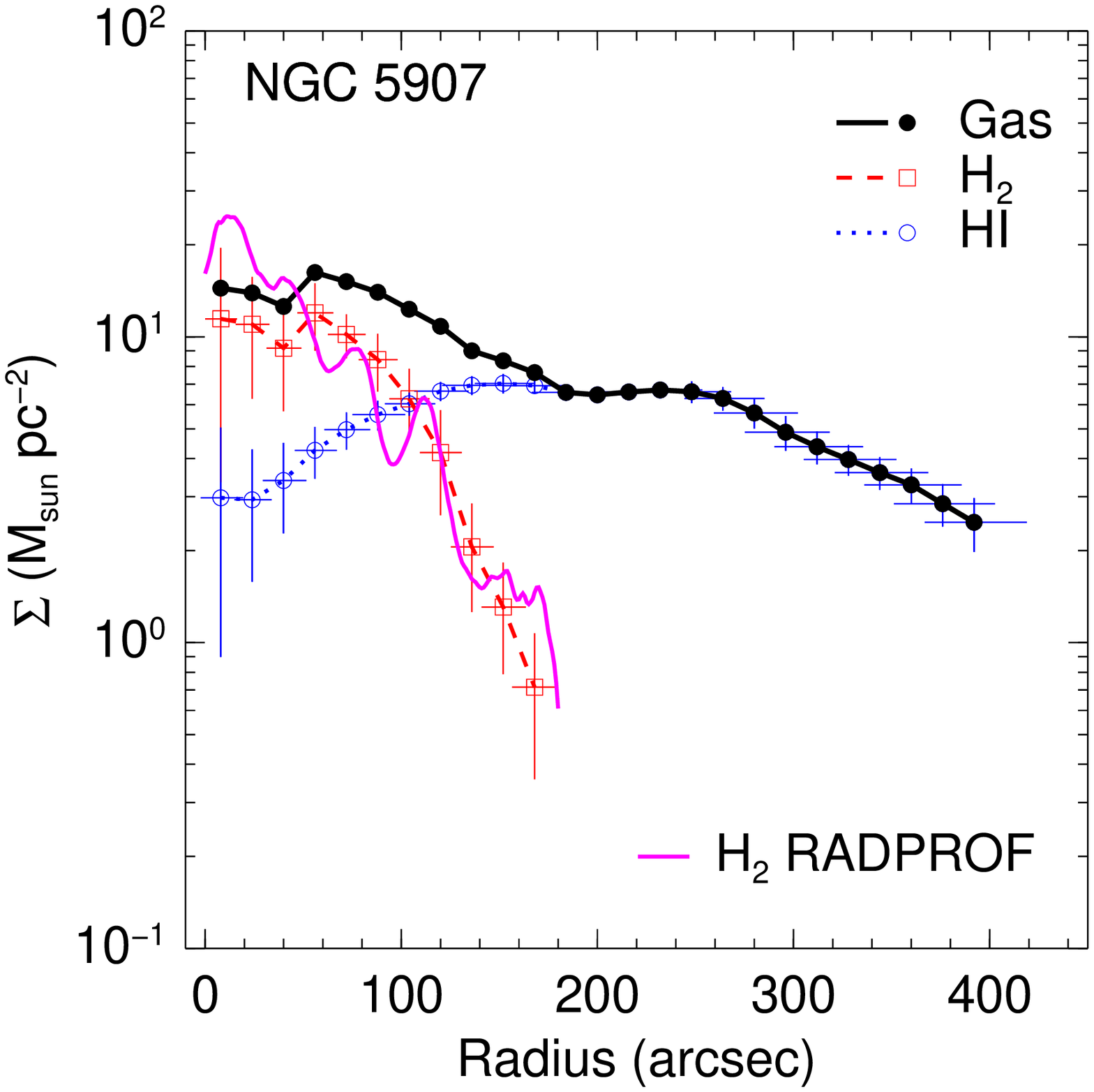}
\end{tabular}
\caption[Radial profiles of \Htwo, \HI, and total gas]{Radial distributions of \Htwo\ (red open squares), \HI\ (blue open circles), and total gas (solid circles) surface density for NGC 4157 ($left$), NGC 4565 ($middle$), and NGC 5907 ($right$). A factor of 1.36 is included in the surface densities for helium.  The magenta line is the \Htwo\ radial profile obtained from RADPROF.
\label{radiprof}}
\end{center}
\end{figure}

In order to derive the radial profile of gas surface density, we employ the position-velocity diagram (PVD) method from Paper I using vertically integrated  position-velocity (p-v) diagrams (Figure \ref{pv1}, \ref{pv2}, and \ref{pv3}) and assuming circular rotation with a flat rotation curve (ignoring the rapidly rising velocity in the central region).  The vertical integration for the p-v diagram extends between 10\ac\ above and below the midplane for CO and $\pm$ 50\ac\ for \HI.
The central region in the p-v diagrams (defined as a solid box in the figures), where contributions from many different radii overlap, is excluded when deriving the radial profile. At each pixel in the position-velocity diagram, the radius is obtained using the position ($x$), the assumed circular speed ($V_{\rm c}$) and the radial velocity ($V_R$):
\begin{equation}
R =  V_{\rm c} \left< \frac{x}{V_{R}-V_{\rm sys}} \right> \quad\quad \rm{with}\ \it |V_{R}-V_{\rm sys}| < V_{\rm c},
\label{rad}
\end{equation}
where the angle brackets indicate the mean value of $x/(V_{R} - V_{\rm sys})$ within the pixel (see Paper I for more details) and the assumed circular speed is 220, 250, 240 \kms\ (indicated as a dashed line in Fig. \ref{pv1}--\ref{pv3}) for NGC 4157, 4565, and 5907, respectively. The circular speed is obtained based on the maximum value of the rotation curve (Figure \ref{rot}) that we derived using the envelope tracing method (\citealt{1996ApJ...458..120S}), which is based on a position-velocity diagram along the midplane. In this method, the projected rotation curve ($V_{\rm proj}$) is derived from the terminal velocity  ($V_{\rm t}$) at each $x$ offset, considering the observational velocity resolution ($\sigma_{\rm obs}$) and the gas velocity dispersion ($\sigma_{\rm g}$): 
\begin{equation}
V_{\rm proj} = V_{\rm t} - \sqrt{\sigma^2_{\rm obs} + \sigma^2_{\rm g}},
\label{Vrot}
\end{equation}
where the terminal velocity is selected as the highest velocity where the intensity exceeds the 3$\sigma$ level. 
The velocity resolutions for our observations are 10 \kms\ for CO data and 20 \kms\ for HI data and the assumed velocity dispersion is 8 \kms. 
Based on the correction term with $\sigma_{\rm obs}$ and $\sigma_{\rm g}$, we adopt an uncertainty for the rotation curve of 13 \kms\ for CO and  $\sim$22 \kms\ for HI. 
The derived surface brightnesses of CO and \HI\ by the PVD method (more details in Section 4.1 of  Paper I) are converted to surface mass densities using the conversion factors shown in Equation (\ref{xco}) for CO (\citealt{1996A&A...308L..21S}; \citealt{2001ApJ...547..792D}) and Equation (\ref{xh1}) for \HI:
\begin{equation}
N(\textrm H_2) \,\rm [cm^{-2}] = 2 \times 10^{20} \,\it I_{\rm CO} \,[\rm K \,\kms].
\label{xco}
\end{equation} 
\begin{equation}
N(\textrm{\HI}) \,\rm [cm^{-2}] = 1.82 \times 10^{18} \,\it I_{\rm HI} \,[\rm K \, \kms].
\label{xh1}
\end{equation}
The derived radial profiles of \sighi, \sightwo, and the total gas (\sighi\ + \sightwo) are shown in Figure \ref{radiprof}. 
The average value of both sides of the disk is used as the surface density profile in the figure and a factor of 1.36 for helium is included in the surface densities. 
For comparison between the molecular and atomic gas and to combine the two components for the total gas, the CO map has been convolved to the resolution of the \HI\ map. Vertical error bars show the standard deviation of the average value (of all data points) in a bin and horizontal error bars represent an uncertainty in the radius, showing obtainable maximum and minimum radii derived from varying $x$ and $V_{R}$ by the angular and the velocity resolutions. 
Some data points in the inner region of NGC 4565 are omitted since they are  below the 3$\sigma$ detection threshold, which is obtained using the RMS noise in the p-v diagram. 
NGC 4565 shows a ring-like morphology as noticed in the CO map; NGC 4157 and 5907 exhibit small, centrally concentrated CO disks. 

For comparison purposes, we also plot the radial profile of \sightwo\ obtained from the GIPSY task RADPROF (see Section \ref{radprof} for details) as a magenta line  in Figure \ref{radiprof}. The RADPROF profile is in good agreement with the PVD profile of \sightwo, although NGC 4565 exhibits differences between them in the central region, where CO emission is deficient and the uncertainties of the PVD profile are large.
As discussed in Paper I, RADPROF is also subject to larger errors in the central region due to emission from outer radii seen in projection.
In addition, it is unable to reproduce sharp gradients in the radial profile, such as those caused by gaps or rings. 

\subsection{Stars and Star Formation Rate}
\label{radprof}
We have used the RADPROF task to obtain radial distributions of the stellar and SFR surface densities since the IR data lack velocity information. The RADPROF task uses an Abel inversion technique which assumes axisymmetry and requires a strip integral for each side of a galaxy, along with position angle, inclination (the values in Table \ref{galprop} are used), beam size, and an initial function for the radial profile  which we assume to be exponentially decreasing for the 3.6 \um\ and a Gaussian function  for the 24 \um\ data. 
Before employing the RADPROF task, all maps are smoothed to the lowest resolution among the dataset.   For better accuracy in comparison between SFR and gas, the 24 \um\ maps were first convolved to a Gaussian beam of 7\ac\ using the IDL code and the kernel provided by \citet{2011PASP..123.1218A} and then convolved to the HI beam except for NGC 4565. Since the HI beam size of NGC 4565 is less than 7\ac, other maps are convolved to the 24 \um\ Gaussian beam. 

\begin{figure}[!tbp]
\begin{center}
\begin{tabular}{c@{\hspace{0.1in}}c@{\hspace{0.1in}}c}
\includegraphics[width=0.32\textwidth]{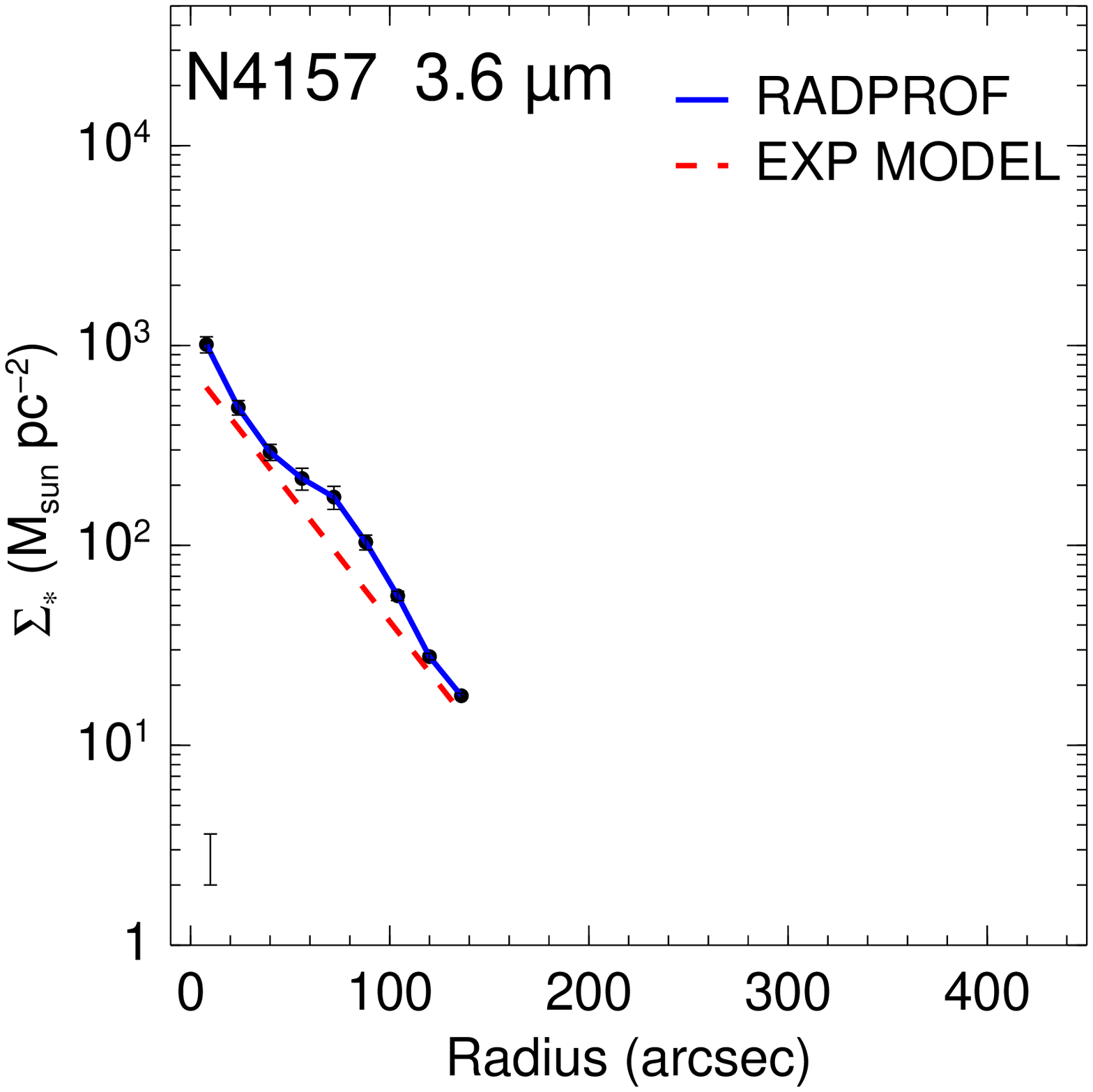}&
\includegraphics[width=0.32\textwidth]{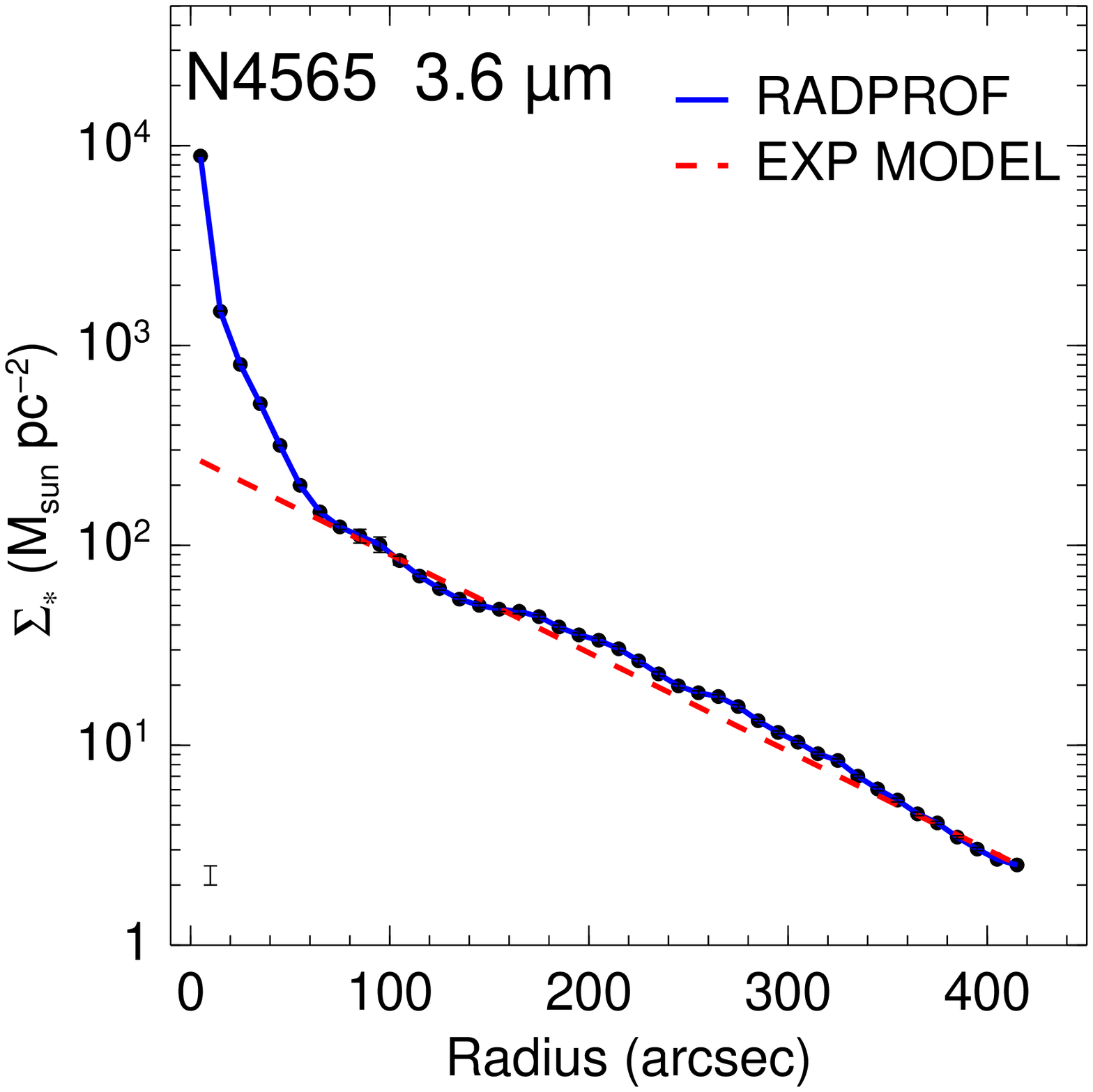}&
\includegraphics[width=0.32\textwidth]{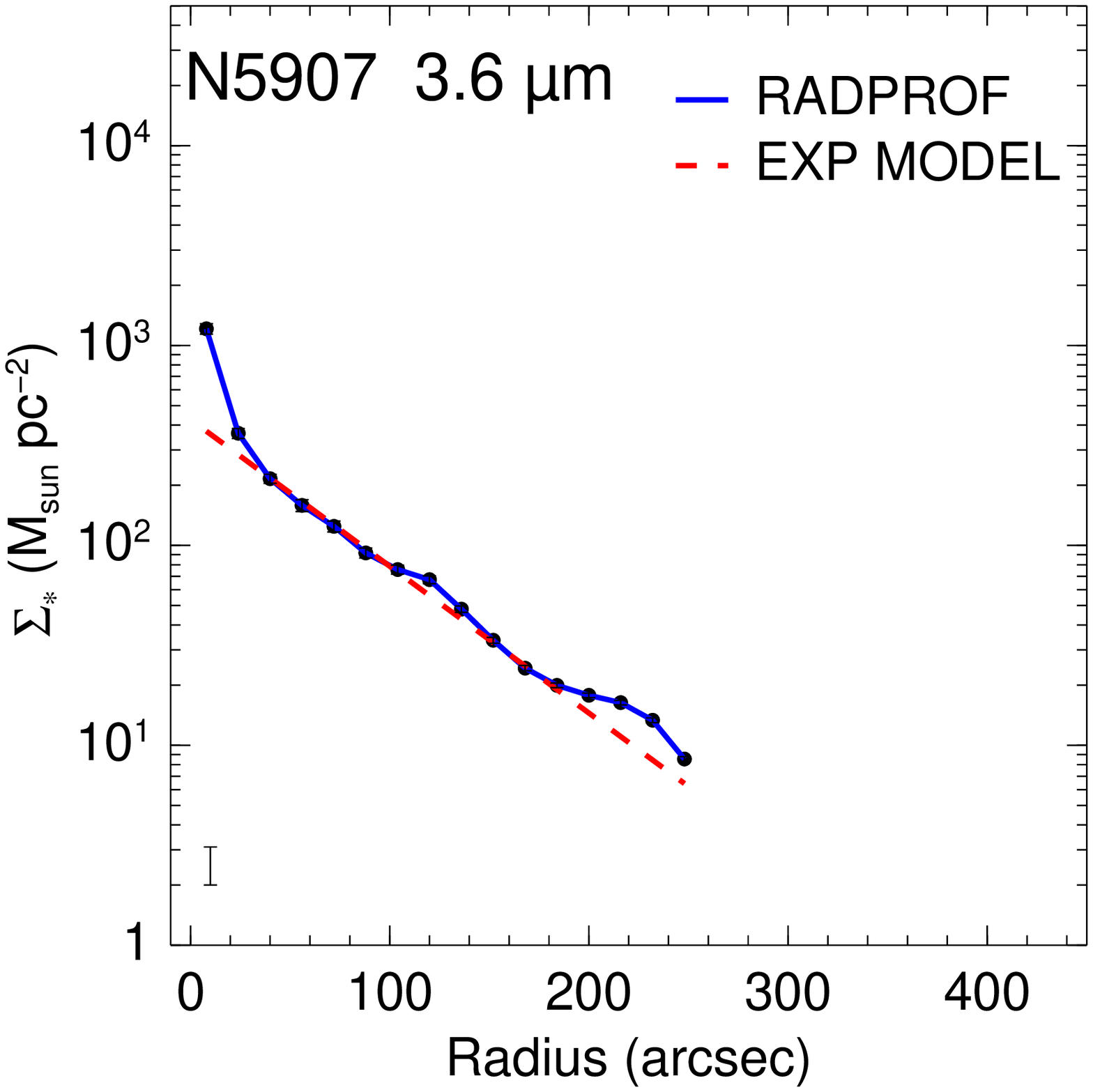}
\end{tabular}
\caption[Stellar surface density profile]{Stellar surface density as a function of radius for NGC 4157 ($left$), NGC 4565 ($middle$), and NGC 5907 ($right$). $Spitzer$ 3.6 \um\ images are used to obtain the stellar radial profile.  The blue solid line is the radial profile from the task RADPROF and the red dashed line shows the exponential disk model. The vertical error bar represents an uncertainty based on the largest difference between the RADPROF and the model profiles except the central regions excluded in the exponential fit.
\label{3p6prof}}
\end{center}
\end{figure}

 
The derived 3.6 \um\ radial profile from RADPROF is converted to stellar mass density using a conversion factor empirically derived by \cite{2008AJ....136.2782L}:
\begin{equation}
\sigstar\, [\surm]=280\, (\cos\,i) \,I\rm_{3.6} \,[MJy \,\,sr^{-1}],
\label{starfact}
\end{equation}
where the inclination $i$ is zero since the 3.6 \um\ intensity ($I_{\rm 3.6}$)  obtained from RADPROF is the face-on surface density.  The assumed mass-to-light ratio ($K$ band) for the conversion factor is 0.5 and the uncertainty quoted by \cite{2008AJ....136.2782L} is about 0.1 -- 0.2 dex depending on galaxy color. 
The stellar radial distributions from RADPROF are shown as a blue line in Figure \ref{3p6prof}. 

In order to estimate the effect of extinction on the stellar surface density, we have estimated the extinction at 3.6 \um\ from the ratio of gas column density $N_{\rm H}$ to color excess $E(B - V)$ given by \citet{1978ApJ...224..132B} and the ratio of the extinction $A_V$ to $E(B - V)$ $\sim$ 3.1 \citep{1985ApJ...288..618R}:
\begin{equation}
\frac{N_{\rm H}}{A_V} \approx 1.87 \times 10^{21} \,\rm{ atoms\, cm^{-2}\, mag^{-1}}.
\end{equation}
Using the ratios $A_{\rm 3.6}/A_K = 0.56$ \citep{2005ApJ...619..931I} and $A_K/A_V = 0.112$ \citep{1985ApJ...288..618R}, the dust extinction at 3.6 \um\ is 
\begin{equation}
A_{\rm 3.6} \approx 3.35 \times 10^{-23} N_{\rm H} \rm{\,mag}. 
\end{equation}
Using the combined map of CO and HI as a proxy for gas column density, we have obtained an extinction corrected map and derived a radial profile from the map. The largest extinctions we measured through the midplane are $\sim$1 mag for all the galaxies.
By comparing the radial profile (Figure \ref{3p6prof}) with the extinction corrected profile, we noticed that the differences between two profiles are a factor of $\sim$2 in the central regions for all the galaxies. 
 In the outer regions, the difference is almost negligible. 
 Note that the extinction correction is not applied to the stellar profile (Figure \ref{3p6prof}).
 

In addition to the RADPROF-derived stellar density profile, we also plot a fitted exponential disk model (red dashed line in the figure). 
The disk model we use is a self-gravitating and isothermal disk model provided by \cite{1981A&A....95..105V}: 
\begin{equation}
L(R,z) = L_0 \,\textrm{e}^{-R/l} \,\textrm{sech}^2\left(\frac{z}{z_*}\right)\;, \label{model}
\end{equation}
where $L_0$ is the space luminosity density at the center and $l$ is the scale length.
The vertically integrated luminosity density $L(R)$ is converted to the mass density using Equation \ref{starfact}. 
For fitting, we use the projected intensity distribution for an edge-on galaxy, obtained by integrating the disk model:
\begin{equation}
\mu(x,z) = \mu(0,0) \left(\frac{x}{l}\right)  K_1\left(\frac{x}{l}\right) \textrm{sech}^2 \left(\frac{z}{z_*}\right), 
\label{expfit}
\end{equation}
where $\mu(0,0)=2lL_0$ and $K_1$ is the modified Bessel function of the second kind of order 1.
When fitting to the 3.6 \um\ maps, the central regions (including the stellar bulge) are excluded: $|x|<10\ac$ for NGC 4157, $|x|<80\ac$ for NGC 4565, and $|x|<30\ac$ for NGC 5907. 
The scale lengths obtained by fitting Equation \ref{expfit} are $\sim$32\ac\ or $\sim$2 kpc (NGC 4157), $\sim$85\ac\ or $\sim$4 kpc (NGC 4565), and $\sim$57\ac\ or $\sim$3 kpc (NGC 5907). The scale heights ($z_*$) are $\sim$12\ac\ or $\sim$780 pc (NGC 4157), $\sim$14\ac\ or $\sim$640 pc (NGC 4565), and $\sim$13\ac\ or $\sim$670 pc (NGC 5907). 
The vertical error bar in the lower left corner represents the uncertainty of the radial profile and is obtained from the largest difference between the RADPROF and the disk model profiles: 0.25 dex (NGC 4157), 0.10 dex (NGC 4565), and 0.19 dex (NGC 5907). Note that the excluded central regions are not considered when finding the largest difference for the uncertainty estimate. 

\begin{figure}[!tbp]
\begin{center}
\begin{tabular}{c@{\hspace{0.1in}}c@{\hspace{0.1in}}c}
\includegraphics[width=0.32\textwidth]{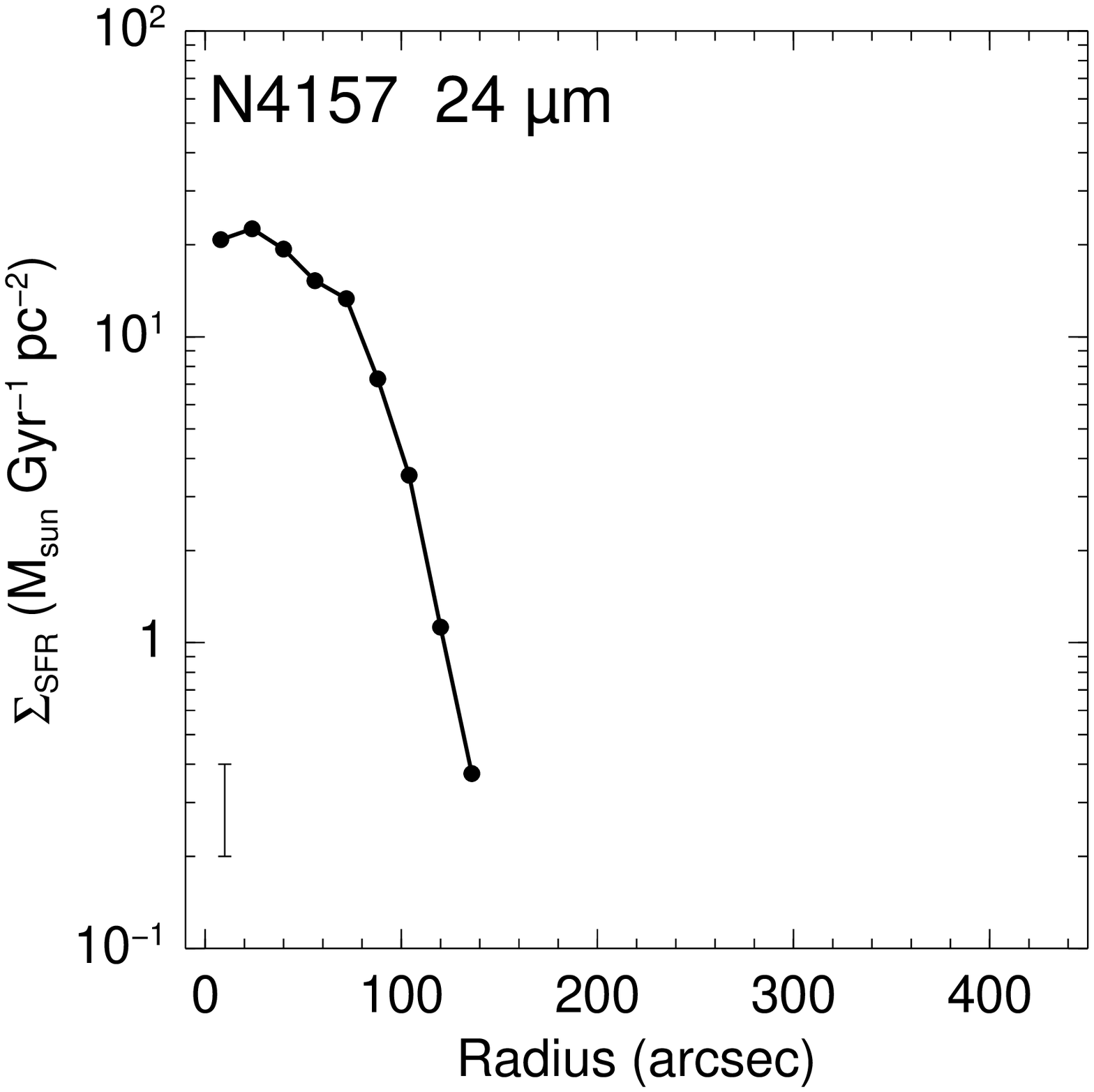}&
\includegraphics[width=0.32\textwidth]{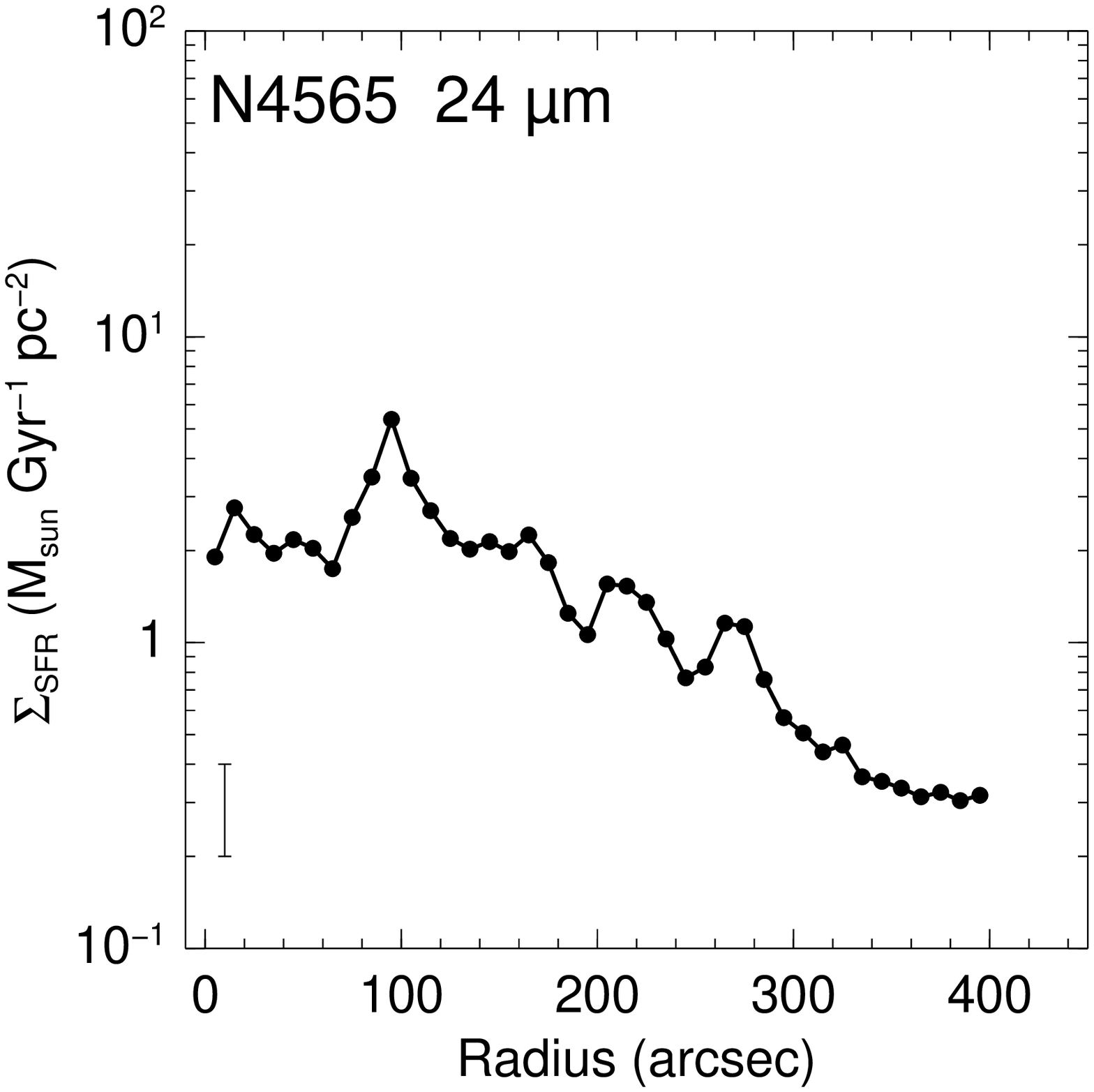}&
\includegraphics[width=0.32\textwidth]{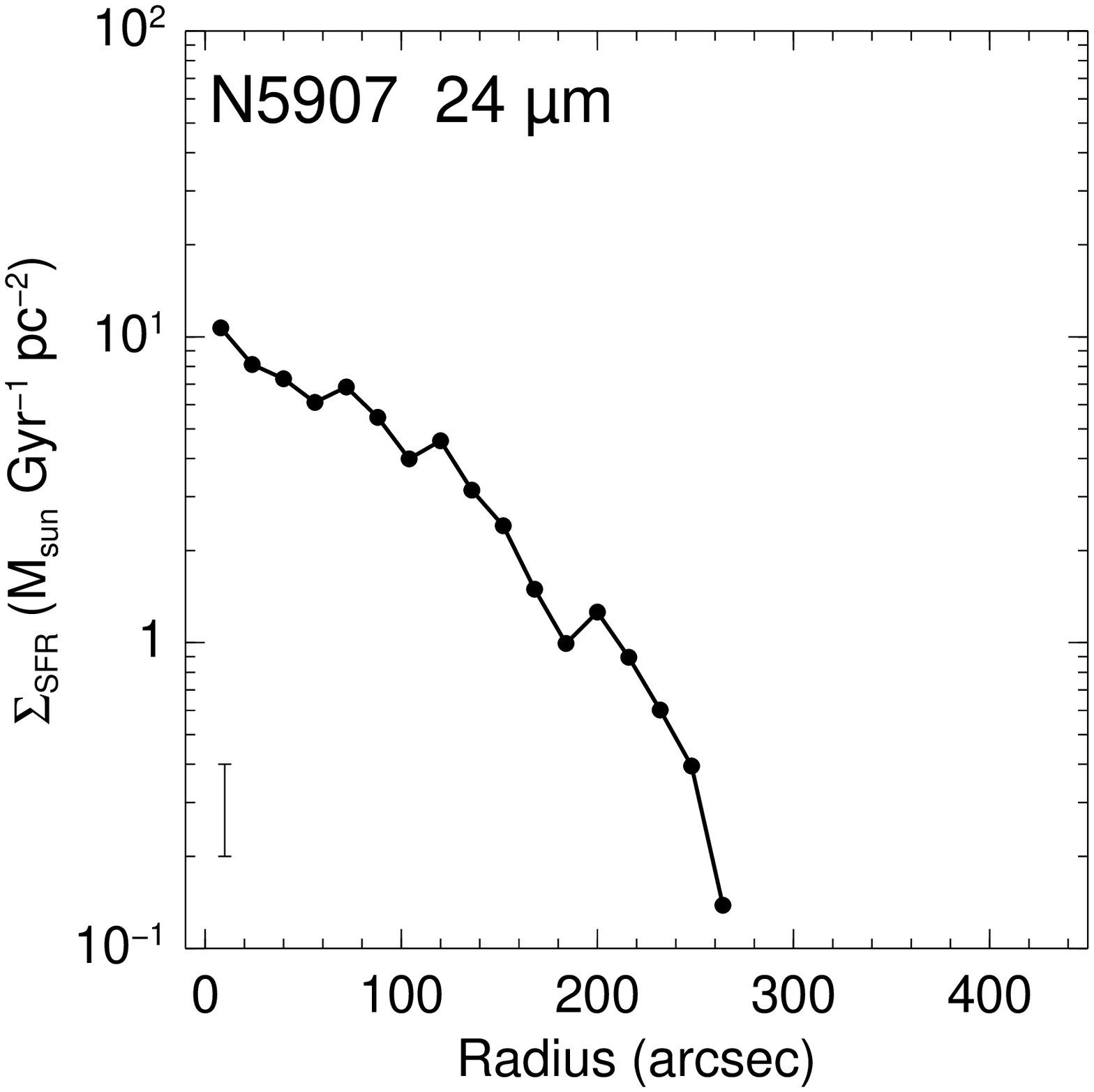}
\end{tabular}
\caption[SFR surface density profile]{SFR surface density as a function of radius for NGC 4157 ($left$), NGC 4565 ($middle$), and NGC 5907 ($right$) obtained from $Spitzer$ 24 \um\ images using the task RADPROF.
\label{24prof}}
\end{center}
\end{figure}

The SFR surface density is determined from the {\it Spitzer} 24 \um\ image by adopting the calibration given by \cite{2007ApJ...666..870C}:
\begin{equation}
\frac{\sigsfr }{\Msol \,\rm yr^{-1}\, kpc^{-2}} = 1.56 \times 10^{-35} \left(\frac{\it S\rm_{24 \mu m}}{\rm{erg \,s^{-1} \,kpc^{-2}}}\right)^{0.8104},
\label{eqSFR}
\end{equation}
where
\begin{equation}
\frac{S\rm_{24 \mu m}}{\rm{erg \,s^{-1} \,kpc^{-2}}} = 1.5 \times 10^{40} \left(\frac{\it I_{\rm 24}}{\rm MJy \,sr^{-1}}\right),
\end{equation}
and $I_{24}$ is the SFR (24 \um) surface brightness derived from RADPROF.

Figure \ref{24prof} shows the SFR radial profiles of the galaxies. 
The error bar, showing a factor of 2 uncertainty, is based on the largest difference between radial profiles from two different methods (RADPROF and ELLINT) for several face-on galaxies (observed by $Spitzer$ at 24 \um) as described in Paper I.
Before deriving the SFR radial profile of NGC 4565, the central compact source (possibly an AGN;  \citealt{2010AJ....140..753L}) in the 24 \um\ map  has been removed using the GIPSY tasks BLOT and PATCH.  The star formation rate in these galaxies seems relatively low compared to the SFR in NGC 891 (Paper I).  
\cite{1997A&A...325..124D} also found that NGC 4565 and 5907 show lower star formation than NGC 891.
The far-infrared luminosity at 40--400 \um\ by IRAS (often used as a tracer for SFR) is 10.18, 10.03, 9.61, 9.80 in log$_{10}(L/L_\odot)$ for NGC 891, 4157, 4565, and 5907, respectively (\citealt{2003AJ....126.1607S}).


\section{Radial Variation in Vertical Structure}
\label{vertical}

\begin{figure}[!tbp]
\begin{center}
\begin{tabular}{c@{\hspace{0.1in}}c@{\hspace{0.1in}}c@{\hspace{0.1in}}c}
\includegraphics[width=0.32\textwidth]{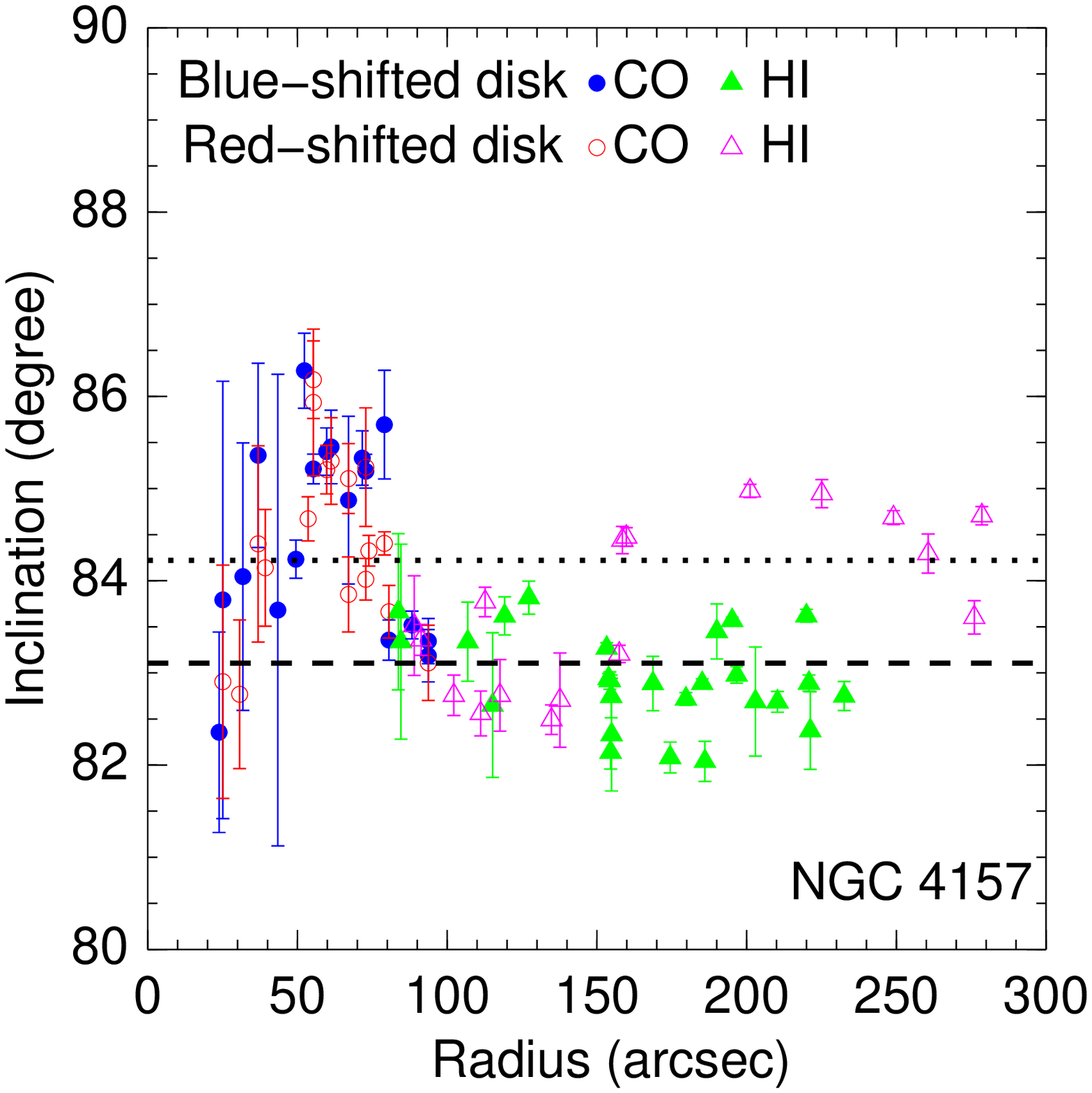}&
\includegraphics[width=0.32\textwidth]{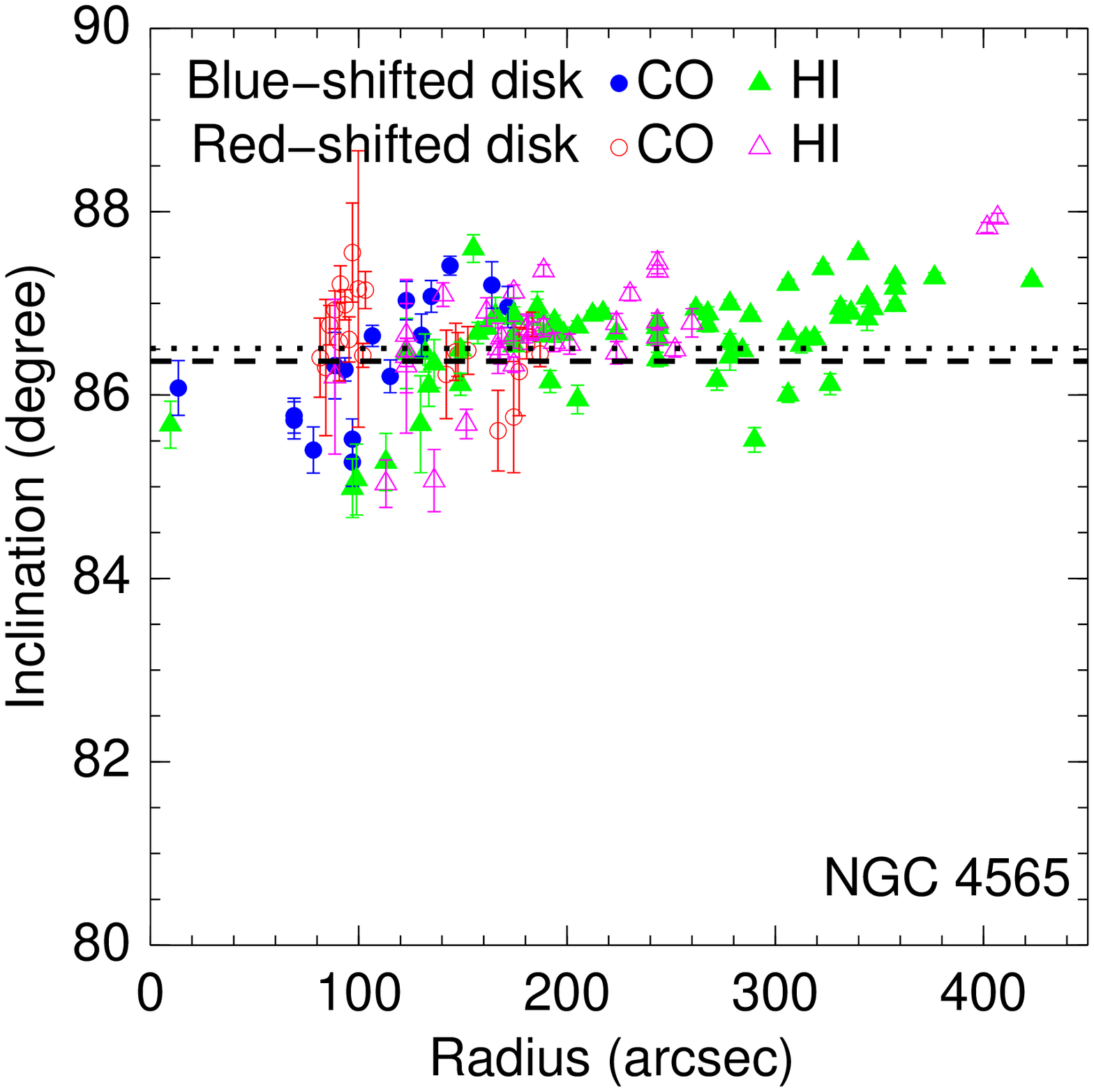}&
\includegraphics[width=0.32\textwidth]{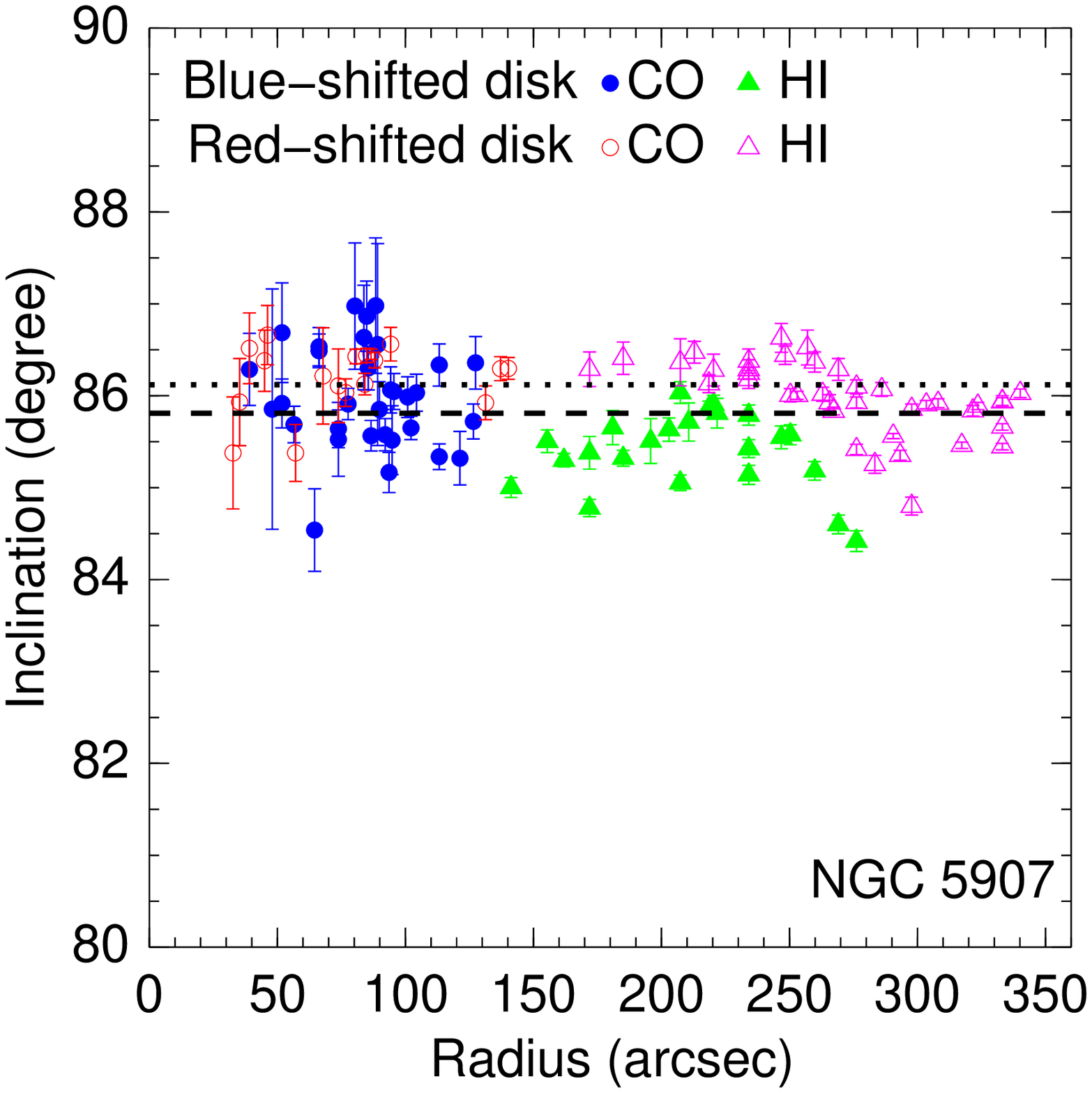}
\end{tabular}
\caption[Inclinations as a function of radius]{Inclinations as a function of radius for NGC 4157 ($Left$), NGC 4565 ($Middle$), and NGC 5907 ($Right$). 
The blue solid and red open circles show the blue-shifted disk and red-shifted disk of CO, respectively. The green solid and magenta open triangles represent the blue and red-shifted disks of \HI, respectively. 
The horizontal dotted and dashed lines represent weighted mean values of the CO and \HI\ inclinations, respectively. 
\label{inclination}}
\end{center}
\end{figure}

\begin{figure}[!tbp]
\begin{center}
\begin{tabular}{c@{\hspace{0.1in}}c@{\hspace{0.1in}}c@{\hspace{0.1in}}c}
\includegraphics[width=0.45\textwidth]{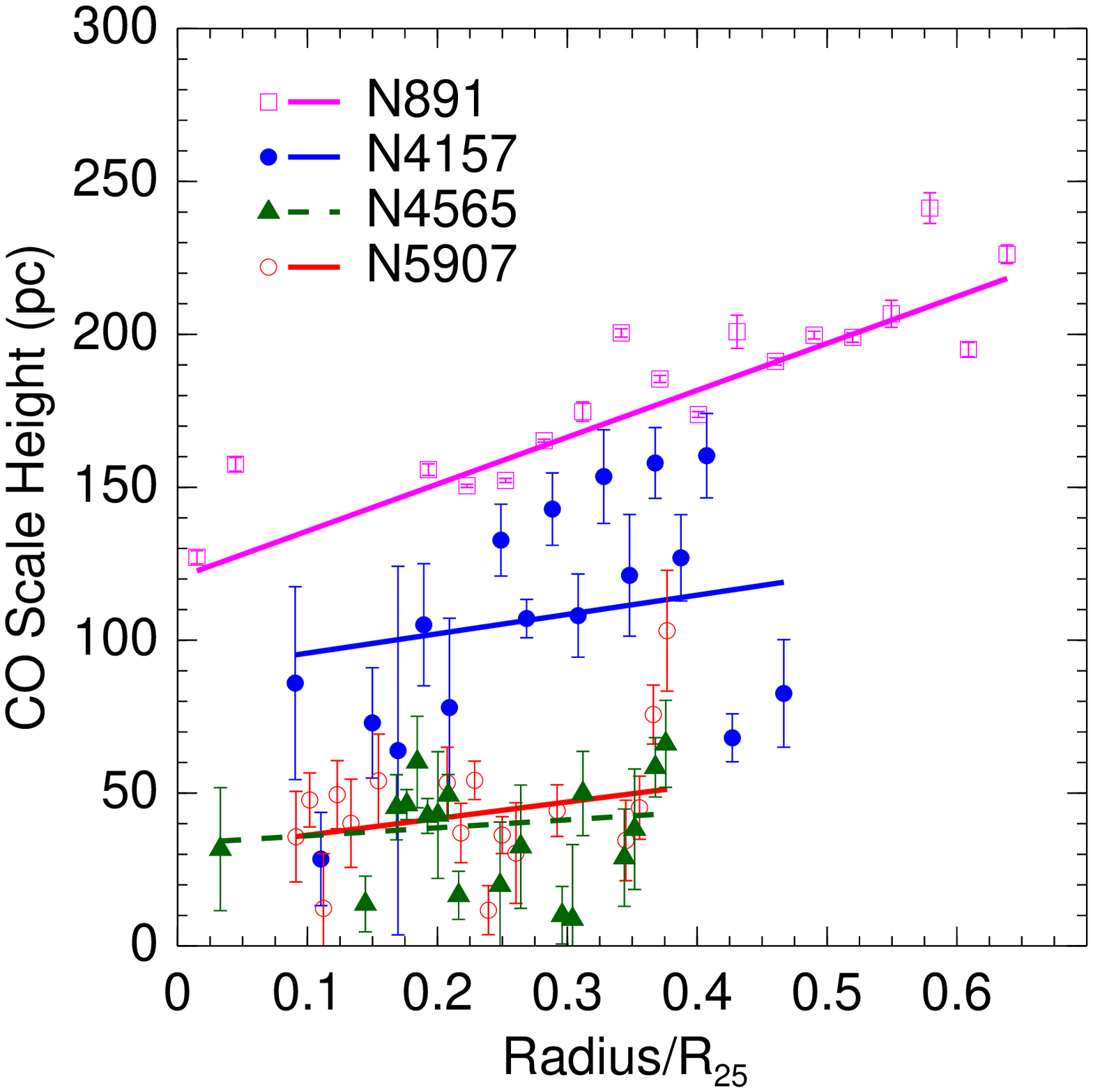}&
\includegraphics[width=0.45\textwidth]{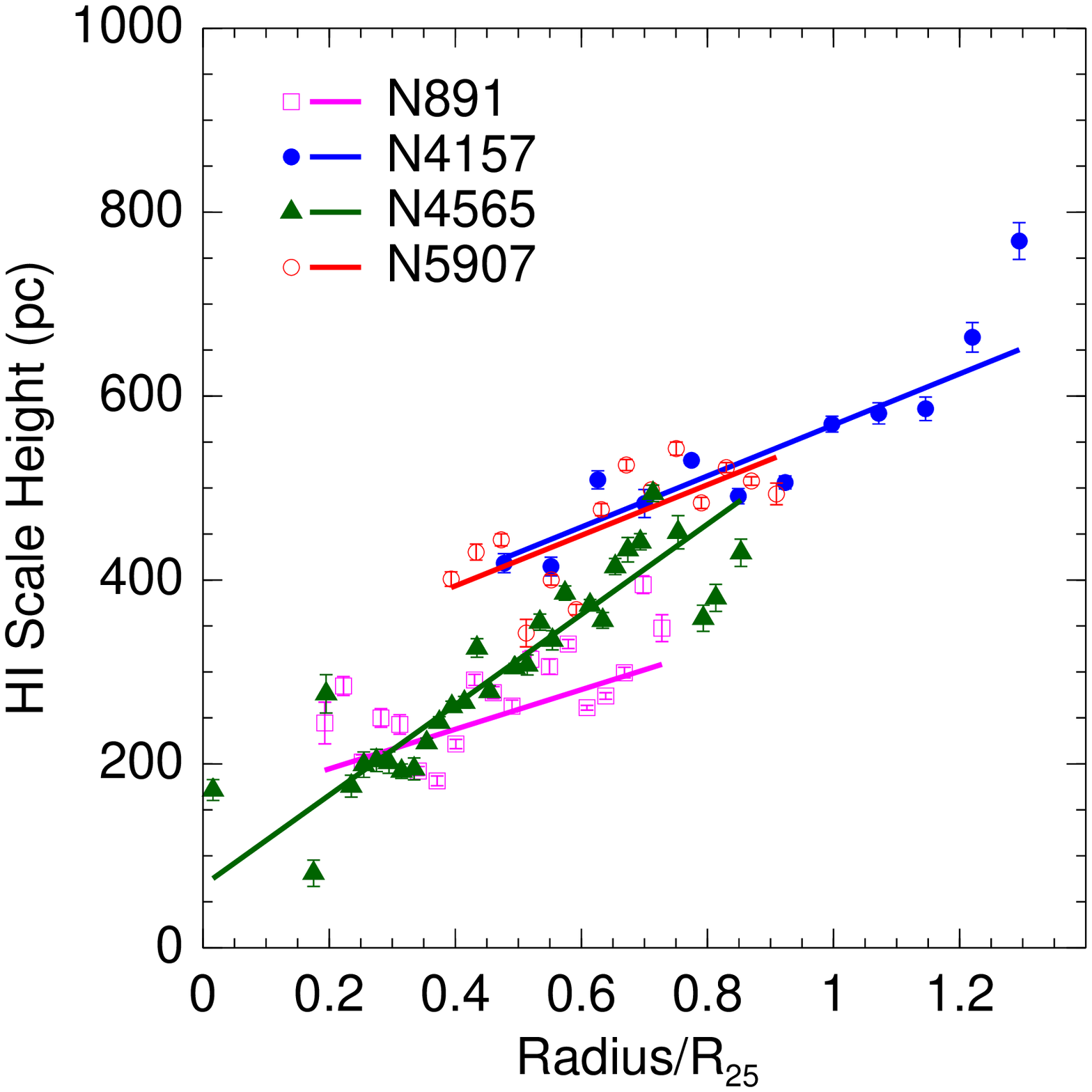}\\
\includegraphics[width=0.45\textwidth]{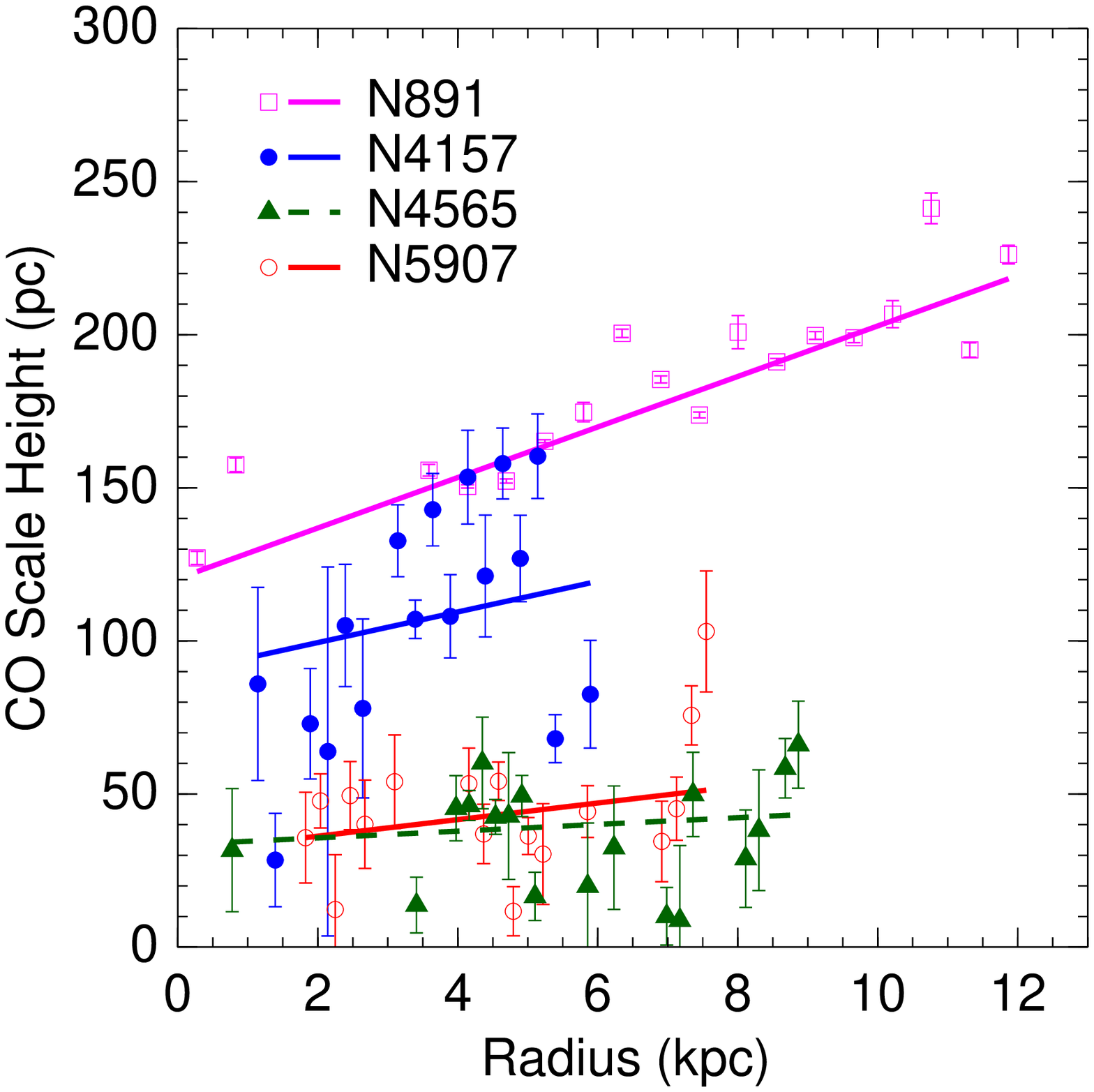}&
\includegraphics[width=0.45\textwidth]{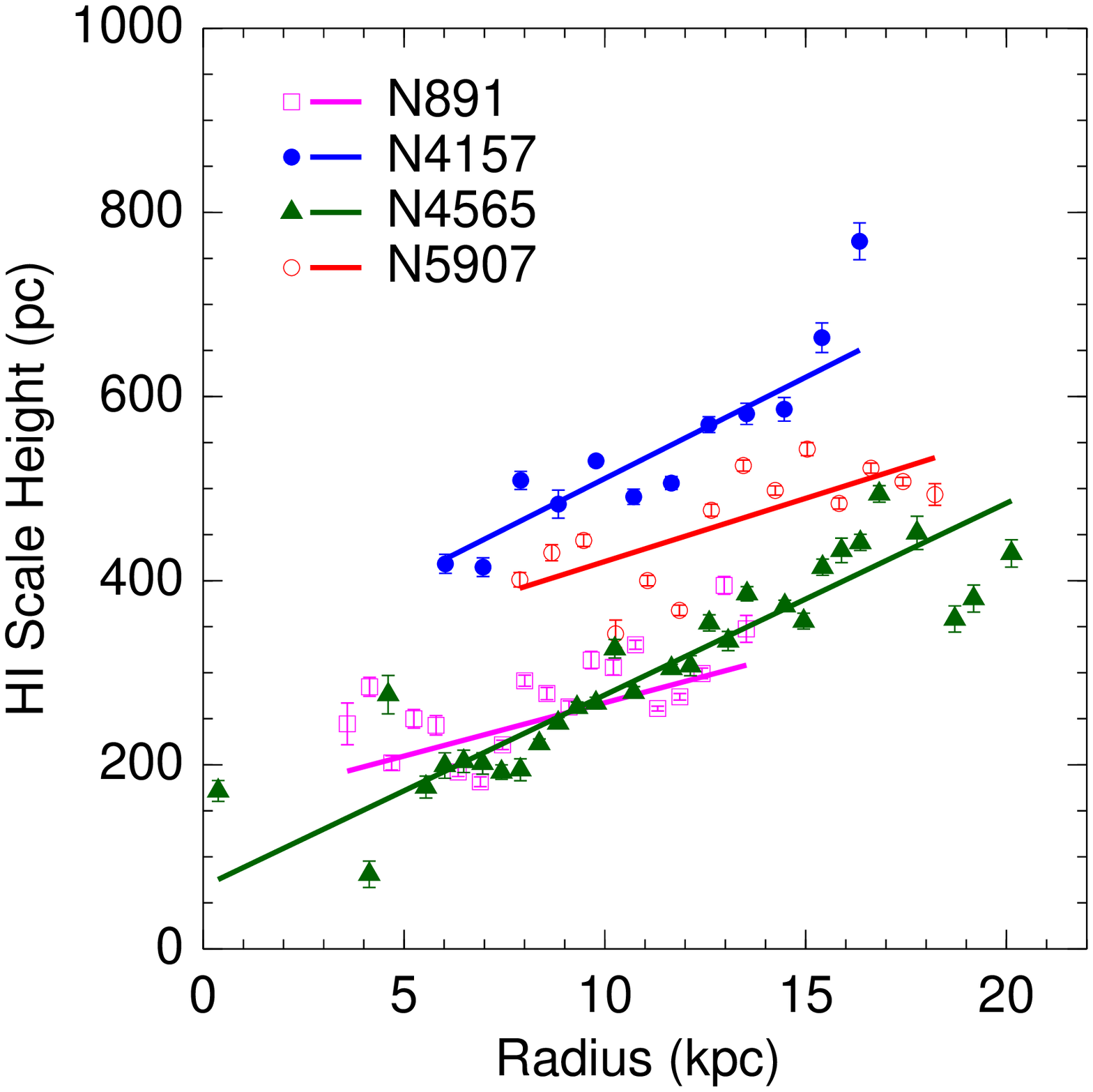}
\end{tabular}
\caption[Scale heights plotted against radius]{Top row: scale heights plotted against radius normalized by the optical radius ($R_{25}$) for CO ($Left$) and \HI\ ($Right$). Bottom row: scale heights plotted against radius in units of kpc for CO and \HI. The lines show linear approximations obtained by least-squares fitting. 
\label{ollingfwhm}}
\end{center}
\end{figure}

\begin{figure}[!tbp]
\begin{center}
\includegraphics[width=0.6\textwidth]{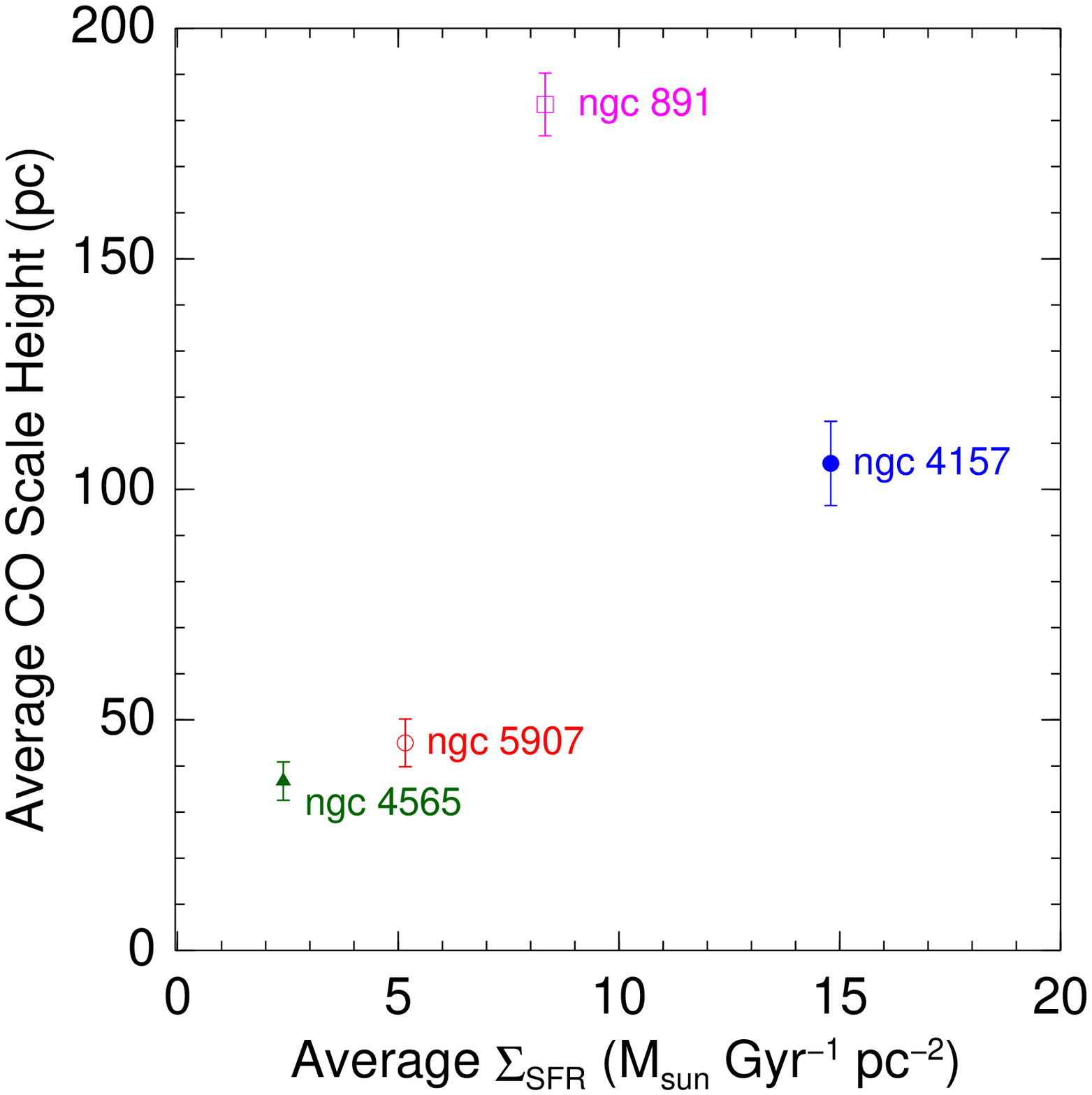}
\caption[Average scale height of CO as a function of average \sigsfr]{Average scale height of CO as a function of average \sigsfr. The name of each galaxy is indicated next to its point. The vertical error bars are the standard deviation of the mean.
\label{sfr_codisk}}
\end{center}
\end{figure}

\subsection{Inclination and Gas Disk Thickness}
\label{inc}

Unlike NGC 891, whose inclination is known to be $> 89\degr$ (\citealt{2007AJ....134.1019O}), the other galaxies in our sample (NGC 4157, 4565, and 5907) appear less inclined. Since the inclination affects the apparent disk thickness, it is important to determine the inclination of the galaxies in order to obtain correct values for the scale height and the velocity dispersion. 

We have derived inclinations and scale heights of these less edge-on galaxies via  Olling's method (\citealt{1996AJ....112..457O}) which is described in detail in Appendix \ref{appen}.
The derived inclinations with radius are shown in Figure \ref{inclination} as blue solid and red open circles for CO and green solid and magenta open triangles for \HI. Different colors indicate different halves of the galaxy. We show a weighted mean value of the inclination points as dotted (CO) and dashed (\HI) lines and adopt a representative value of $i = 86.5\degr$ for NGC 4565 and $i = 86\degr$ for NGC 5907 since their mean CO and \HI\ values are very close. However, in the case of NGC 4157, the CO (84\degr) and \HI\ (83\degr) values show a difference, so we adopt the CO inclination value since the \HI\ derived value may be affected by the outer warp.  Even though there is the 1$\degr$ difference between CO and \HI\ inclinations in NGC 4157, the values match well in the region where both tracers can be analyzed. We also note that the rotation curves of CO and \HI\ in this galaxy show a mismatch (especially in the region of 30\ac--70\ac) unlike the other galaxies.  Since the inclinations are determined using the distance $y_{\rm off}$ (see Appendix~\ref{appen}) which depends on the velocity field map, the mismatch between the rotation curves may contribute to the differences between the inclinations derived from CO and \HI. 

The measured scale heights (see Appendix \ref{appen}) considering the projection effects for the less edge-on galaxies are shown in Figure \ref{ollingfwhm}. We use the Gaussian width (0.42 $\times$ FWHM) for the scale height. The scale heights for NGC 891 presented in Paper I have been included in this figure for comparison purposes.  The lines show the linear least-squares fits to the data points and we use the best fit line to derive the radial variation in the volume density and velocity dispersion in Section \ref{volden} and \ref{veldisp}. 
For the inner and outer regions where no data points are available, we assume a constant scale height within the innermost measured point when deriving the total midplane density and the total gas dispersion for the gravitational instability parameter $Q$ (Section \ref{gravQ}) and the interstellar pressure (Section \ref{rmol}). 

The thickness of the gas disk may be related to the star formation activity (\citealt{1997A&A...325..124D}, and references therein). From comparison between the CO disk thickness and the SFR surface density profile (Figure \ref{24prof}), we notice that the galaxies with higher SFR (NGC 891 and NGC 4157) show a thicker CO disk. In order to investigate the relationship between the CO disk thickness and SFR, we plot the average CO scale height as a function of average \sigsfr\ in Figure \ref{sfr_codisk}. The average \sigsfr\ values are obtained by integrating \sigsfr\ only over the CO disk range (i.e., the radial range over which estimates of the CO scale height are available)  and normalizing by the disk area. This figure suggests a weak correlation between the {\it galaxy-averaged} CO disk thickness and the SFR. 
Within a galaxy, however, the two values are anti-correlated, with the CO disk being thicker in the outer galaxy where the SFR is lower.

\subsection{Stellar Disk Thickness}
\label{stdisk}
Visual inspection of the 3.6 \um\ images does not reveal clear indications of an increase in the scale height with radius.  This may be due, however, to the rapidly decreasing surface brightness with radius and a viewing angle that is not exactly edge-on.  We have therefore adopted an iterative approach to modeling the variation in stellar scale height. 
 
\subsubsection{Initial estimates}\label{stmodinit}

Olling's method is not applicable for deriving the stellar scale height due to the lack of velocity information.
In the case of NGC 891 (Paper I), we assumed a sech$^2(z/h_*)$ function to derive the variation of the stellar scale height ($h_*$) with radius. The vertical distributions at each radius was obtained by employing RADPROF to obtain radial distributions at many different vertical heights. 
For the less inclined galaxies studied here, however, projection effects must be taken into account.
As an initial estimate for $h_*(R)$, we first employ the same RADPROF approach as in Paper I, then apply a multiplicative factor of $\sin i$ and make a least-squares linear fit to the resulting data points.
The fitted line is then used to generate an initial disk model.

\subsubsection{Iterative Modeling}
\label{stmod}

In order to derive more accurate stellar scale heights, we generated a series of models in GIPSY task GALMOD, fixing the inclination and radial surface density profile to derived values and allowing the scale height to vary for different models. 
We obtained minor axis profiles of the 3.6 \um\ image and the models at various major axis ($x$) offsets to derive the line-of-sight (LOS) projected scale height by fitting an exponential function to the vertical profiles. (An exponential function is what the sech$^2$  function approaches at high $z$.)
The central regions (including bulge) are excluded from the fitting. 
For galaxies except NGC 891, we use the 3.6 \um\ images from the S$^4$G archive \citep{2010PASP..122.1397S} in this procedure, and use the accompanying masks to exclude from the fitting any regions contaminated by point sources. 

First, we generated an initial model as described in Section \ref{stmodinit}. Next, we compared the projected scale heights of the model cube and the 3.6 \um\ map finding the modeled scale height to be an overestimate except for the almost edge-on galaxy NGC 891. We then input a new scale height into GALMOD estimated from the comparison between the initial model and the data. The new scale height has a much lower central value (about three times lower than the original value) and a similar slope. After comparing this second model to the data, we adjusted the parameters for the central value and the slope to obtain a subsequent model. This iteration was carried out until the modeled scale height reproduces the fit to the data. Following this scheme, we could find a reasonable model within seven trials. 
The stellar scale height could be affected by uncertainties in the radial profile (which goes into the GALMOD modeling) so we checked the effects of extinction by starting with an extinction corrected radial profile. The new scale height from the extinction corrected density profile is increased by a factor of $\sim$1.1 for all the galaxies; extinction does not appear to have a major effect on the scale height.  But we are only considering the effect of extinction on the radial profiles, not the vertical profiles.
The variation in the LOS-integrated scale heights have been compared with what has been observed in other edge-on galaxies given by \citet{1997A&A...320L..21D}; the gradients in our measurements are in the range of their sample.

From a comparison between the initial scale height based on the RADPROF method and the scale height of the best model we note the following. 
For the less edge-on galaxies, the initial central values from the data seem to be overestimated by a factor of $\sim$2--3 compared to the values of the best models. The initial slopes of the scale heights are higher by a factor of $\sim$2 for NGC 4157 and 5907 and $\sim$3 for NGC 4565 than those of the best model. 
On the other hand, the almost edge-on galaxy NGC 891 shows similar values between the best model and the data. 
We also verified that a constant scale height model produced LOS-integrated scale heights that were in significant disagreement with the data.
We will use these scale heights obtained by the modeling when deriving the midplane volume density and vertical velocity dispersion of stars in the following sections.

\begin{table}[!tb]\footnotesize
\begin{center}
\caption{Parameters from fitting the radial variation of the scale heights \label{fitpara}}
\begin{tabular}{cccccccc}
\tableline\tableline
Galaxy&Adopted $i$ [$\degr$]&\multicolumn{3}{c}{Central Value ($a$) [pc]}&\multicolumn{3}{c}{Gradient ($b$) [pc/kpc]}\\
& &CO&\HI&Stars&CO&\HI&Stars\\
\tableline
NGC 891&89&120$\pm$1&151$\pm$5&161&8.3$\pm$0.1&11.6$\pm$0.5&22.0\\
NGC 4157&84&89$\pm$11&291$\pm$11&150&5.0$\pm$2.7&22.0$\pm$1.1&41.5\\
NGC 4565&86.5&33$\pm$8&68$\pm$5&165&1.1$\pm$1.4&20.8$\pm$0.4&20.0\\
NGC 5907&86&31$\pm$8&284$\pm$8&133&2.7$\pm$1.5&13.7$\pm$0.6&25.0\\
\tableline
\end{tabular}
\end{center}
\end{table}


\subsection{Summary of Disk Thicknesses}

We characterize the radial variation in scale height using a linear function $h(R)$ expressed as:
\begin{equation}
h = a + b \times R, 
\label{scalefit}
\end{equation}
where the intercept ($a$) and the slope ($b$) with their uncertainties are summarized in Table \ref{fitpara}. 
For stars, we show the parameters used to generate the best model. 
For ease of presentation, the slope is quoted in pc kpc$^{-1}$ rather than in dimensionless form.
Unlike the scale heights of \HI, the uncertainties in the slopes of the CO scale heights are very large in the three galaxies NGC 4157, 4565, and 5907 and in the case of NGC 4565 the measured gradient is not significant.
However, most galaxies show increases with radius in the scale heights of CO, \HI, and stars. 

\begin{figure}[!tbp]
\begin{center}
\begin{tabular}{c@{\hspace{0.1in}}c@{\hspace{0.1in}}c}
\includegraphics[width=0.32\textwidth]{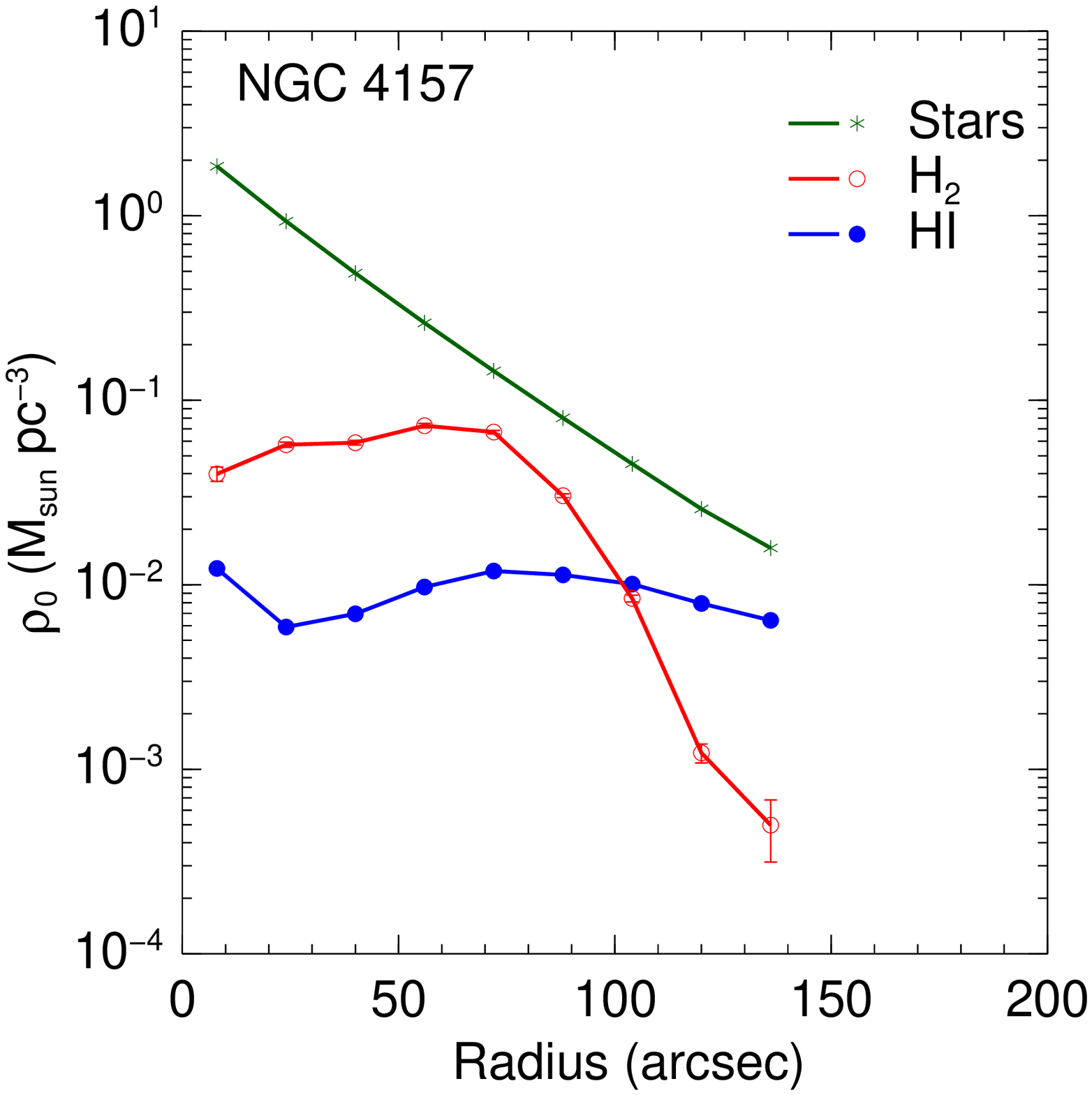}&
\includegraphics[width=0.32\textwidth]{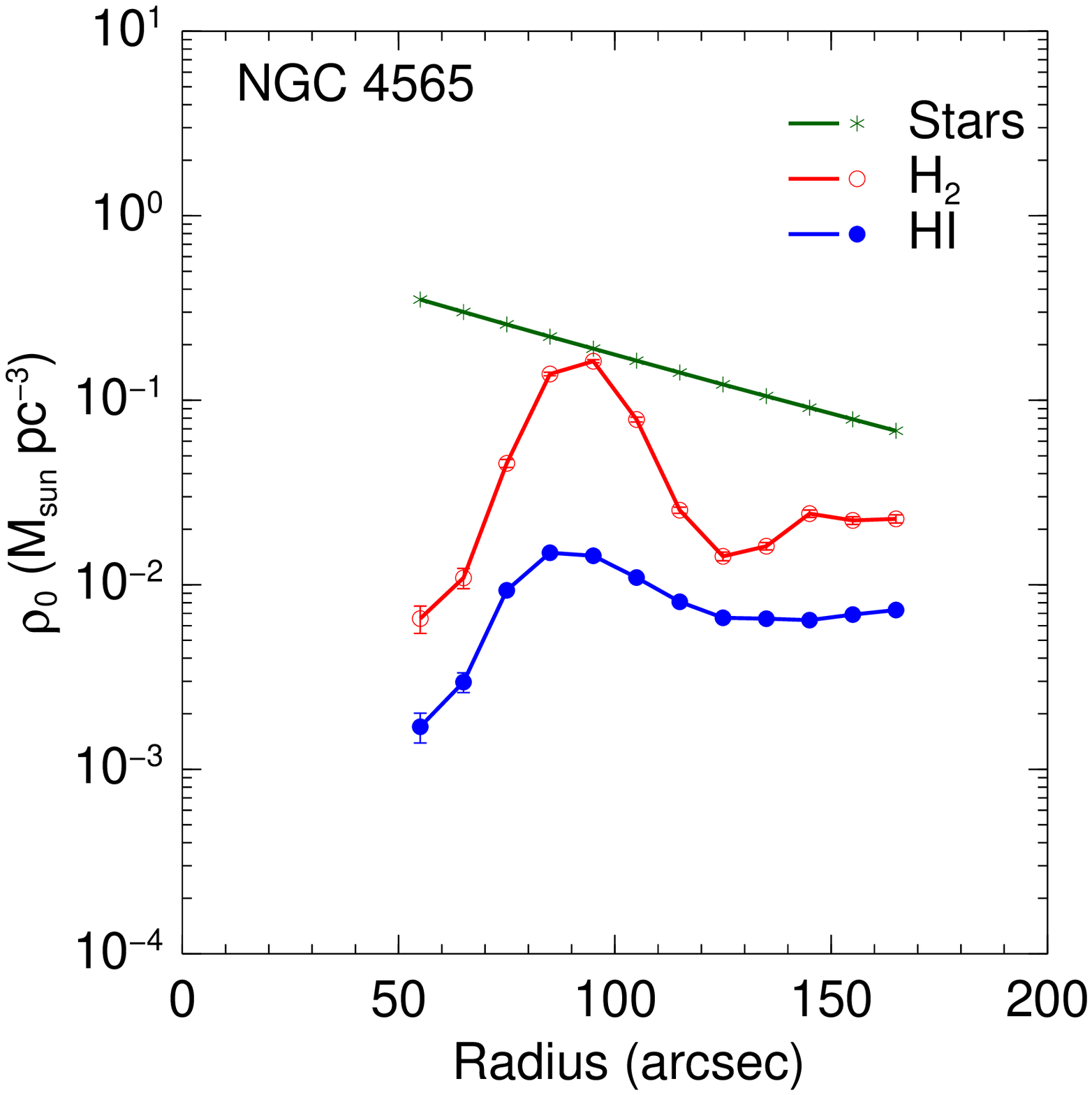}&
\includegraphics[width=0.32\textwidth]{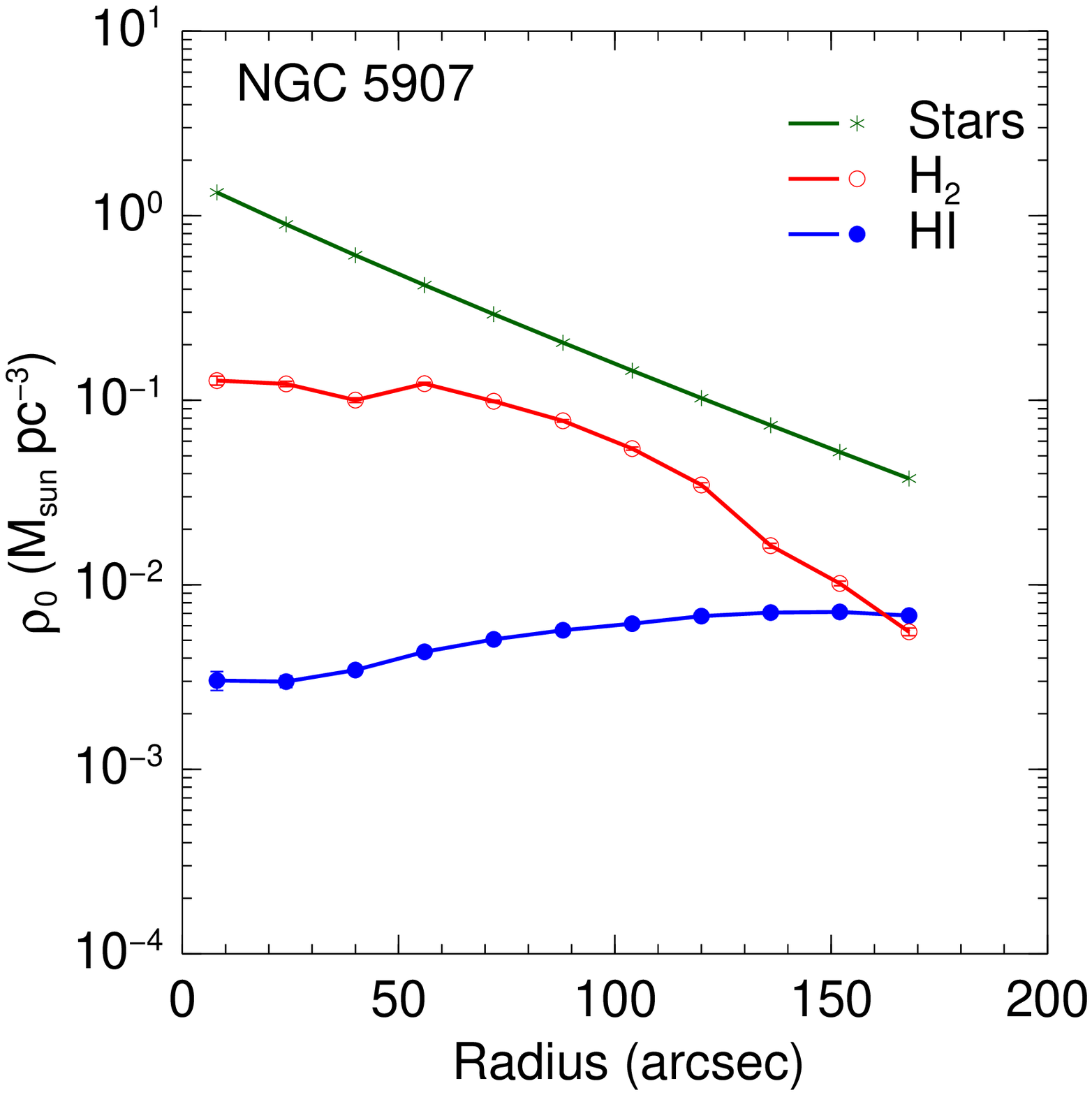}
\end{tabular}
\caption[Midplane volume density against radius]{Midplane volume densities of \Htwo\ (red open circles), \HI\ (blue solid circles), and stars (green asterisks) as a function of radius. 
\label{midvol}}
\end{center}
\end{figure}

\subsection{Midplane Volume Densities}
\label{volden}

Using the derived surface mass densities and the scale heights from the previous sections, we have obtained the midplane volume density ($\rho_0$) profiles with radius:
\begin{equation}
\rho_{\rm 0H_2} = \frac{\sightwo}{h_{\rm H_2}\sqrt{2\pi}},  \qquad 
\rho_{\rm 0HI} = \frac{\sighi}{h_{\rm HI}\sqrt{2\pi}},  \qquad 
\rho_{0*} =  \frac{\sigstar}{2 h_*}.
\end{equation}
Here we assume a Gaussian distribution for the gas (\Htwo\ and \HI) and a sech$^2(z/h_*)$ function for stars. The estimated volume densities are shown in Figure \ref{midvol}. The uncertainties shown in the figure include errors in the radial profile but not in the scale height or the CO-to-H$_2$ conversion factor.  A factor of 2 variation in the $X_{\rm CO}$ factor would lead to a similar variation in the volume density.
This midplane density profile allows us to explore the role that the turbulent interstellar pressure ($\rho_0 \sigma_{\rm g}^2$) plays in controlling the molecular to atomic (volume) gas density ratio, in comparison with the relationship between the hydrostatic midplane pressure and the \Htwo/\HI\ ratio based on surface mass density. In addition, the density profile will allow us to investigate the star formation law in terms of volume density instead of surface density, although we defer this analysis to the next paper in this series.


\begin{figure}[!tbp]
\begin{center}
\begin{tabular}{c@{\hspace{0.1in}}c@{\hspace{0.1in}}c}
\includegraphics[width=0.32\textwidth]{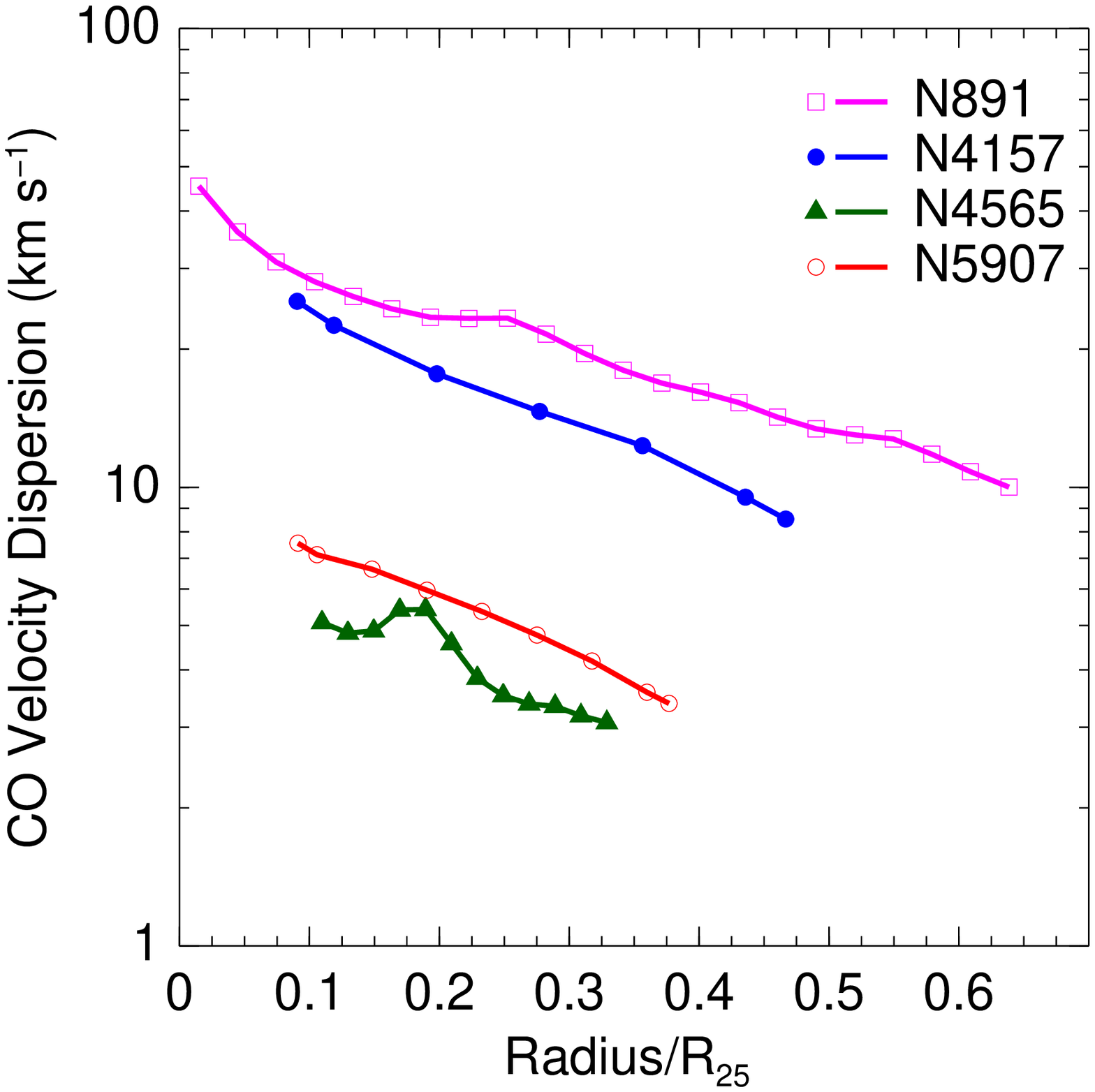}&
\includegraphics[width=0.32\textwidth]{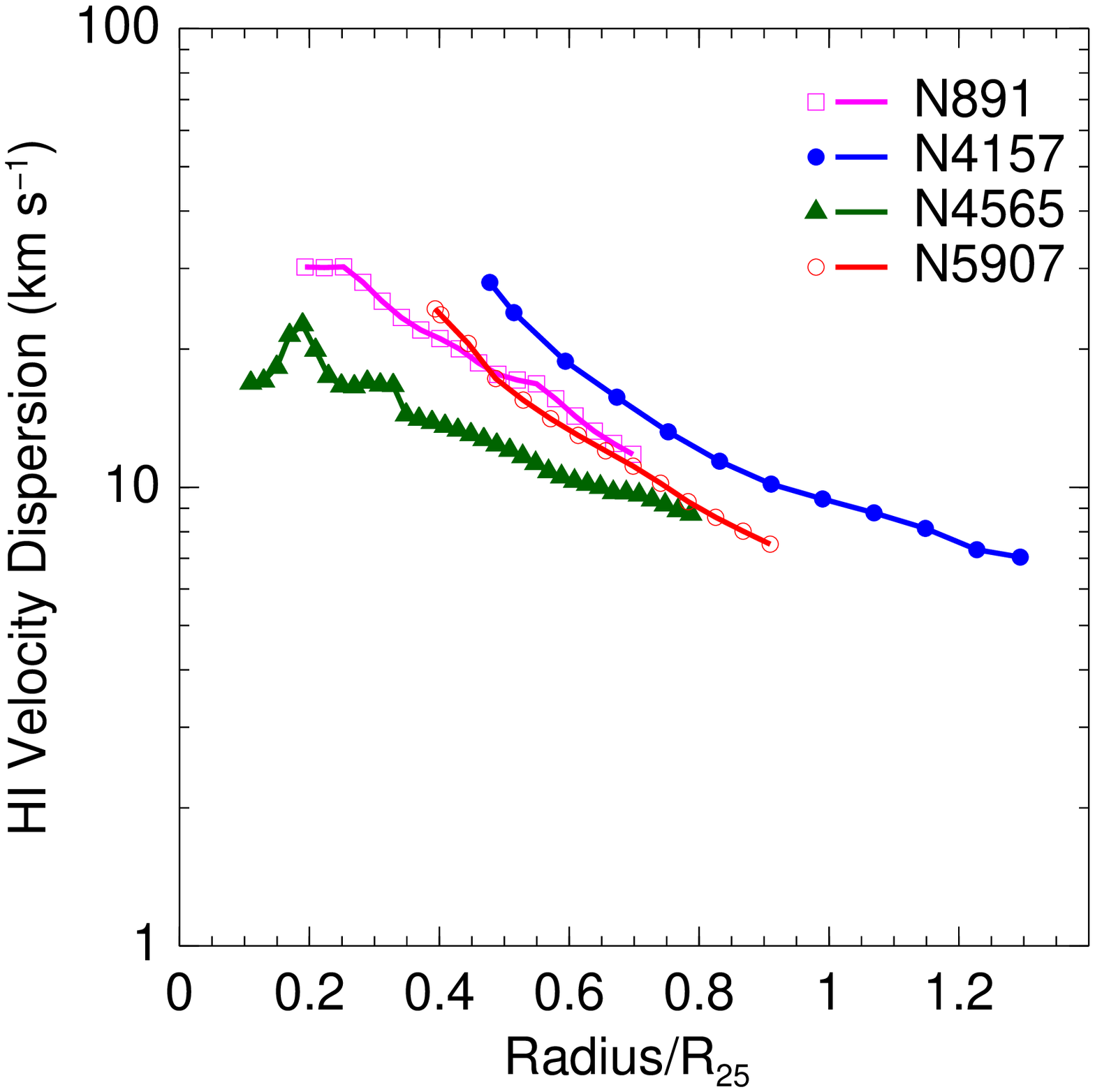}&
\includegraphics[width=0.32\textwidth]{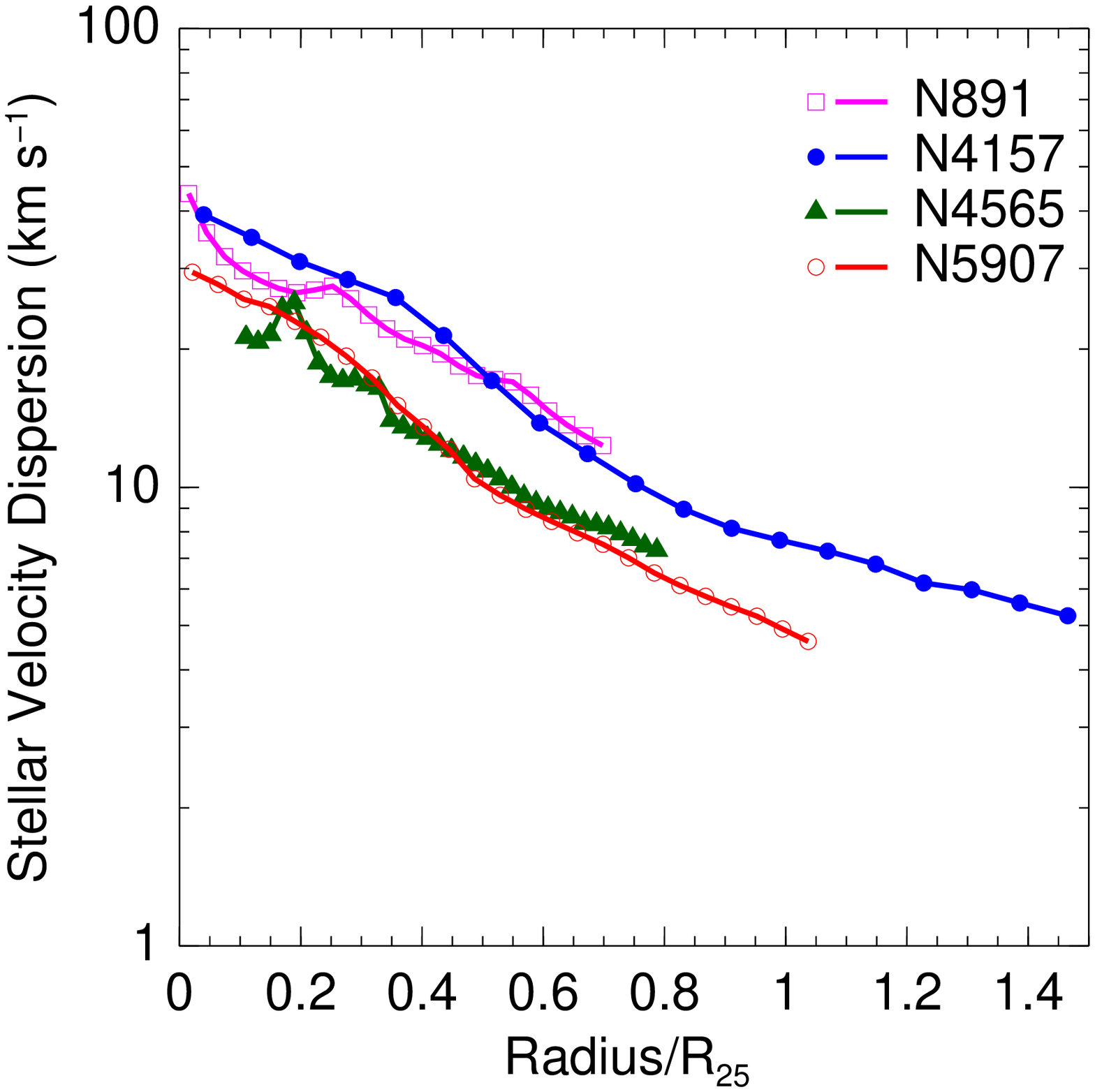}
\end{tabular}
\caption[Vertical velocity dispersion against radius]{Vertical velocity dispersions against radius normalized by the optical radius ($R_{25}$) for CO ($Left$), \HI\ ($Middle$), and stars ($Right$). 
\label{vdisp}}
\end{center}
\end{figure}

\subsection{Vertical Velocity Dispersion}
\label{veldisp}

Since it is not easy to measure a gaseous vertical velocity dispersion directly, some earlier studies had suggested fairly constant values (\citealt{1997A&A...326..554C}; \citealt{2007AJ....134.1952P}) . As an alternative, we have inferred the velocity dispersions as a function of radius using a numerical solution to the Poisson equation for a multi-component disk (\citealt{2002A&A...394...89N}):
\begin{eqnarray}
\sigma_i^2 = \frac{4\pi G\rho_{\rm 0,tot}\rho_{0i}}{- (d^2\rho_i/dz^2)_{z=0}},
\label{poissong}
\end{eqnarray}
where the subscript $i$ can refer to either g (gas) or $*$ (stars),
\begin{eqnarray}
d^2\rho_{\rm g}/dz^2 &=& - \rho_{\rm 0g}/h_{\rm g}^2,\qquad
d^2\rho_*/dz^2 = - 2\rho_{0*}/h_*^2\quad {\rm at}\ z=0,\\
\Rightarrow \sigma_{\rm g} &=& \sqrt{4\pi G h_{\rm g}^2 \rho_{\rm 0,tot}},\qquad
\sigma_* = \sqrt{2\pi G h_*^2 \rho_{\rm 0,tot}}.
\label{eqvdis}
\end{eqnarray}
Here the total midplane density is, 
\begin{equation}
\rho_{\rm 0,tot} = \rho_{\rm 0H_2} + \rho_{\rm 0HI} + \rho_{0*}.
\end{equation}
 We ignore the dark matter halo under the assumption that the dark matter density in the disk is relatively small compared to that of the stars. 
The boundary conditions we have used are $\rho_i = \rho_{0i}$ and $d\rho_i/dz = 0$ at the midplane ($z=0$). 

The velocity dispersions of CO, \HI\, and stars for the galaxies including NGC 891 are shown in Figure \ref{vdisp}. 
The two galaxies NGC 4565 and 5907 show very low values of the CO velocity dispersion  compared to the other galaxies. The two main reasons for the low velocity dispersions are because of  low values in (1)  the total midplane density and (2) the scale height of CO (see Equation \ref{eqvdis}). 
On the other hand, the velocity dispersions of \HI\ and stars show similar values (within a factor of $\sim$2) and trends for all galaxies. 
Note that when we estimate the midplane total density to derive the \HI\ and stellar velocity dispersions, the molecular densities in the regions outside the CO field of view are assumed to be zero---a reasonable approximation because these regions are strongly \HI\ dominated (with the possible exception of NGC 4565).
In general, our results show that the velocity dispersions of CO, \HI, and stars decrease as a function of radius. 

\begin{figure}[!tbp]
\begin{center}
\includegraphics[width=0.7\textwidth,angle=270]{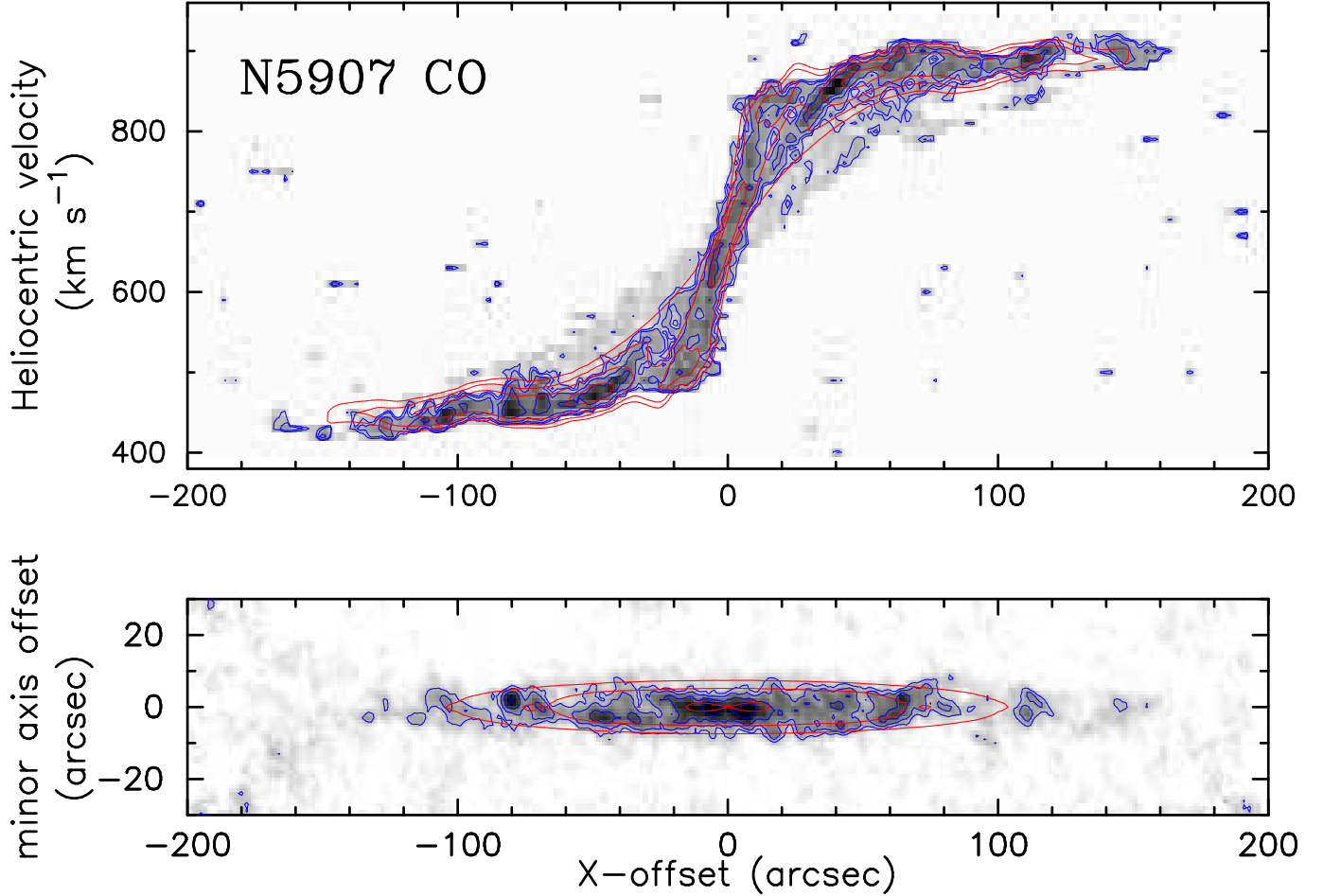}
\caption{CO position-velocity diagram along the major axis ($top$) and integrated intensity map of NGC 5907 ($bottom$). TiRiFiC model contours (red) are overlaid on the map contours (blue). Both contour levels are  $0.2 \times 1.9^n$ K arcsec, with n=0, 1, 2, 3 ($top$) and $26 \times 1.4^n$ K \kms, with n=0, 1, 2, 3 ($bottom$).
\label{cotirific}}
\end{center}
\end{figure}


\subsection{TiRiFiC Modeling}
In order to verify the derived scale heights and velocity dispersions of the gas, we have built CO and HI models of NGC 5907 (as a representative example)  using the Tilted Ring Fitting Code (TiRiFiC) \citep{2007A&A...468..731J} and compared the models with the data. The main input parameters we used for modeling in TiRiFiC are  the derived rotation curve, density profile, scale height, velocity dispersion, and inclination.  From the comparisons of the midplane  and $z$-integrated p-v diagrams and the integrated intensity maps (Figure \ref{cotirific}), we confirmed that the gas models match the data reasonably well although the model does not include the HI warp. It suggests that  the derived CO and \HI\ scale heights and velocity dispersions as well as the rotation curves are consistent with the data.  

\section{Star Formation}
\label{sflaw}
In this section, we examine the Kennicutt-Schmidt (K-S) law (star formation law in terms of surface density) in our edge-on galaxy sample and compare the obtained K-S law indices with values from face-on systems found in other studies. We also test the role of the gravitational instability parameter $Q$ in determining the region where stars form. 
We are unable to derive volume density estimates for the SFR from the present data so we defer discussion of the volumetric star formation law (SFL) to a future study. 

\begin{figure}[!tbp]
\begin{center}
\begin{tabular}{c@{\hspace{0.1in}}c@{\hspace{0.1in}}}
\includegraphics[width=0.45\textwidth]{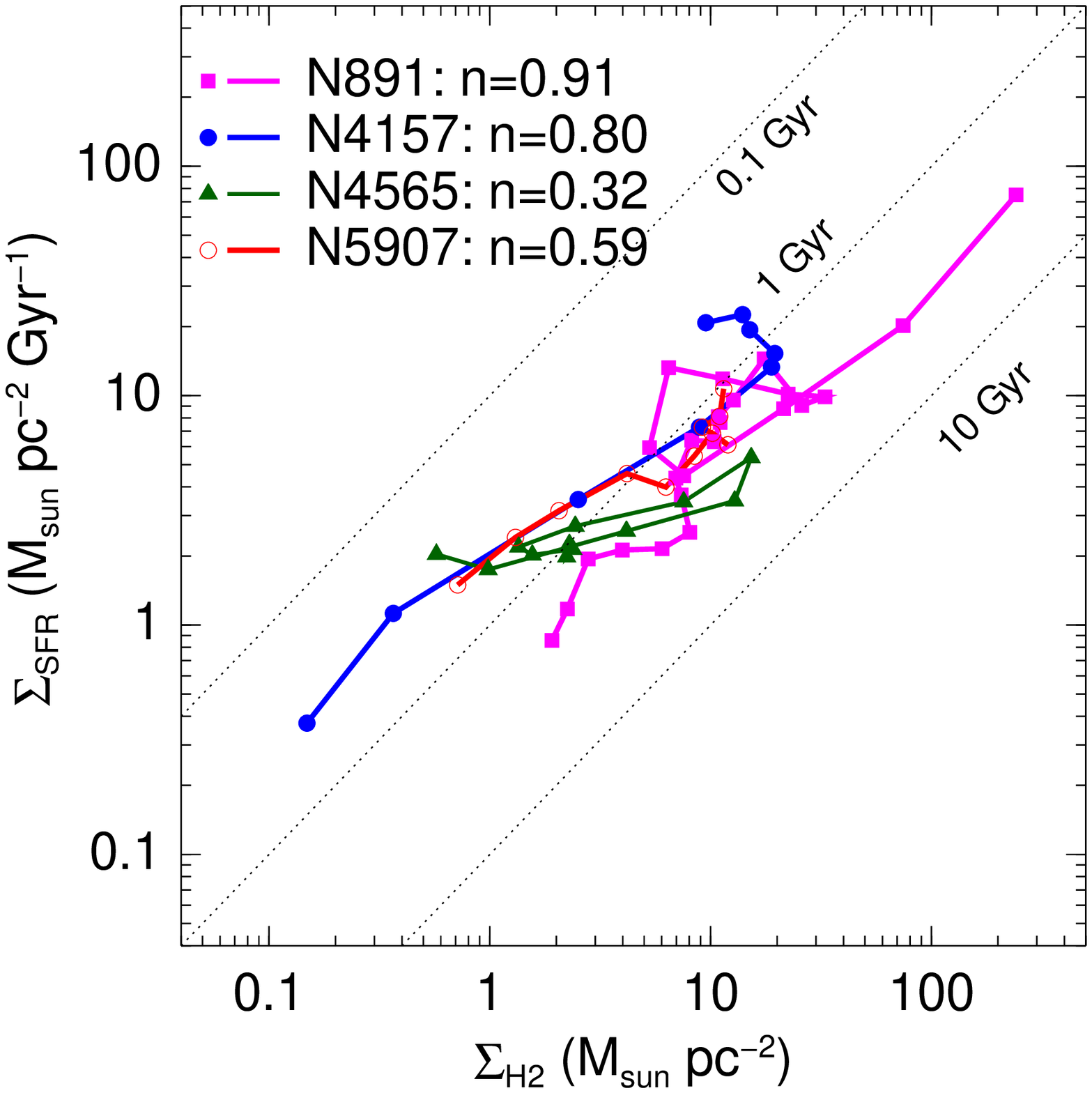}&
\includegraphics[width=0.45\textwidth]{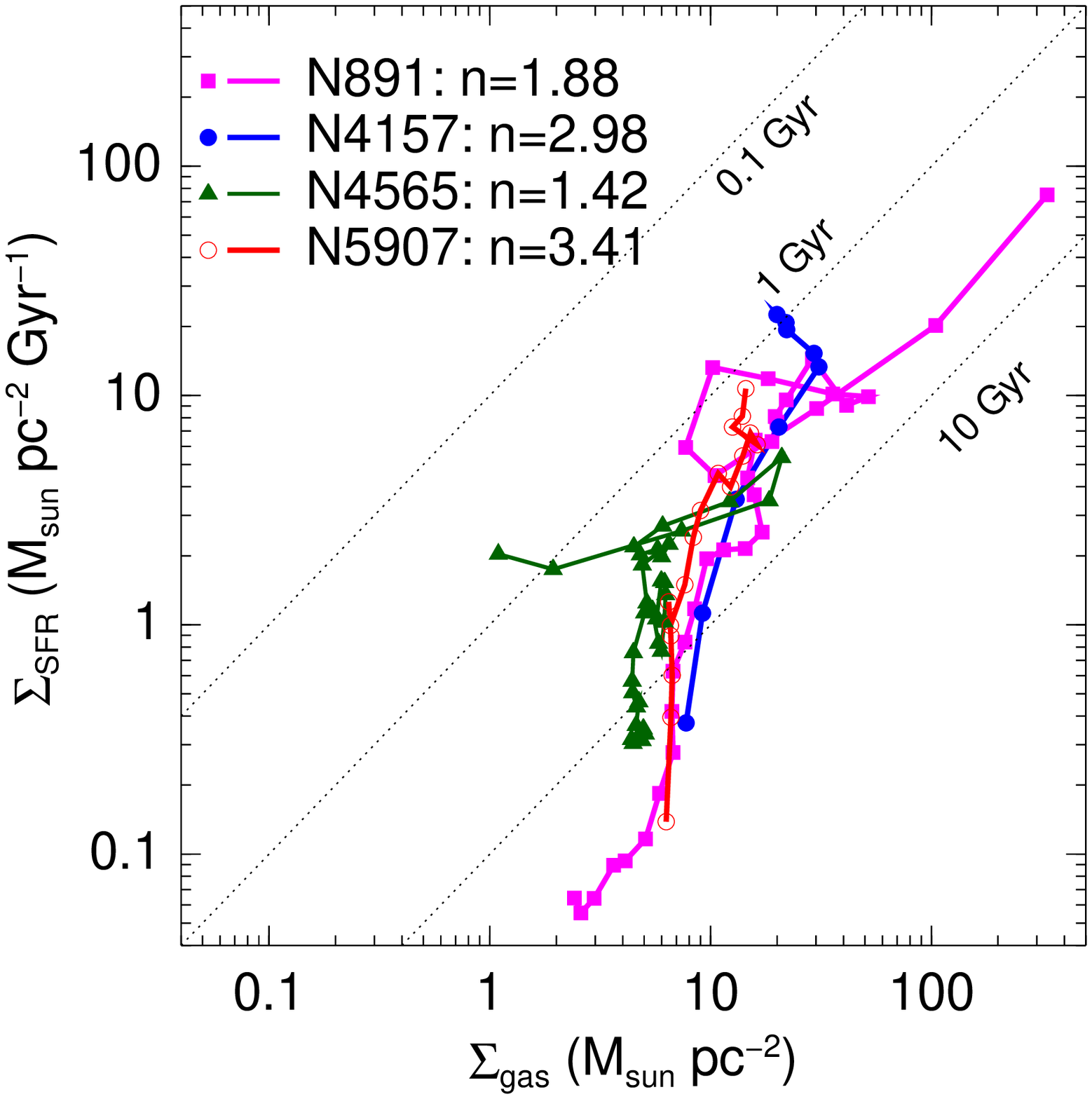}
\end{tabular}
\caption[Plots of \sigsfr\ vs. \sightwo\ and \sigsfr\ vs. \siggas]{The relationship between \sigsfr\ and \sightwo\ ($left$) and between \sigsfr\ and \siggas\ ($right$) for NGC 891, 4157, 4565, and 5907. The power-law index (Schmidt law index) is indicated in the upper-left corner. The dashed line shows constant SFE, with the gas depletion time (1/SFE) labeled.  
\label{sfrlaw}}
\end{center}
\end{figure}

\subsection{Kennicutt-Schmidt Law}
In Figure \ref{sfrlaw}, we plot \sigsfr\ versus  \sightwo\ (left) and \siggas\ (right) for our galaxy sample including NGC 891  in order to examine
the strength of the relationships and compare their slopes. 
The uncertainties on the SFR and the gas surface densities are shown in their radial profiles in Section \ref{radial}. In addition, about a factor of 2 uncertainty in the CO-to-\Htwo\ conversion factor (e.g., \citealt{2013ARA&A..51..207B}) and the 24 \um\ SFR calibration (e.g., \citealt{2012ApJ...761...97M}) are suggested by the literature. 
We use the ordinary least-squares (OLS) bisector (\citealt{1990ApJ...364..104I}) to fit the relationships between log quantities.
The obtained power-law index and RMS scatter (dex) around the fit for all galaxies are presented in Table \ref{table_sflaw}. 
Note that the index of NGC 891 is a bit different from Paper I since we use the OLS bisector instead of the least-squares fitting used in Paper I. 
In addition, the index of NGC 891 for the total gas is much higher (by a factor of 2) than the index in Paper I because we use gas data points up to $R = 400$\ac, instead of 300\ac\ in Paper I. 
Our results suggest that the correlation between \sigsfr\ and \sightwo\ is better than the correlation between \sigsfr\ and \siggas\ when considering the RMS scatter around the relationship. In addition, most individual relations between \sigsfr\ and \siggas\ appear to show distinct slopes at the low and high ends; this is most noticeable in NGC 4565. 

The most frequently cited relation given by \citet{1998ApJ...498..541K} suggests an index of 1.4: \sigsfr\ $\propto$ \siggas$^{1.4}$.
\citet{2002ApJ...569..157W} obtained weighted average indices of 0.78 (molecular gas) and 1.12 (total gas) from seven spiral galaxies. 
\citet{2005ApJ...625..763L} found an index of 1.3 for the relationship between SFR (traced by the radio continuum) and \Htwo\ in dwarf galaxies. 
\cite{2008AJ....136.2846B} derived average values of 0.96 for \Htwo\ and 1.85 for \HI\ + \Htwo\ from seven spiral galaxies. 
\cite{2008AJ....136.2846B} also obtained a smaller scatter for the molecular gas relation (0.2 dex) than for the total gas relation (0.3 dex). 
The overall tendency towards a steeper index in the correlation between  total gas and SFR compared to the  ``molecular star formation law" (\sigsfr\ $\propto$ \sightwo$^{\rm n}$) is also shown in our results since the relatively constant values in the atomic gas density profile steepen the total gas star formation law (\sigsfr\ $\propto$ \siggas$^{\rm n}$). 
However, the derived ``molecular" SFL indices are generally smaller (0.32--0.91) than found in previous studies, especially in NGC 4565 and 5907. 
Among the reasons for the shallow slopes in our results might be optical depth effects in the 24 \um\ emission which decreases the 24 \um\ brightness in bright regions or contribution from diffuse starlight heating which increases the 24 \um\ brightness in faint regions. 
As mentioned in Section \ref{radprof}, we have explored the opacity issue on 3.6 \um\ and obtained the uncertainty of a factor of 2 by the extinction effects. Since the 24um opacity should be less than the 3.6 \um\ opacity, we expect less than a factor of 2 increase in the central region at the 24 \um\ radial profile. When we applied the uncertainty by the extinction at 3.6 \um\ to the SFR radial profile, the indices for \Htwo\ are increased by a factor of $\sim$1.2 (NGC 4157), $\sim$1.4 (NGC 4565), and $\sim$1.2 (NGC 5907).
Nevertheless, the index of NGC 4565 is still smaller, but it is not much unexpected since the CO emission of NGC 4565 is deficient in the central regions as shown in the CO map (ring-like morphology) and the gradient of the 24 \um\ profile  is flatter than the others. In addition, NGC 4565 has an AGN, which can heat the dust to produce 24 \um\ even without much star formation. 
\citet{2011ApJ...730...72R,2012ApJ...745..183R}
have suggested that one possible reason for the flatter slopes in some galaxies among their sample might be a greater contribution from diffuse emission to the 24 \um\  emission. 
For this to explain the trends we see in NGC 4565 and NGC 5907, they would need to suffer a factor of $\sim$2 more diffuse 24 \um\ contamination than the other galaxies whose slopes are consistent with that of NGC 4254 without diffuse emission  \citep{2011ApJ...730...72R}. 

\begin{table}[!tb]\footnotesize
\begin{center}
\caption{Power-law indices for the star formation law \label{table_sflaw}}
\begin{tabular}{ccccc}
\tableline\tableline
Galaxy&\multicolumn{2}{c}{Molecular Gas}&\multicolumn{2}{c}{Total Gas}\\
 &Index&RMS scatter&Index&RMS scatter\\
\tableline
NGC 891&0.91&0.19&1.88&0.42\\
NGC 4157&0.80&0.15&2.98&0.25\\
NGC 4565&0.32&0.06&1.42&0.36\\
NGC 5907&0.59&0.08&3.41&0.23\\
\tableline
\end{tabular}
\end{center}
\end{table}

\begin{figure}[!tbp]
\begin{center}
\begin{tabular}{c@{\hspace{0.1in}}c@{\hspace{0.1in}}}
\includegraphics[width=0.45\textwidth]{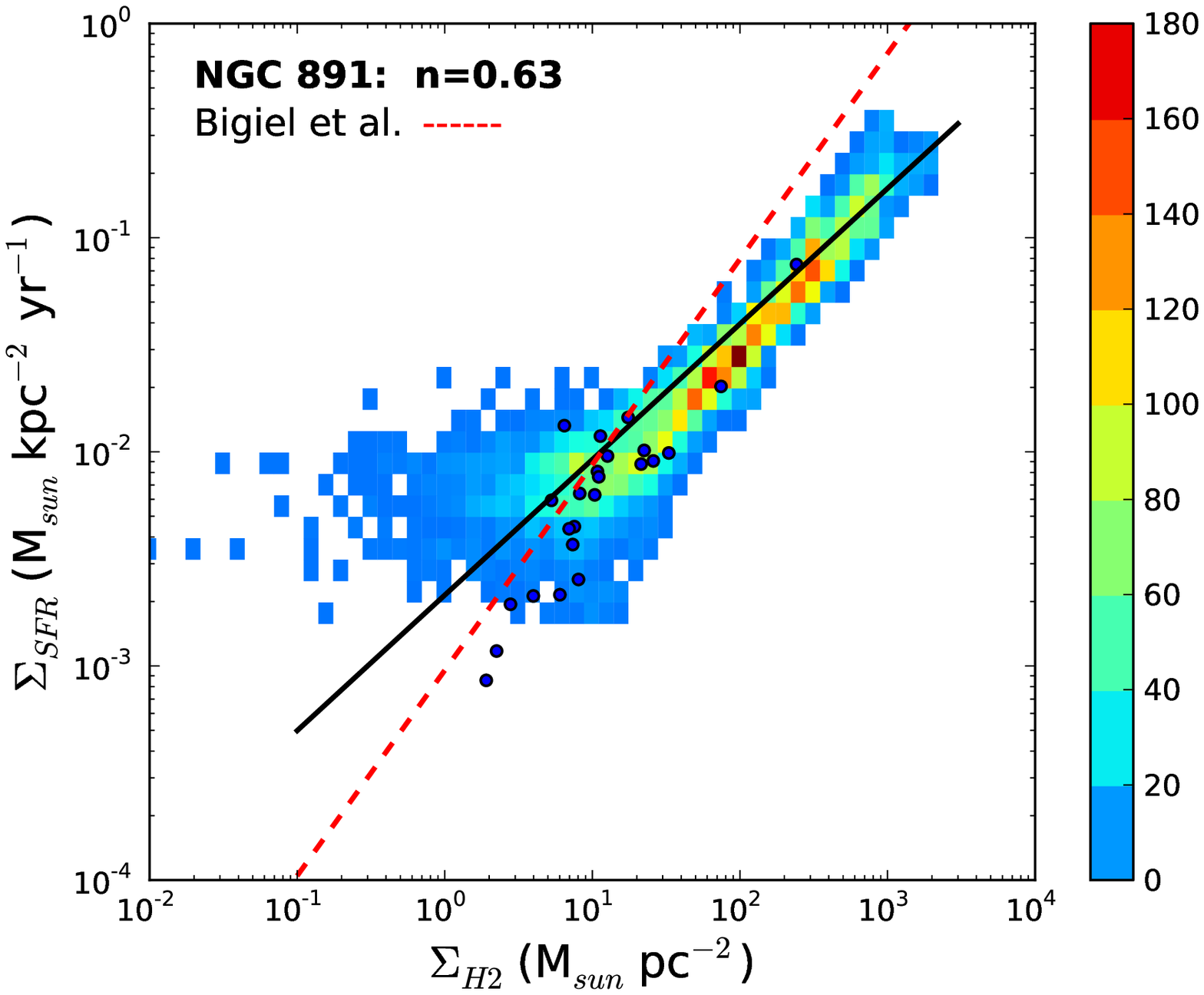}&
\includegraphics[width=0.45\textwidth]{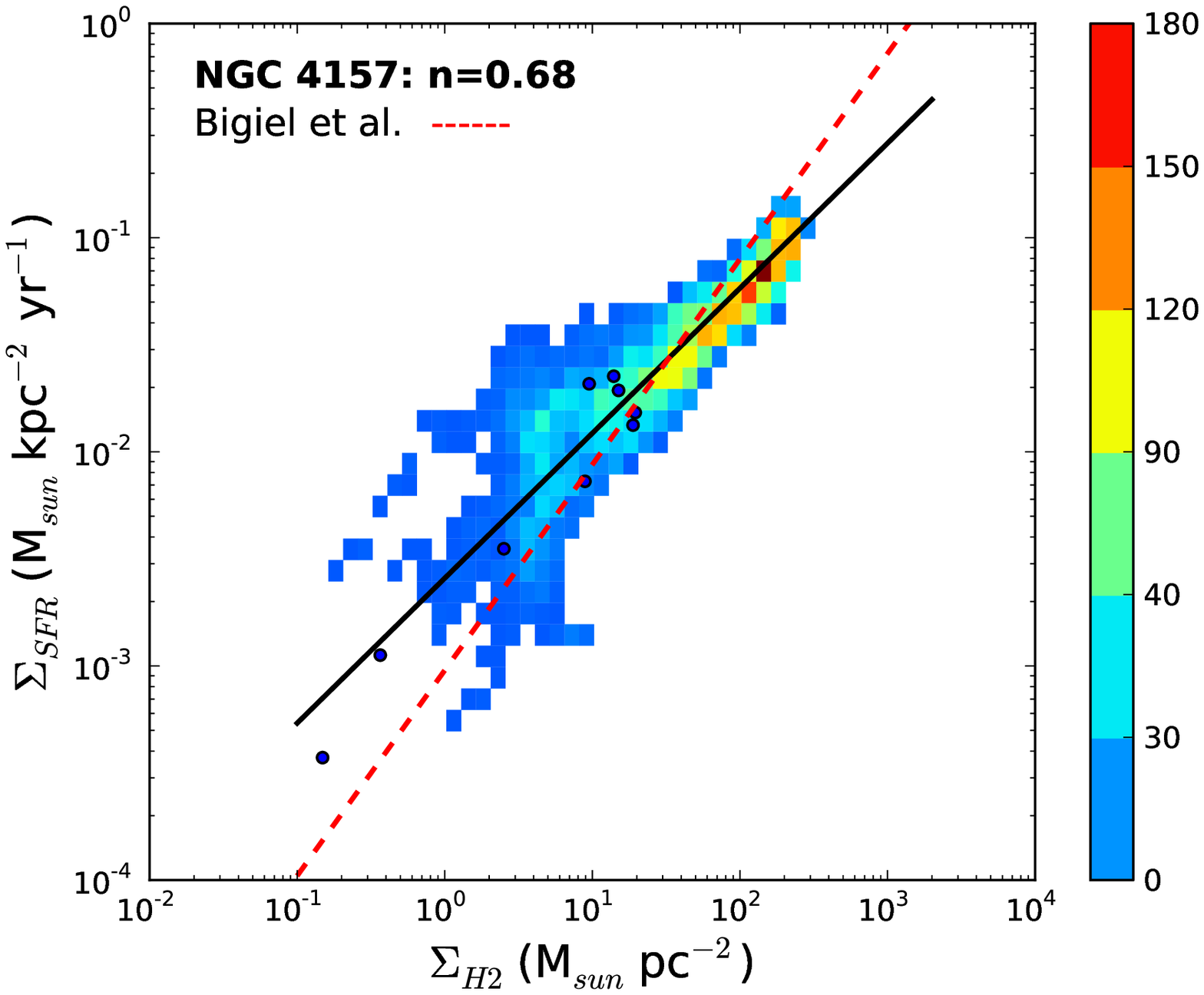}\\
\includegraphics[width=0.45\textwidth]{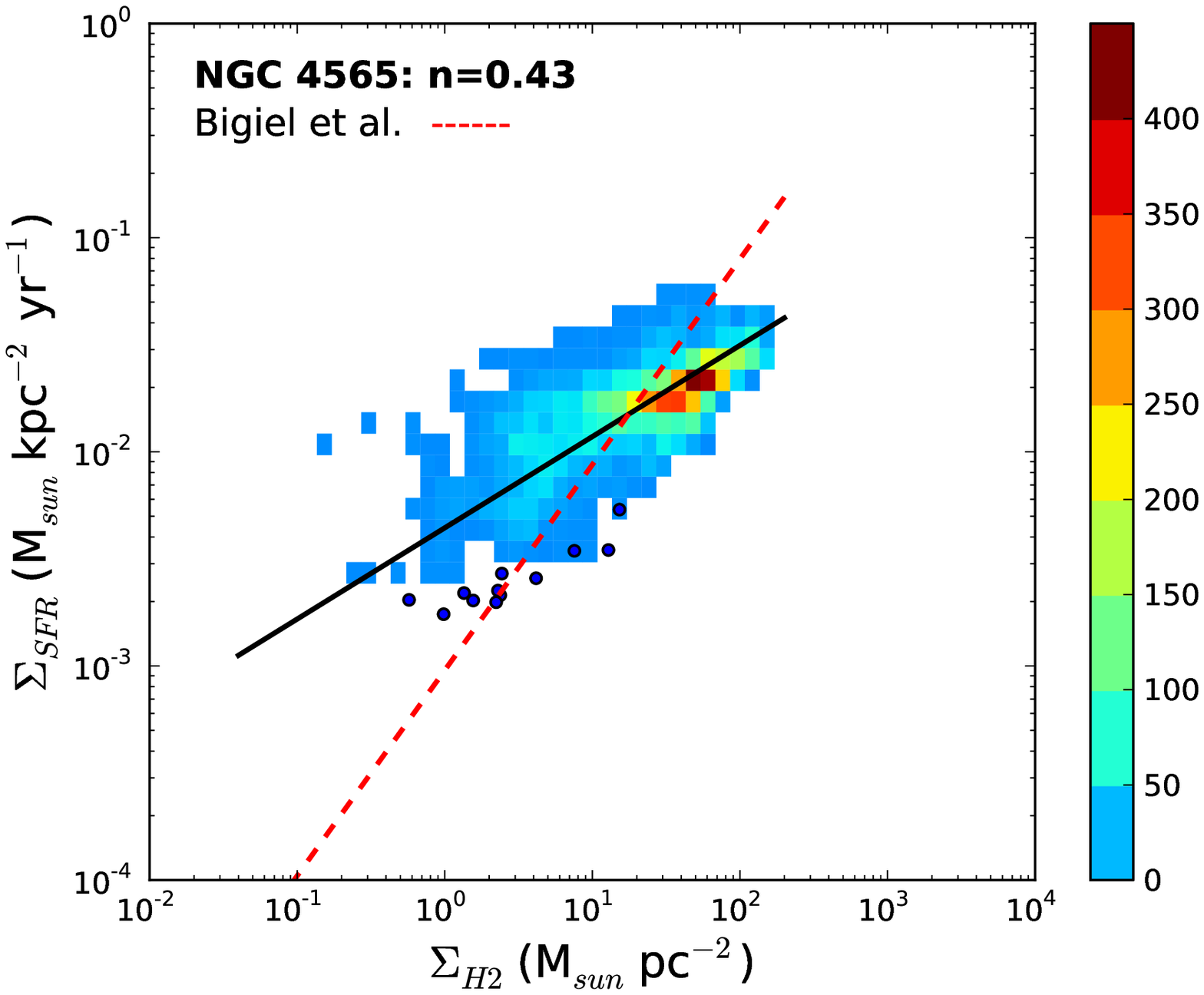}&
\includegraphics[width=0.45\textwidth]{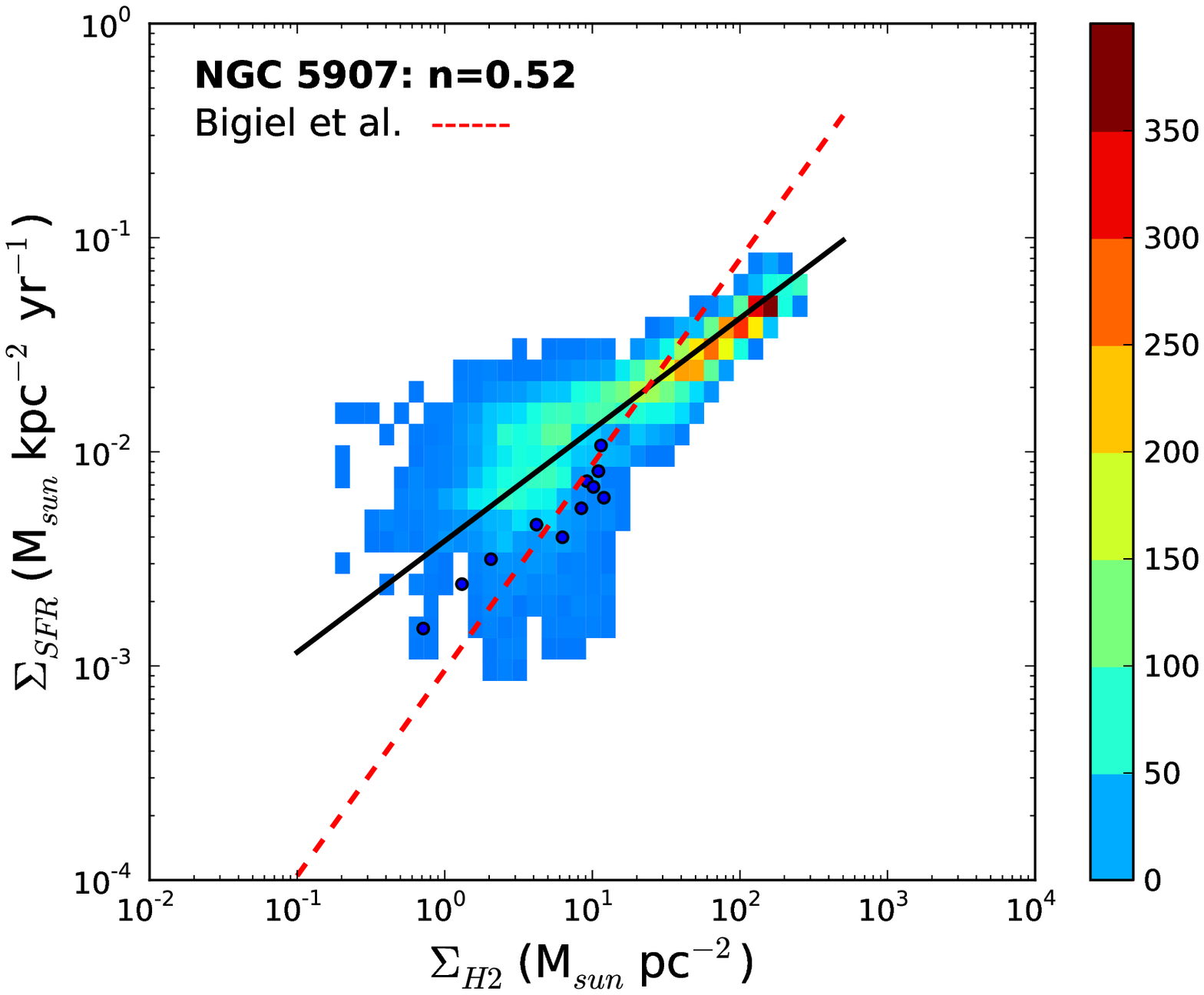}
\end{tabular}
\caption[Plot of \sigsfr\ vs. \sightwo\ based on a  pixel-by-pixel analysis]{The relationship between \sigsfr\ and \sightwo\ based on a  pixel-by-pixel analysis. Color contours represent the numbers of the data points in a pixel (0.1 dex by 0.1 dex). The blue poionts show the relationship based on the radial profile shown in Fig. \ref{sfrlaw} (left). The solid line is the best fit to the pixel-by-pixel data using the OLS bisector. The red dashed line shows the ``molecular gas" Schmidt law index of 0.96 based on the pixel-by-pixel analysis of \cite{2008AJ....136.2846B}.
\label{p2p_co}}
\end{center}
\end{figure}

\begin{figure}[!tbp]
\begin{center}
\begin{tabular}{c@{\hspace{0.1in}}c@{\hspace{0.1in}}}
\includegraphics[width=0.45\textwidth]{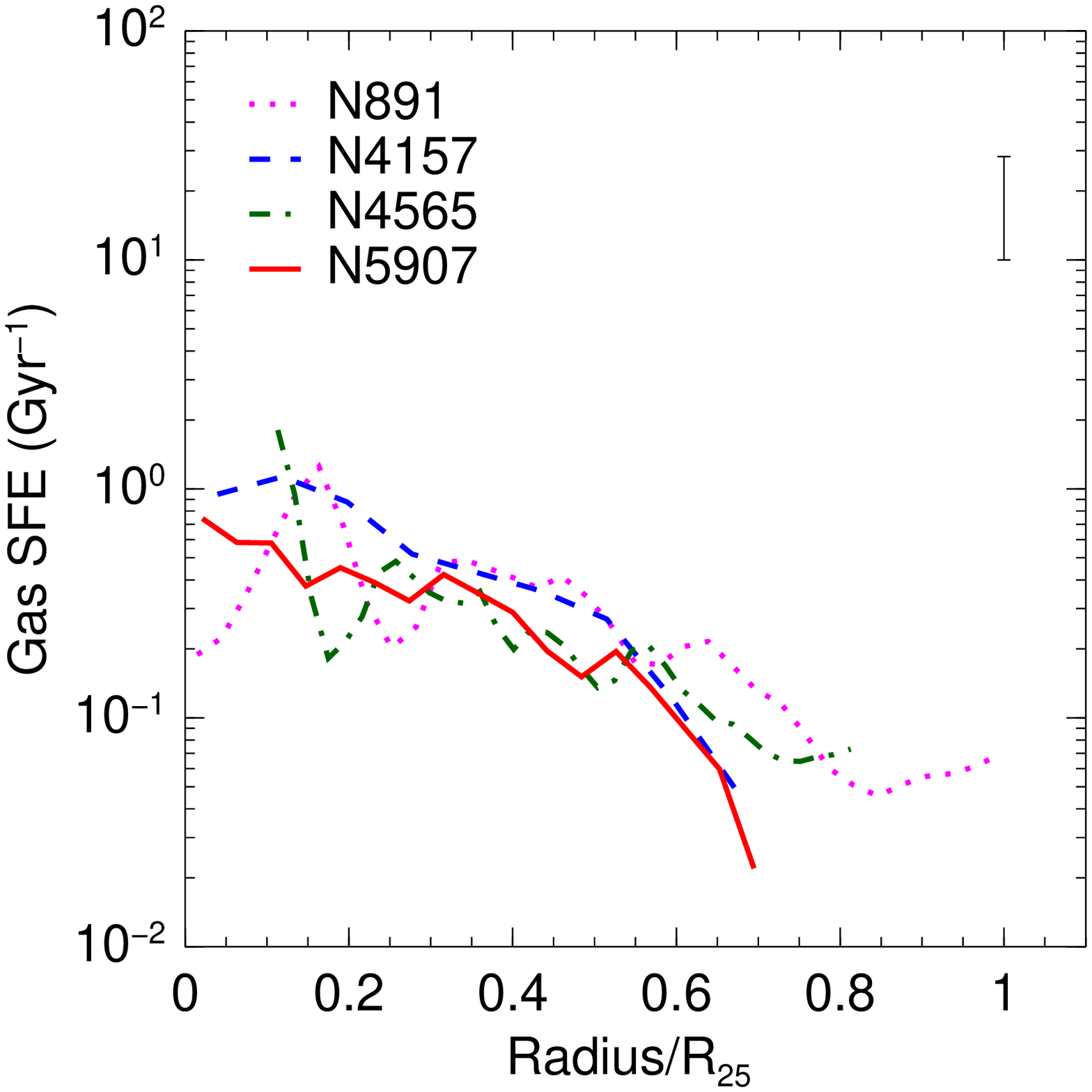}&
\includegraphics[width=0.45\textwidth]{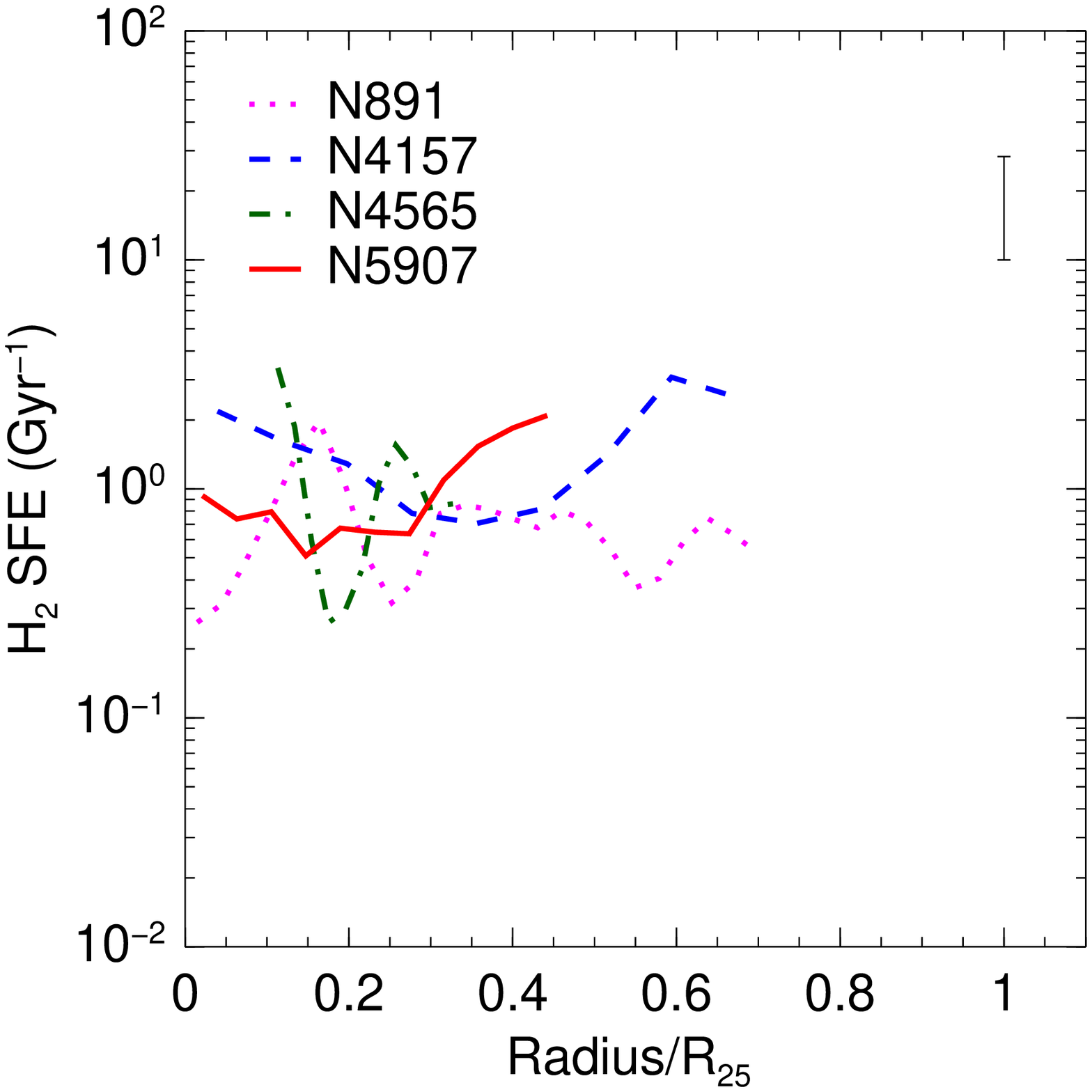}
\end{tabular}
\caption[Star formation efficiency as a function of radius]{Star formation efficiency as a function of radius normalized by $R_{25}$ in terms of total gas ($left$) and the molecular gas ($right$) for NGC 891, 4157, 4565, and 5907. The representative error bar of the SFE is shown in the upper right corner.
\label{sfe}}
\end{center}
\end{figure}


In order to check whether the difference between RADPROF (SFR) and PVD (CO) methods is responsible for the flatter slope in the SFR vs.\ molecular gas relation, we can also investigate the relationship between \sigsfr\ and \sightwo\ on a pixel-by-pixel basis. 
For the pixel-by-pixel comparison, we use the masked CO intensity maps. 
Since most of the galaxies have very low resolution in \HI\ compared to CO, we do not attempt a pixel-by-pixel comparison between \sigsfr\ and \siggas\ and focus on the ``molecular star formation law" with the CO map  convolved to the 24 \um\ Gaussian beam (7\ac) for NGC 4157, 4565, and 5907. The CO beam of NGC 891 is 7\ac.
Using the beam matched maps, we extract data on a  pixel-by-pixel basis using the MIRIAD task IMCMP. All the extracted data are placed in 0.1 dex bins and the number of data points in each bin is mapped to the color scale in Figure \ref{p2p_co}.
The blue solid points in the figures show the correlations from the radial profile analysis for comparison purposes. 
Most of the radial profile points for each galaxy follow the mean trends defined by the pixel-by-pixel plot (Fig. \ref{p2p_co}), but their values are generally lower compared to the contours since the projection effect of an edge-on galaxy leads to higher values in the pixel-by-pixel method. In particular, NGC 4565 shows much lower values for the radial profiles, due to its ring-like morphology leading to lower surface densities, especially in the central region. Note in our pixel-by-pixel analysis, unlike most K-S analyses, no deprojection is done. 

The best-fit line in the figures is obtained  using the OLS bisector and the fitted power-law index (K-S law index) is shown in the upper-left corner of the figure. 
The obtained indices from the pixel-by-pixel analysis are in the range of 0.43 -- 0.68 for the molecular star formation law. Those values are similar to the indices from the radial profile analysis. This is in contrast to the results of \cite{2008AJ....136.2846B} who obtained  a power-law index of about unity (shown as red dashed line in Fig. \ref{p2p_co}) for the \Htwo\ star formation relation based on a pixel-by-pixel comparison for seven spiral galaxies.  

Figure \ref{sfe}  shows the star formation efficiency (SFE) as a function of radius (normalized by $R_{25}$) in terms of total gas (SFE = \sigsfr/\siggas) and the molecular gas (\sigsfr/\sightwo). 
The inverse of this quantity (SFE$^{-1}$) represents the gas depletion time.
Overall, most galaxies show declining SFE (in terms of total gas) with radius even though fluctuations in SFE exist in some galaxies. On the other hand, the molecular SFE in Figure \ref{sfe} (right) appears to increase with radius in some galaxies. That is caused by a steeper drop in the \sightwo\ profile compared to that of SFR  and this situation could be due in part to the 24 \um\ opacity and diffuse emission issues. 
Once again this contrasts with previous results which find a roughly constant SFE in the molecular gas. 
\citet{1999AJ....118..670R} observed the constancy of SFE in a study of the SFE within CO-emitting galaxies. \citet{2008AJ....136.2846B} and \citet{2008AJ....136.2782L} showed a constant SFE in the \Htwo\ dominated region and decreasing SFE with radius in the \HI\ dominated region.


\begin{figure}[!tbp]
\begin{center}
\begin{tabular}{c@{\hspace{0.1in}}c@{\hspace{0.1in}}}
\includegraphics[width=0.45\textwidth]{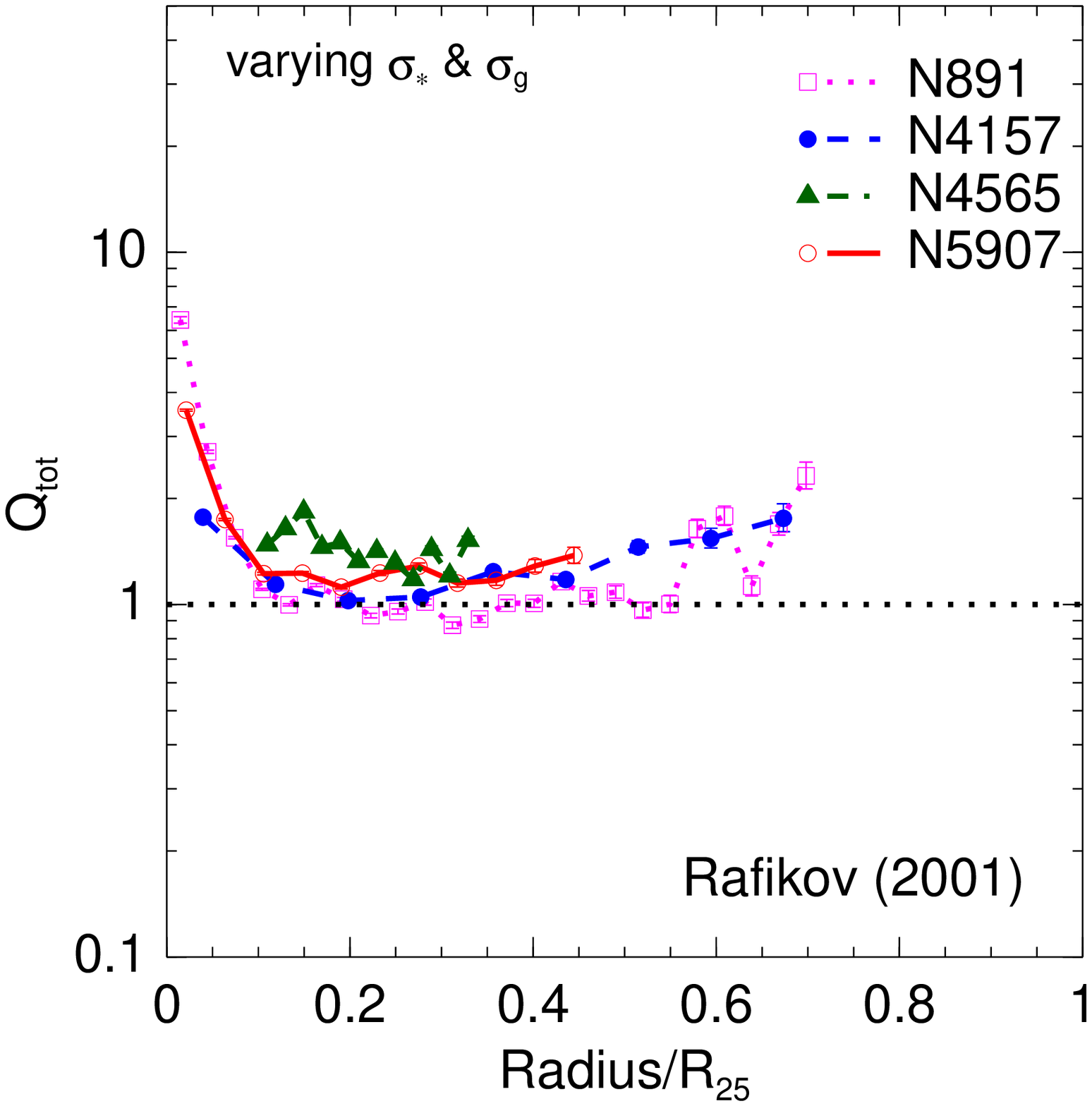}&
\includegraphics[width=0.45\textwidth]{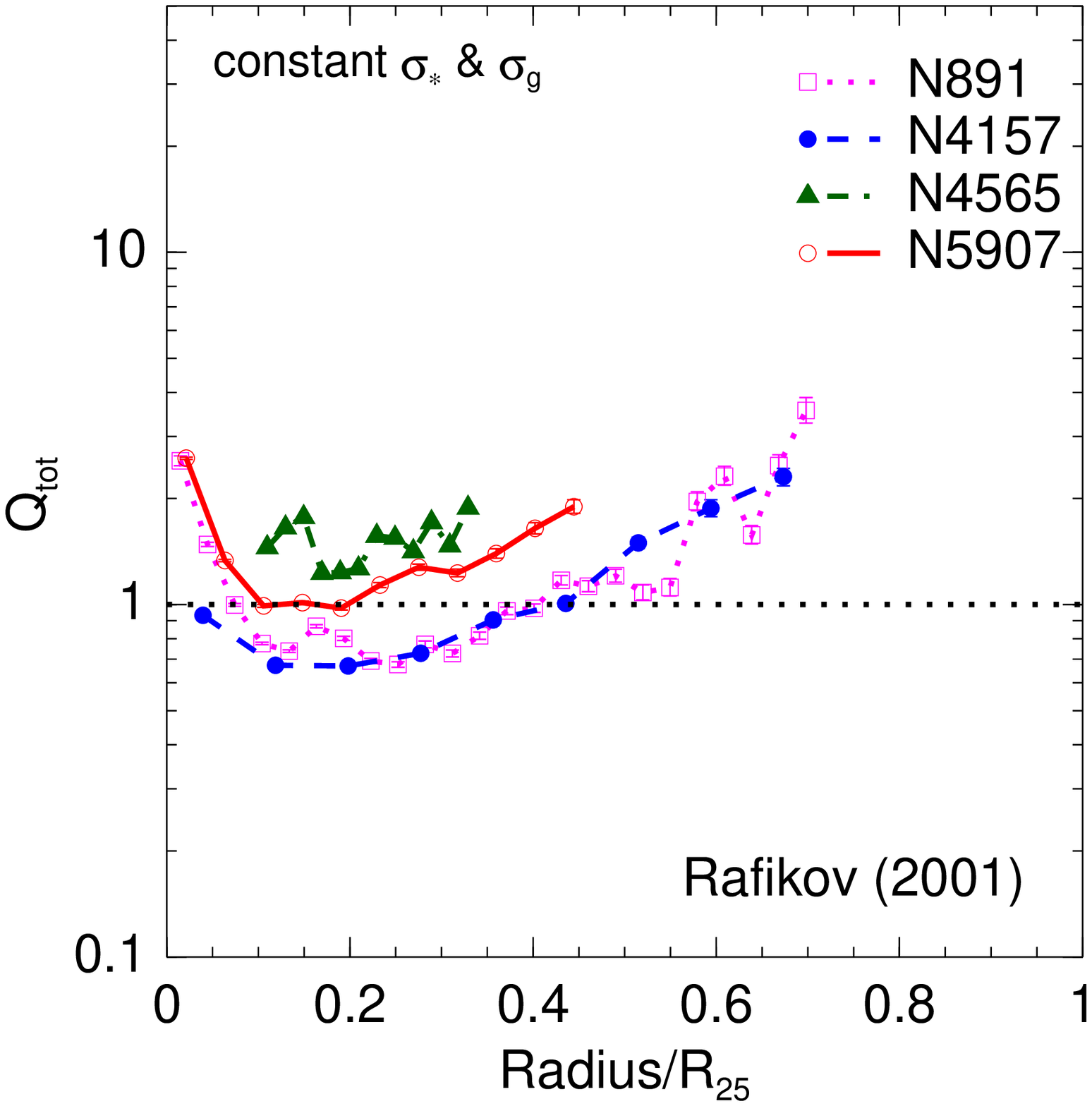}\\
\includegraphics[width=0.45\textwidth]{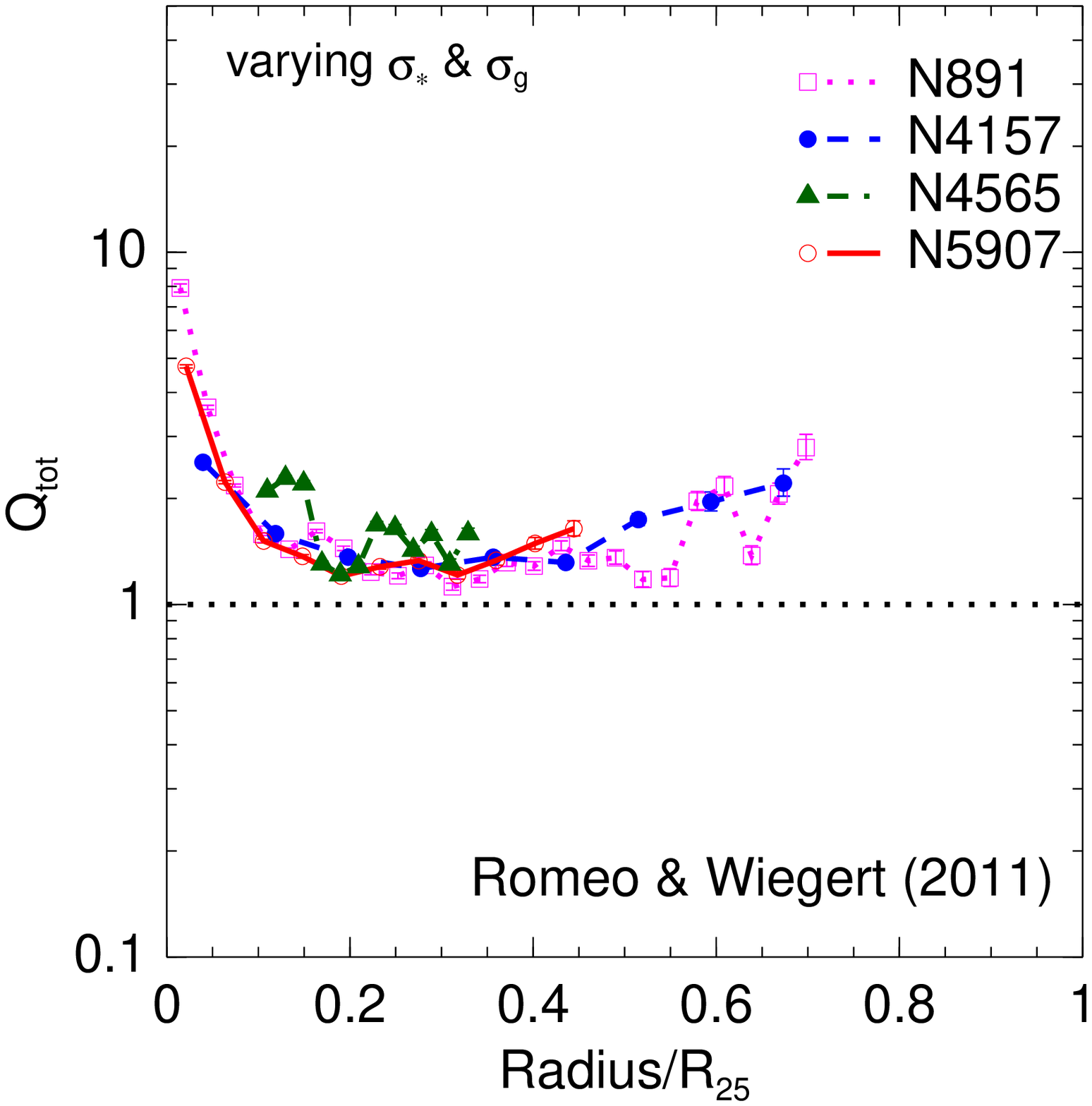}&
\includegraphics[width=0.45\textwidth]{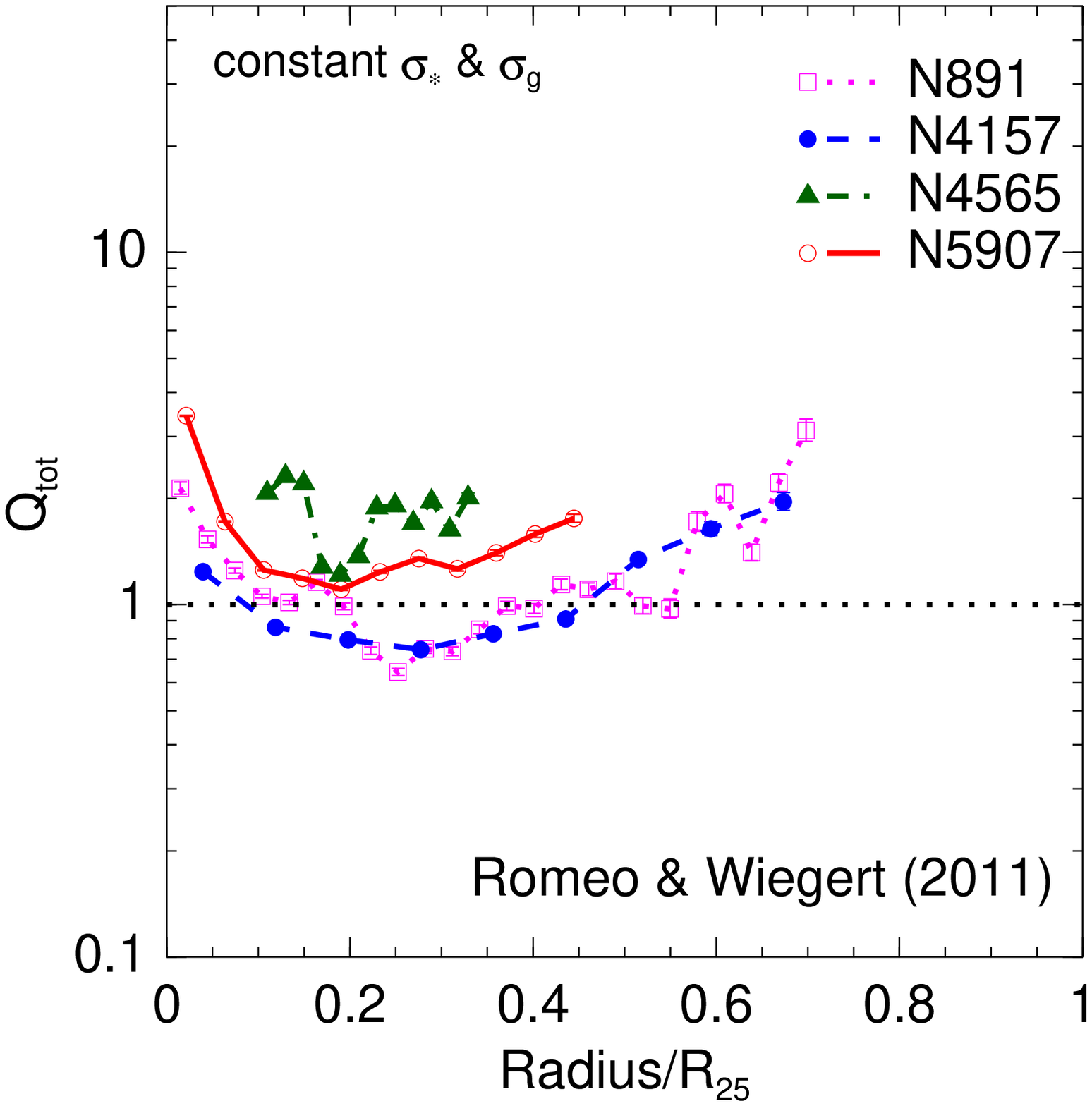}
\end{tabular}
\caption[$Q_{\rm tot}$ as a function of radius]{$Q_{\rm tot}$ as a function of radius normalized by $R_{25}$ for varying velocity dispersions (left panels) and constant velocity dispersions (right panels). Unstable regions lie below the horizontal dotted lines. 
The top panels are obtained using $Q_{\rm tot}$ provided by \citet{2001MNRAS.323..445R} and the bottom panels use the \citet{2011MNRAS.416.1191R} approximation. 
\label{Qpara}}
\end{center}
\end{figure}

\subsection{Gravitational Instability}
\label{gravQ}

Since both gas and stars can contribute to the gravitational instability of a galactic disk, many proposals have been made to evaluate stability using a single parameter $Q_{\rm tot}$ analogous to the \citet{1964ApJ...139.1217T} $Q$ parameter but including the properties of both gaseous and stellar disks (e.g. \citealt{1984ApJ...276..114J}; \citealt{2001MNRAS.323..445R}; \citealt{2011MNRAS.416.1191R}).  A link between gravitational instability and star formation has been frequently suggested (\citealt{1989ApJ...344..685K}; \citealt{2001ApJ...555..301M}; \citealt{2005ApJ...620L..19L}) but has been difficult to confirm observationally (\citealt{2002ApJ...569..157W}; \citealt{2008AJ....136.2782L}).  Recently, \citet{2011ApJ...737...10E} has suggested that $Q_{\rm tot}$ primarily governs the spiral structure on large scales and has little direct influence on local collapse leading to star formation.  He also suggests that gas dissipation can increase the stability threshold from $\sim$1 to 2--3 and allow small-scale instability to persist even at large $Q_{\rm tot}$.
In order to investigate the relationship between the gravitational instability parameter $Q$ and massive star formation, 
we obtain the parameter $Q_{\rm tot}$ using a modified version (to treat \HI\ and \Htwo\ separately) of the equation provided by \citet{2001MNRAS.323..445R}, assuming two components of collisional gas and collisionless stars in a thin rotating galactic disk:
\begin{equation}
\frac{1}{Q_{\rm tot}} = \frac{2}{Q_{\rm HI}}\Gamma_{\sigma_{\rm HI}} \frac{q}{1+q^2\Gamma_{\sigma_{\rm HI}}^2} + \frac{2}{Q_{\rm H_2}}\Gamma_{\sigma_{\rm H_2}} \frac{q}{1+q^2\Gamma_{\sigma_{\rm H_2}}^2} + \frac{2}{Q_{\rm star}}\frac{1}{q}[1-e^{-q^2}I_0(q^2)] > 1,
\label{Qgs}
\end{equation}
where
\begin{eqnarray}
Q_{\rm HI} = \frac{\kappa \sigma_{\rm HI}}{\pi G \sighi}, \qquad 
Q_{\rm H_2} = \frac{\kappa \sigma_{\rm H_2}}{\pi G \sightwo}, \qquad 
Q_{\rm star} = \frac{\kappa \sigma_{*,R}}{\pi G \sigstar},\\
\Gamma_{\sigma_{\rm HI}}  = \sigma_{\rm HI}/\sigma_{*,R}, \qquad \Gamma_{\sigma_{\rm H_2}}  = \sigma_{\rm H_2}/\sigma_{*,R},\\
q = \frac{k \sigma_{*,R}}{\kappa},\qquad 
\kappa = \frac{V_{\rm rot}}{R} \sqrt{2\left(1 + \frac{R}{V_{\rm rot}}\frac{dV_{\rm rot}}{dR}\right)}.  
\end{eqnarray}
Here $\sigma_{*,R}$ is the stellar velocity dispersion in the radial direction,  estimated from the vertical velocity dispersion:  $\sigma_{*,R} = \sigma_*$/0.6 (\citealt{1993A&A...275...16B}). $I_0$ is the Bessel function of zero order, $k$ is the wavenumber ($2\pi/\lambda$), $\kappa$ is the epicyclic frequency, and $V_{\rm rot}$ is the rotational velocity, obtained using Equation \ref{Vrot} with a correction for the inclination: $V_{\rm rot} = V_{\rm proj}/\sin i$. 
Note that $Q_{\rm tot}$ is the minimum among those values we can obtain with a  range of $\lambda$; usually the selected $\lambda$ (providing the minimum $Q_{\rm tot}$) is up to about 4 kpc. 
In this procedure, we use the inferred vertical velocity dispersion as a function of radius (Figure \ref{vdisp}) to derive the parameter $Q_{\rm tot}$ shown in the left panel of Figure \ref{Qpara} for the galaxies including NGC 891 from Paper I. 
The error bars of $Q_{\rm tot}$ show maximum and minimum values implied by the uncertainties in \sightwo, \sighi, and \sigstar. Since the differences between the maximum and minimum values are very small, the error bars are not very noticeable. 
The unstable condition is $Q_{\rm tot} < 1$. The results show that most galaxies are marginally stable ($Q \sim 1$) in most regions, supporting the self-regulation of star formation (e.g., \citealt{2010ApJ...721..975O}; \citealt{2011ApJ...743...25K}). 
The $Q_{\rm tot}$ value is increasing toward the center, contradicting the general expectation of an  unstable central region.
This tendency is also shown in a study of  \citet{2008AJ....136.2782L} even though they used a constant velocity dispersion for gas (11 \kms) and varying velocity dispersion of stars proportional to \sigstar$^{0.5}$. 
The increase of $Q$ toward the center appears to be due to the epicyclic frequency 
associated with the steeply rising slope of the rotation curve.
It is well known that a strong central mass concentration makes $Q$ increase in the central regions; this is called the ``$Q$ barrier" and is discussed by \citet{1985MNRAS.217..127S} and others.
Generally, the $Q_{\rm tot}$ profiles beyond the central regions are consistent with the total gas SFE; $Q$ is rising as the SFE decreases.  

In Figure \ref{Qpara} (top right), we obtain another $Q_{\rm tot}$ profile with the vertical velocity dispersions assumed to be constant ($\sigma_*$ = 21 \kms\ and $\sigma_{\rm g}$ = 8 \kms), which is commonly assumed in the literature (e.g., \citealt{2001MNRAS.323..445R}; \citealt{1997A&A...326..554C}).
In this regime, two galaxies (NGC 891 and 4157) show unstable regions predicting star formation in the inner disk, while the other galaxies (NGC 4565 and 5907) show a marginally stable disk. 
Considering the relatively low SFR in NGC 4565 and 5907 compared to the SFR in NGC 891 and 4157,  the $Q_{\rm tot}$ profile with constant velocity dispersions shows a reasonable prediction for the massive star formation. However, the assumed constant values are in conflict with the inferred radial variation in the velocity dispersion. 
The increase toward the center is also shown in this profile.

The different $Q_{\rm tot}$ profiles for constant and varying velocity dispersions show how much the adopted velocity dispersions affect the parameter. 
The variable velocity dispersions bring the $Q$ vs. $R$ trends into better agreement galaxy-to-galaxy. 
By plotting individual $Q$ profiles for gas and stars in both cases, it appears that $Q_{\rm tot}$ is affected mostly by $Q_{\rm star}$. The range and values of $Q_{\rm gas}$ are very large compared to $Q_{\rm star}$ and the profile shape of $Q_{\rm tot}$ follows that of $Q_{\rm star}$. One can see the individual $Q$ profiles in Paper I (Figure 15) for NGC 891.

For comparison purposes, we also applied the recent approximation considering realistically thick disks given by  \citet{2011MNRAS.416.1191R}:
 \begin{displaymath}
\frac{1}{Q_{\rm tot}} = \left\{
\begin{array}{ll} 
\frac{\textrm {$W$}}{\textrm{$T_*Q_{\rm star}$}} + \frac{\textrm{1}}{\textrm{$T_{\rm g}Q_{\rm gas}$}} & \textrm{if\,\, $T_*Q_{\rm star} \ge T_{\rm g}Q_{\rm gas}$},\\
\\
\frac{\textrm{1}}{\textrm{$T_*Q_{\rm star}$}} + \frac{\textrm{$W$}}{\textrm{$T_{\rm  g}Q_{\rm gas}$}} & \textrm{ if\,\, $T_{\rm g}Q_{\rm gas} \ge T_*Q_{\rm star}$}, 
\end{array} \right.
\end{displaymath}
where
\begin{eqnarray}
W = \frac{2s}{1+s^2}, \qquad  s = \frac{\sigma_{\rm g}}{\sigma_{*,R}}, \qquad
T \approx 0.8 + 0.7 \left( \frac{\sigma_z}{\sigma_R} \right). 
\end{eqnarray}
Here the ratio of vertical to radial velocity dispersion  ($\sigma_z/\sigma_R$)  is 0.6 for stars and 1 for gas. 
The $Q_{\rm tot}$ profiles using varying and constant $\sigma$ are shown in the bottom panels of Figure \ref{Qpara}. Even though there is not much difference between the two different approaches provided by \citet{2001MNRAS.323..445R} and  \citet{2011MNRAS.416.1191R}, the Rafikov $Q_{\rm tot}$ values using varying $\sigma$ seem consistently lower than the Romeo and Wiegert values by enough that the Rafikov values are $\sim$1 but the Romeo and Wiegert values are usually $\geq$1. 
Overall, they show similar behaviour in the profiles although the values are slightly different.

\begin{figure}[!tbp]
\begin{center}
\begin{tabular}{c@{\hspace{0.1in}}c@{\hspace{0.1in}}}
\includegraphics[width=0.45\textwidth]{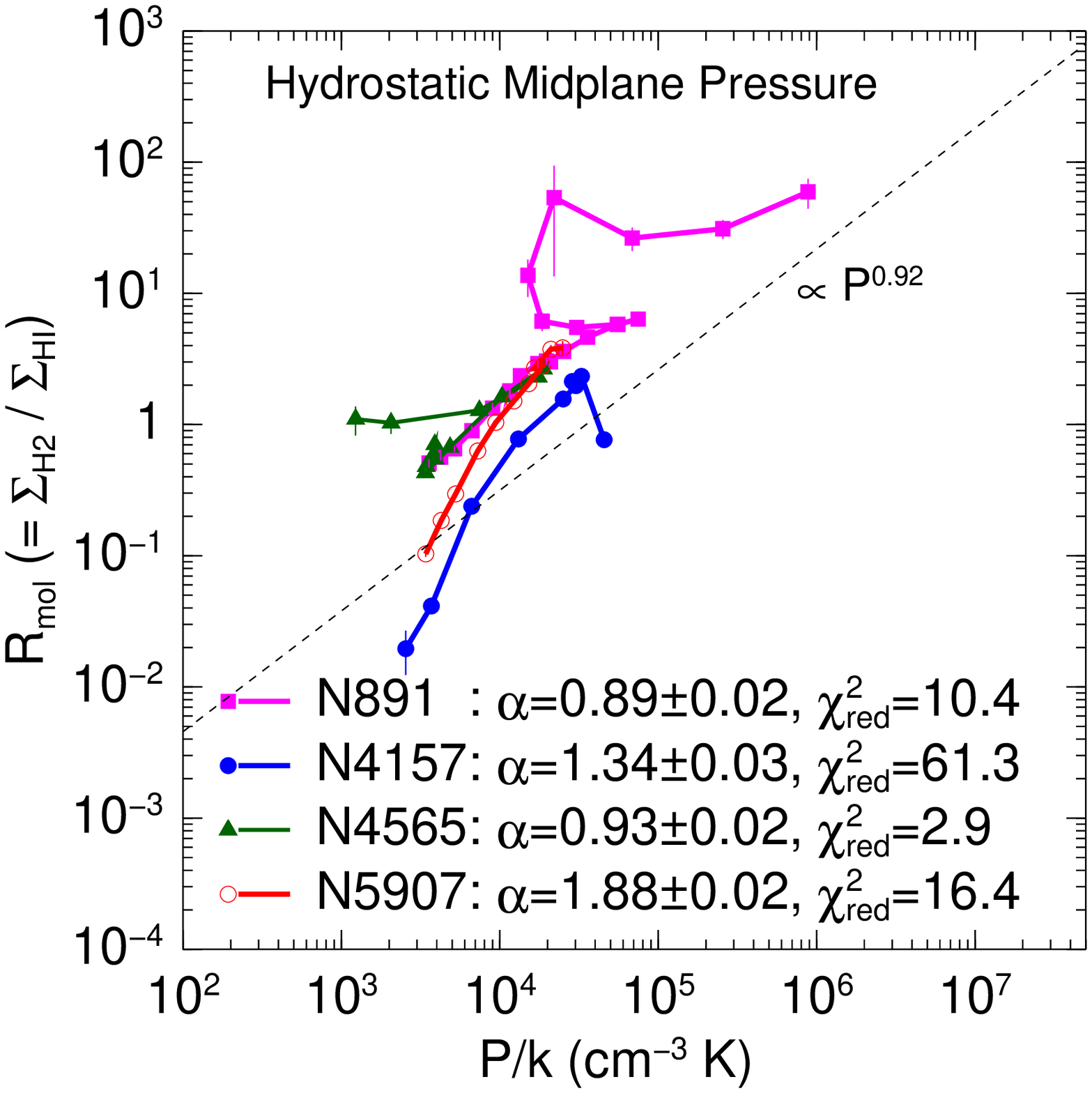}&
\includegraphics[width=0.45\textwidth]{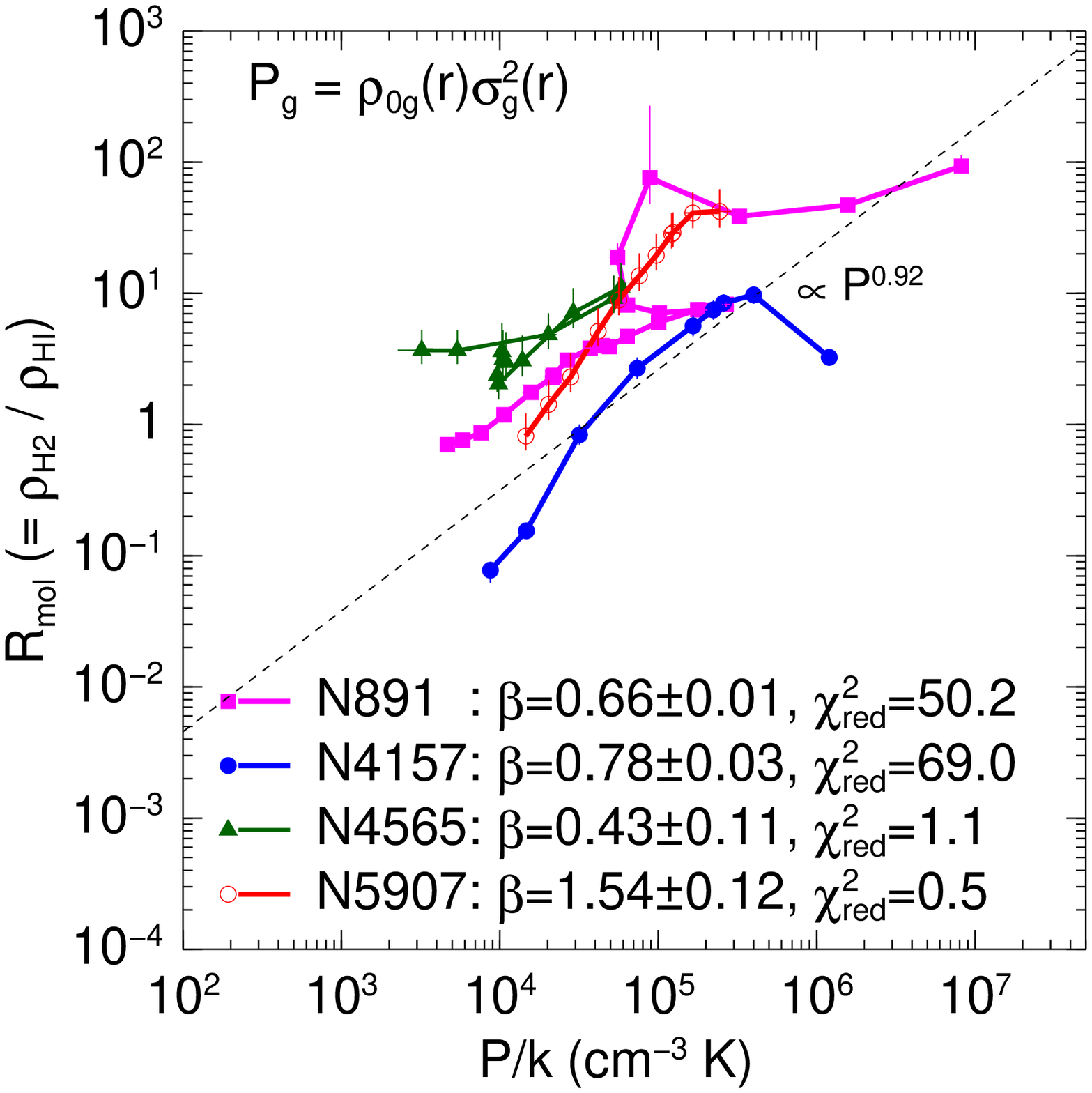}
\end{tabular}
\caption[Plot of \Htwo/\HI\ vs. pressure]{$Left$: \sightwo/\sighi\ as a function of the hydrostatic midplane pressure. $Right$: \rhohtwo/\rhohi\ as a function of the interstellar gas pressure $P_{\rm g}$ using the derived volume density at the midplane and varying velocity dispersion. The obtained power-law slopes are shown in the bottom. The dotted line represents the power-law relationship with slope of 0.92 from \citet{2006ApJ...650..933B} for comparison. 
\label{P_rmol}}
\end{center}
\end{figure}

\section{Molecular to Atomic Gas Ratio}
\label{rmol}

The ratio of molecular to atomic gas density is expected to scale with the interstellar gas pressure (e.g., \citealt{1993ApJ...411..170E}).
Since SFR is strongly correlated with the molecular gas, the interstellar pressure may provide a prescription for the star formation rate in a galaxy, if there is a strong correlation between the interstellar pressure and the ratio. 
Therefore, we examine whether the correlation is applicable in our sample of galaxies.
First, we have investigated the relationship between the hydrostatic midplane pressure and the molecular to atomic gas surface density ratio (\sightwo/\sighi). In order to obtain the hydrostatic pressure ($P_0$), we use the equation given by Paper I:
\begin{equation}
P_{0} = 0.89 (G\sigstar)^{0.5} \siggas \frac{\sigma_{\rm g}}{z_*^{0.5}}\;.
\label{Ph}
\end{equation}
This equation is very close to that derived by \citet{2004ApJ...612L..29B}. We assume the gas velocity dispersion $\sigma_{\rm g}$ = 8 \kms\ following \citet{2004ApJ...612L..29B}. For the constant stellar scale height ($z_*$), the values obtained by fitting the exponential model to 3.6 \um\ maps in Section \ref{radprof}  are used: 780 pc, 640 pc, and 670 pc for NGC 4157, 4565, and 5907, respectively. 
In order to compare with the power-law relationship found by \citet{2006ApJ...650..933B}, we plot the ratio \sightwo/\sighi\ against the hydrostatic pressure in Figure \ref{P_rmol} (left panel) and obtain a power-law slope ($\alpha$) by least-squares fitting:
\begin{equation}
\frac{\sightwo}{\sighi} = \left(\frac{P_0}{P_{\rm tr}}\right)^\alpha, 
\label{blitz}
\end{equation}
where $P_{\rm tr}$ is the hydrostatic pressure at the transition radius where \sightwo/\sighi\ = 1 in the fit.
The obtained power-law indices and reduced $\chi^2$ (goodness of fit) values are shown in the bottom of the figure.
The dotted line in the figure represents a power-law relationship with index of 0.92 as found by \citet{2006ApJ...650..933B} for comparison. 
It seems that most of galaxies are well fitted by the power-law relation 
even though the range of slopes is large.
However, the index of 0.92 is an average value and the range of slopes for the  sample of galaxies in the study by  \citet{2006ApJ...650..933B} is 0.58 -- 1.64.
The relation in two galaxies (NGC 4157 and 5907) appears to steepen in the outer disk due to the rapid decrease in \sightwo\ in the region.

The midplane hydrostatic pressure ($P_0$) is obtained by assuming constant values for the stellar scale height and gaseous velocity dispersion. However, those values vary with radius as shown in Figure \ref{ollingfwhm} and \ref{vdisp}. Therefore, we have derived the interstellar gas pressure using the derived volume density at the midplane and radially dependent vertical velocity dispersion in order to examine whether a power-law relationship is still valid for the interstellar gas pressure ($P_{\rm g}$) at the midplane: 
\begin{equation}
P_{\rm g} = \rho_{\rm 0g}(R)\, \sigma_{\rm g}^2(R) \approx \rho_{\rm 0H_2}(R)\,\sigma_{\rm H_2}^2(R) +  \rho_{\rm 0HI}(R)\,\sigma_{\rm HI}^2(R).
\label{Pg}
\end{equation}
We assume the partial pressures contributed by the \HI\ and \Htwo\ are additive. 
We find that a power-law relationship between the midplane  $\rho_{\rm 0H_2}/\rho_{\rm 0HI}$ ratio  and the interstellar pressure seems to roughly hold as shown by Figure \ref{P_rmol} (right panel). 
The vertical and horizontal error bars in Figure \ref{P_rmol} show  uncertainties in $R_{\rm mol}$ and pressure based on maximum and minimum values for volume density and velocity dispersion that were estimated from uncertainties in the surface densities and scale heights. 
In addition, the uncertainty (a factor of 2) of the CO-to-H$_2$ conversion factor would affect the pressure using varying $\sigma_{\rm g} (R)$ by a factor of $\sim$20\% for NGC 4157, $\sim$30\% for NGC 4565, and $\sim$25\% for NGC 5907.
The power-law indices are a bit smaller compared with using constant values for $\sigma_{\rm g}$ and $z_*$ (left panel) but remain in good agreement with \citet{2006ApJ...650..933B}.
The reason why the relation based on volume density and varying velocity dispersion still shows a good correlation is that both $R_{\rm mol}$ and $P$ are shifted to higher values: $R_{mol}$ is higher at the midplane because the CO disk is thinner than the \HI, but the midplane pressure is also higher since the inferred velocity dispersion is much higher than the assumed value of 8 \kms. The two effects together tend to move the points up and to the right, so the correlation is roughly preserved. However, as noted in Paper I, the correlation is not preserved out of the midplane (at least for NGC 891; see Figure 17 of that paper).

\section{Summary and Conclusions}
\label{sum}


We have measured the thicknesses of CO, \HI, and stellar disk for a sample of edge-on galaxies (NGC 4157, 4565, and 5907) in order to derive the midplane volume density as a function of radius. 
In addition, we have inferred the vertical velocity dispersion from the measured disk thickness and surface density by solving the Poisson equation. Using the estimated volume density and inferred velocity dispersion as functions of radius, we have derived the interstellar gas pressure ($\rho_{\rm 0g} \sigma_{\rm g}^2$) in order to investigate the role of the pressure in controlling the \Htwo/\HI\ ratio. Our conclusions are summarized below. 

1. We have obtained the inclinations using Olling's (1996) method  to consider the projection effect of the less edge-on galaxies when we derive the disk thickness. The measured inclinations of NGC 4157 show a small difference between CO (84$\degr$) and \HI\ (83$\degr$) disks, while NGC 4565 and 5907 show a consistent value for both disks:  86.6$\degr$ for NGC 4565 and 86$\degr$ for NGC 5907. 

2. Using the inclinations, we have derived the gas disk thickness from the FWHM values of the observational data and the model at many different channel maps showing a ``butterfly" shape in Appendix \ref{appen}. 
Our results show that the scale heights of CO, \HI\, and stars increase with radius over the disk although the flaring of the  CO disk in NGC 4565 and 5907 is not clearly visible. 
The strongly star-forming galaxies (NGC 891 and NGC 4157) show a thicker CO disk compared to the other galaxies, suggesting a correlation between the disk thickness and star formation activity.
The vertical velocity dispersions of CO, \HI\, and stars inferred using the scale height and the surface density are declining as a function of radius. 

3. We have obtained the Schmidt law index for the total gas and the molecular gas using both the radial profile and the pixel-by-pixel analyses. For the molecular star formation law, the fitted indices for all galaxies range from 0.32 to 0.91 in the radial profile method and from 0.43 to 0.68 in the pixel-by-pixel method.
For the total gas star formation law, the range of the index is from 1.42 to 3.41 in the radial profile analysis.
The correlation between \sigsfr\ and \sightwo\ appears to be stronger than the relationship between \sigsfr\ and \siggas\ when taking into account the RMS scatter and the curvature in the relationship for the total gas. 

4. We have derived the gravitational instability parameter $Q_{\rm tot}$ in order to investigate the relationship between the parameter and massive star formation. In addition, we have compared the $Q_{\rm tot}$ profiles using constant and varying velocity dispersions of gas and stars. Our results show that $Q_{\rm tot}$ using the constant velocity dispersions seems to predict the star forming region in the inner disk,  while $Q_{\rm tot}$ using the varying velocity dispersions does not show a clear relationship with massive star formation. 
In addition, both cases are not able to predict massive star formation near the center.
While the gravitational instability parameter does not appear to predict the locus of massive star formation, the parameter using the varying velocity dispersions is suggestive of self-regulated star formation ($Q \sim 1$) even though the values are a bit larger than unity in most of galaxies.

5. We have obtained the power-law index of the relationship between the hydrostatic midplane pressure and the molecular to atomic gas surface density ratio. The range of the fitted index for the galaxies in this study is 0.89--1.88.  In addition, we have derived the interstellar gas pressure using the derived volume density and varying velocity dispersion $\sigma_{\rm g} (R)$ in order to examine the power-law relationship between the gas pressure and the ratio of \Htwo\ to \HI\ density at the midplane ($\rho_{\rm 0H_2}/\rho_{\rm 0HI}$). The  power-law correlation between the pressure and the ratio is still valid at the midplane even if we use the varying scale heights and velocity dispersions. The power-law index for the relationship based on $\sigma_{\rm g} (R)$ ranges from 0.43 to 1.54.

\acknowledgments

We thank the anonymous referee for useful comments and suggestions that improved this paper. 
K.Y. thanks Laura Zschaechner and Gyula J\'{o}zsa for help with using Tirific. 
This study is supported by the National Science Foundation under cooperative agreement AST-0838226 and by a Spitzer Cycle-5 data analysis award from NASA. 
Support for CARMA construction was derived from the states of California, Illinois, and Maryland, the James S. McDonnell Foundation, the Gordon and Betty Moore Foundation, the Kenneth T. and Eileen L. Norris Foundation, the University of Chicago, the Associates of the California Institute of Technology, and the National Science Foundation. Ongoing CARMA development and operations are supported by the National Science Foundation under a cooperative agreement, and by the CARMA partner universities. 
K.Y. and J.M.H. acknowledge support from the European Research Council under the European Union's Seventh Framework Programme (FP/2007-2013) / ERC Grant Agreement nr. 291531.
The National Radio Astronomy Observatory is a facility of the National Science Foundation operated under cooperative agreement by Associated Universities, Inc. 
\clearpage


\clearpage

\appendix

\section{Inclination and Scale Height using Olling's Method}
\label{appen}
 
In order to derive the inclination and the scale height for less edge-on galaxies, we have employed the method by \citet{1996AJ....112..457O} and we describe it as follows. The inclination ($i$) and the disk thickness (FWHM) are obtained from the apparent disk width (FWHM$_{\rm obs}$) measured at each velocity by taking into account the size of the in-plane region that contributes emission at that velocity. 

First, we have created model emission cubes using the rotation curve (Figure \ref{rot}) that we derived in Section \ref{gasprofile}, the surface density profile (Figure \ref{radiprof}), and an assumed constant velocity dispersion (8 \kms) for both CO and HI maps.
For NGC 4157 and 5907, we use the CO rotation curve (red dashed line in Fig. \ref{rot}) for the CO model and the \HI\ rotation curve (blue solid line in the figure) for the \HI\ model. However, in case of NGC 4565, we use the \HI\ rotation curve for both CO and \HI\ models because of a lack of CO emission in the inner region of the galaxy. Since the rotation curves of CO and \HI\ in this galaxy are very consistent, using the \HI\ curve for the CO model is justified.

In this procedure, we used a modified version of the MIRIAD task VELMODEL to produce a model of the in-plane distribution of gas at each observed radial velocity.  From this task, we obtain a face-on view of the emission contributing to each observed velocity channel, using the rotation curve and the systemic velocity. 
The structure of the velocity field depends on the adopted rotation curve.  
In addition, we used the MIRIAD task IMGEN to make a velocity dispersion model (assuming $\sigma_{\rm g}$ = 8 \kms) and a modified version of the task ELLINT to built a surface density model (using our derived surface density profile). 
For combining all three 1-D models (velocity field, velocity dispersion, and surface density models) together to make the model cube, the task VELIMAGE has been employed. 

\begin{figure}
\figurenum{A1}
\begin{center}
\begin{tabular}{c@{\hspace{0.1in}}c@{\hspace{0.1in}}c}
\includegraphics[width=0.5\textwidth,angle=270]{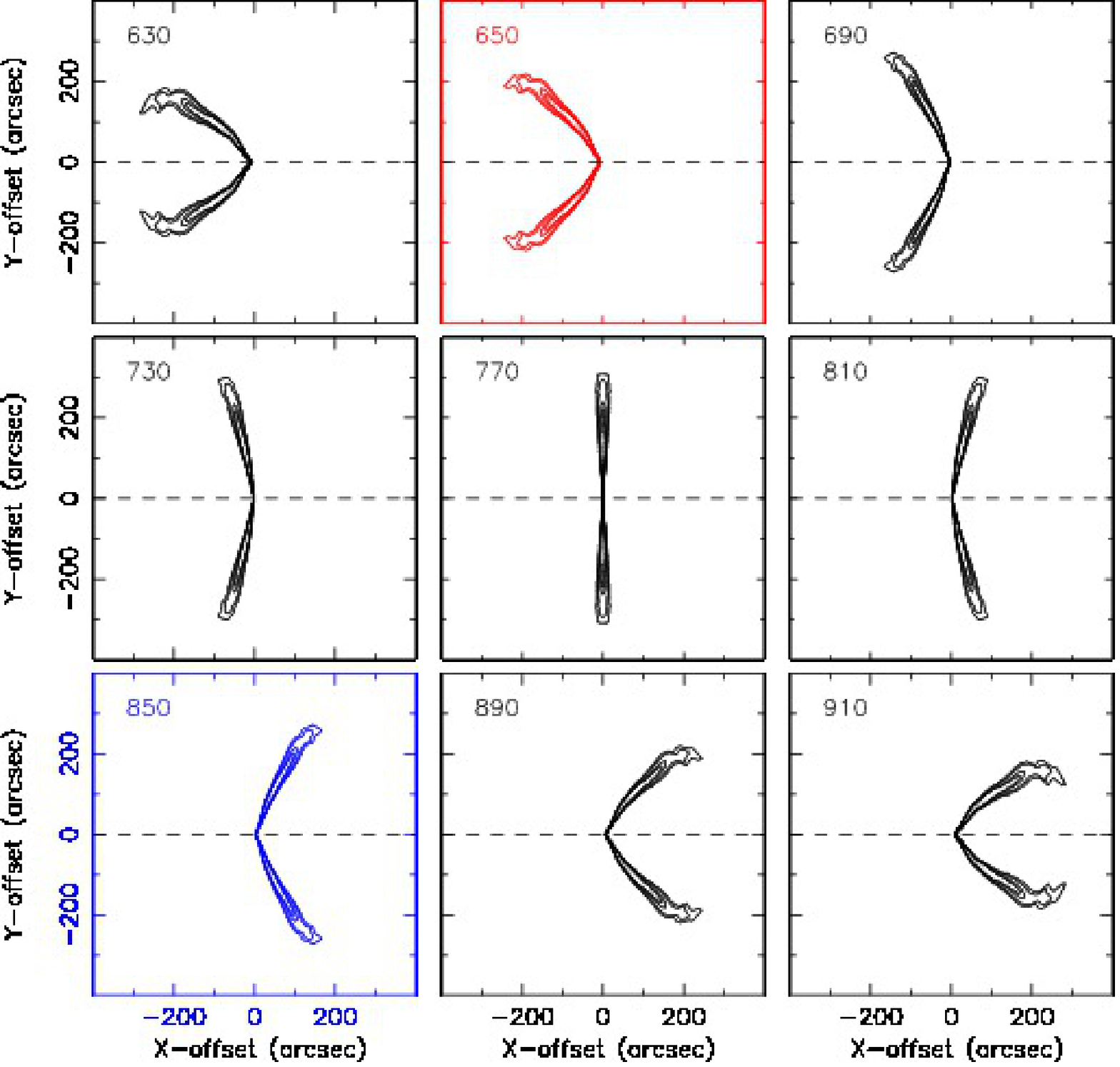}&
\includegraphics[width=0.4\textwidth,angle=270]{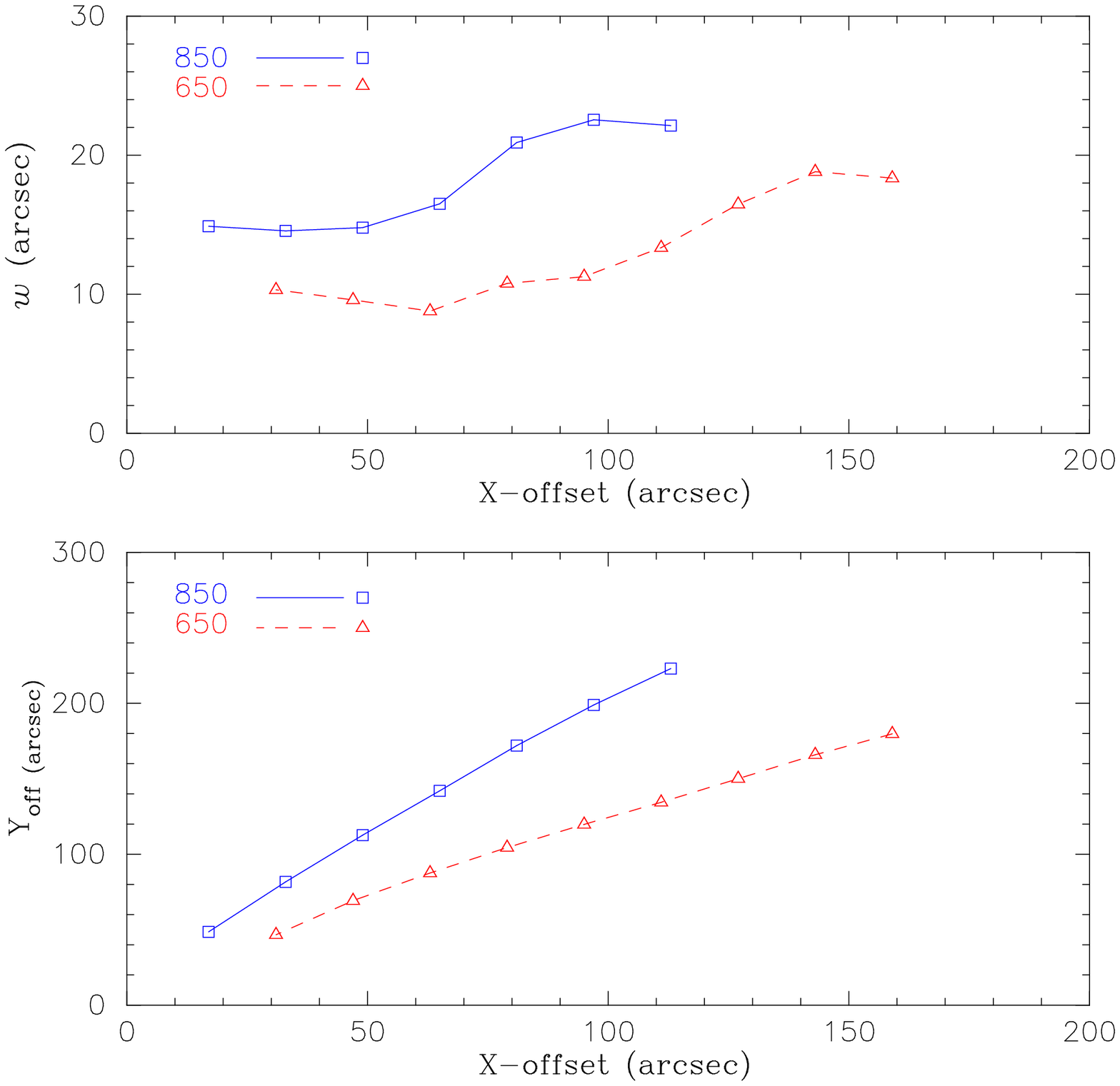}
\end{tabular}
\caption[Deprojected channel maps of a model galaxy]{$Left$: Deprojected channel maps of a model galaxy generated using the velocity field map, assumed velocity dispersion of 8 \kms, and the HI surface density profile of NGC 4157.  The radial velocity of each channel map is indicated in the upper left corner.
The systemic velocity  is  770 \kms. $Right\, Top$: Gaussian width ($w$) against $x$-offset from the channel maps at 650 and 850 \kms. $Right\, Bottom$: Minor axis offset from the midplane ($y_{\rm off}$) as a function of $x$-offset from the channel maps at 650 and 850 \kms.
\label{csdd}}
\end{center}
\end{figure}

Second, we have fitted a double Gaussian profile ($e^{-0.5((y-y_{\rm off})/w)^2} + e^{-0.5((y+y_{\rm off})/w)^2} $) to the minor axis distribution at each $x$-offset for a given channel map of the model cube in order to obtain the Gaussian width ($w$) and the vertical offset ($y_{\rm off}$) from the line of nodes ($y$ = 0). 
Some channel maps of the model are shown in Figure \ref{csdd} (left) and the Gaussian width and the vertical offset from the midplane obtained from the velocity channel map of  650 \kms\ and  850 \kms\ are shown in the figure (right). 
The planar width (FWHM${\rm_p}$) of the model is given by:
\begin{equation}
{\rm FWHM_p}(x, V_{R})  \approx 2.35 \sqrt {w^2(x, V_R) + y_{\rm off}^2(x, V_{R})},
\label{fwhmp}
\end{equation}
where the minor axis offset $y_{\rm off}$ is ignored when the minor axis profile can be fit with two Gaussians.

\begin{figure}
\figurenum{A2}
\begin{center}
\begin{tabular}{c@{\hspace{0.1in}}c@{\hspace{0.1in}}c@{\hspace{0.1in}}c}
\includegraphics[width=0.3\textwidth]{fA2a.eps}&
\includegraphics[width=0.28\textwidth]{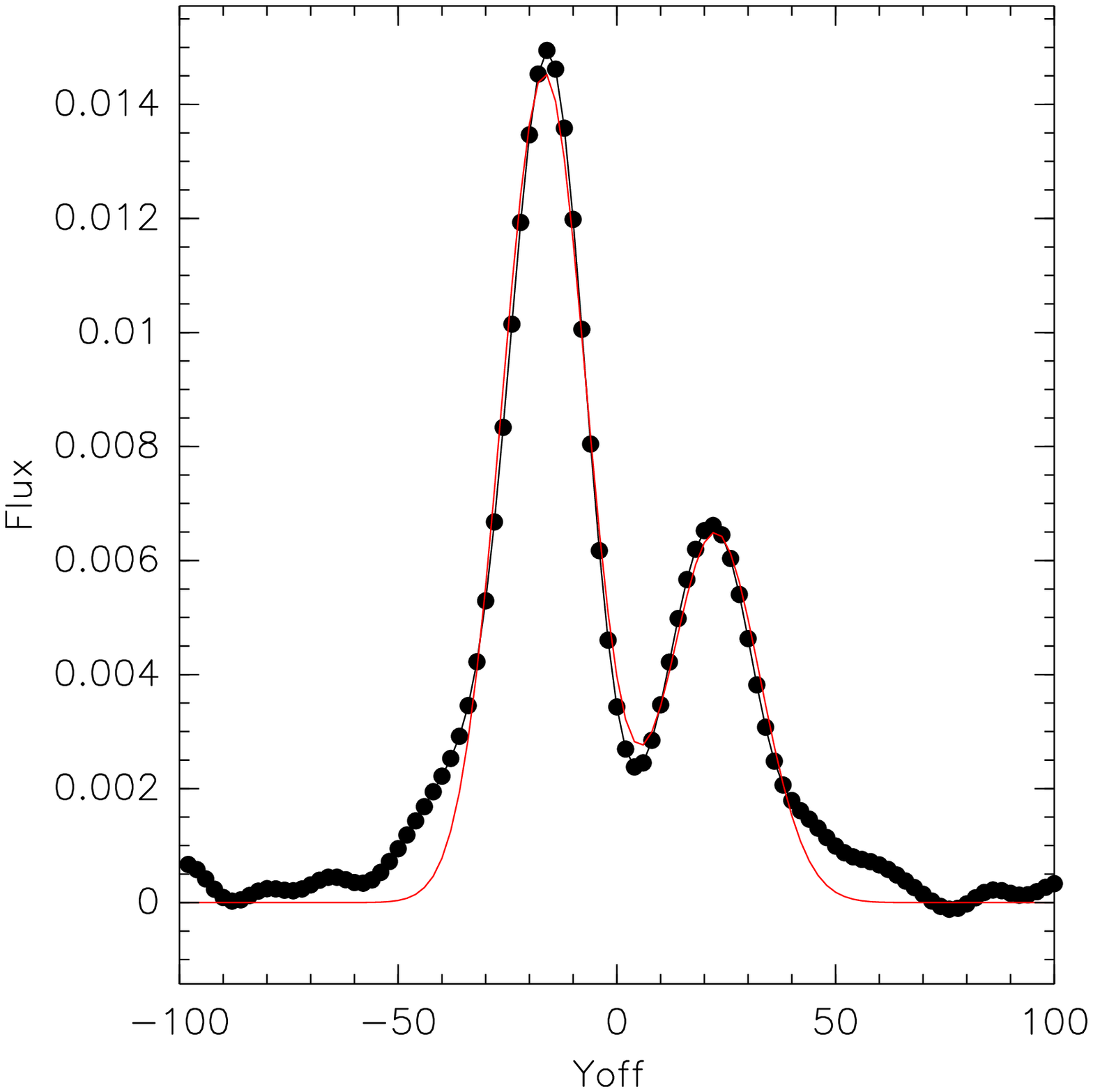}&
\includegraphics[width=0.28\textwidth]{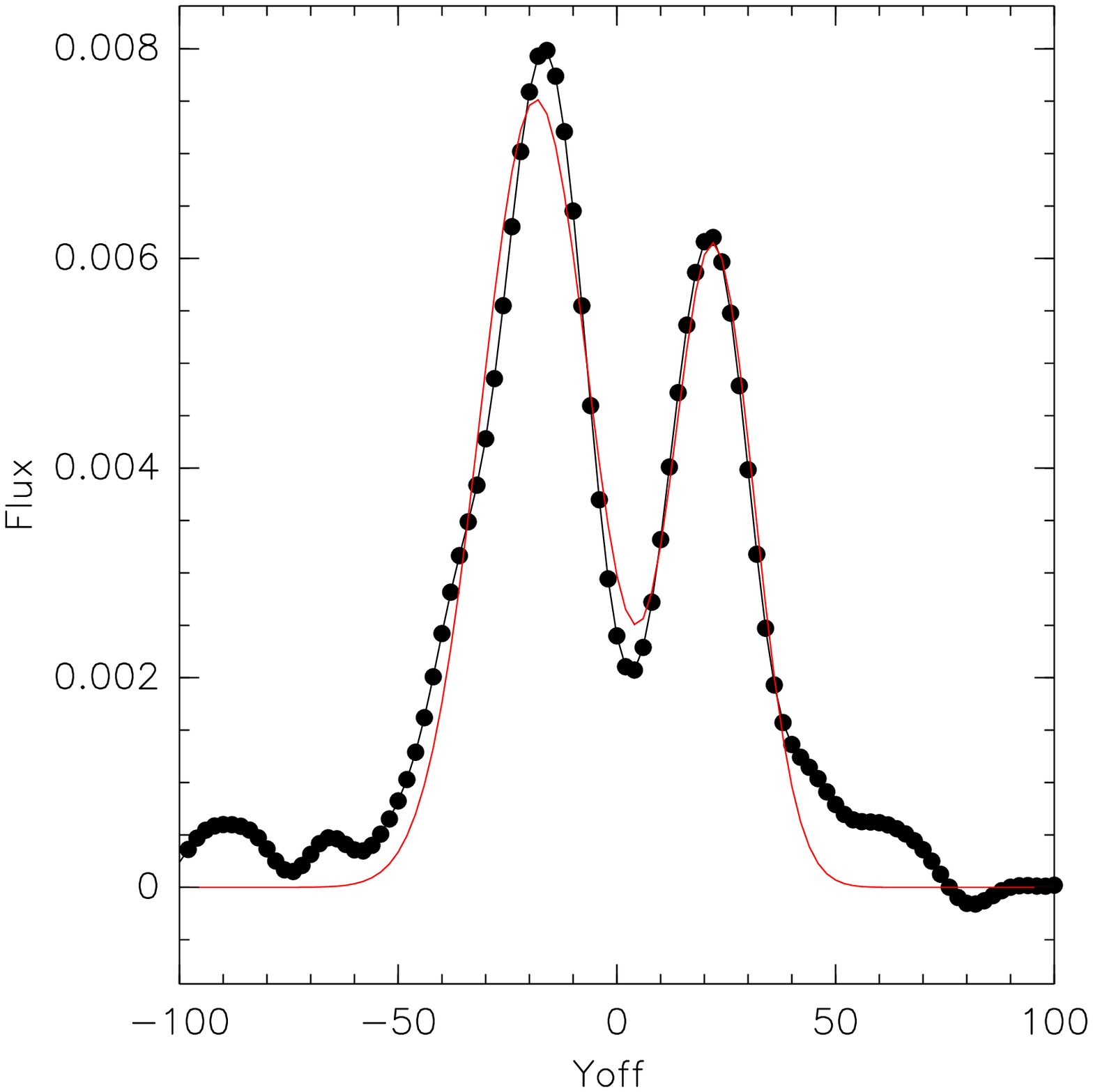}&
\end{tabular}
\caption[Double Gaussian fit to the minor axis profile]{Velocity channel map (taken at $V_R = 670$ \kms) for the NGC 4157 HI data $(Left)$. Double Gaussian fit (red line) to the minor axis profile (black line) at $x = -95$\ac $(Middle)$  and at $x = -111$\ac $(Right)$. 
\label{butterfly}}
\end{center}
\end{figure}

Third, returning to the observed data cubes, we have determined distances between the two peaks of the minor axis profile ($\Delta d$) and the FWHM$\rm_{obs}$ thicknesses (2.35 $\times$ Gaussian width) as functions of $x$ and the radial velocity ($V_{R}$) by fitting a double Gaussian profile to many velocity channel maps. 
The channel maps of observational data are selected when they show clear a ``butterfly" shape, so they can be fitted by a double Gaussian profile. 
In addition, among many fitted $x$-values in the selected channel map, the final parameters we use for deriving the inclination and the disk thickness are chosen when at least one of two peaks  is greater than twice the rms noise of the channel map and $\Delta d$ is greater than the synthesized beam (FWHM$_{\rm beam}$).
One of the channel maps we have used for fitting is shown in Figure \ref{butterfly} and examples of fitting results at two different $x$ offsets in the channel map are also shown in the figure.

\begin{figure}
\figurenum{A3}
\begin{center}
\includegraphics[width=0.6\textwidth]{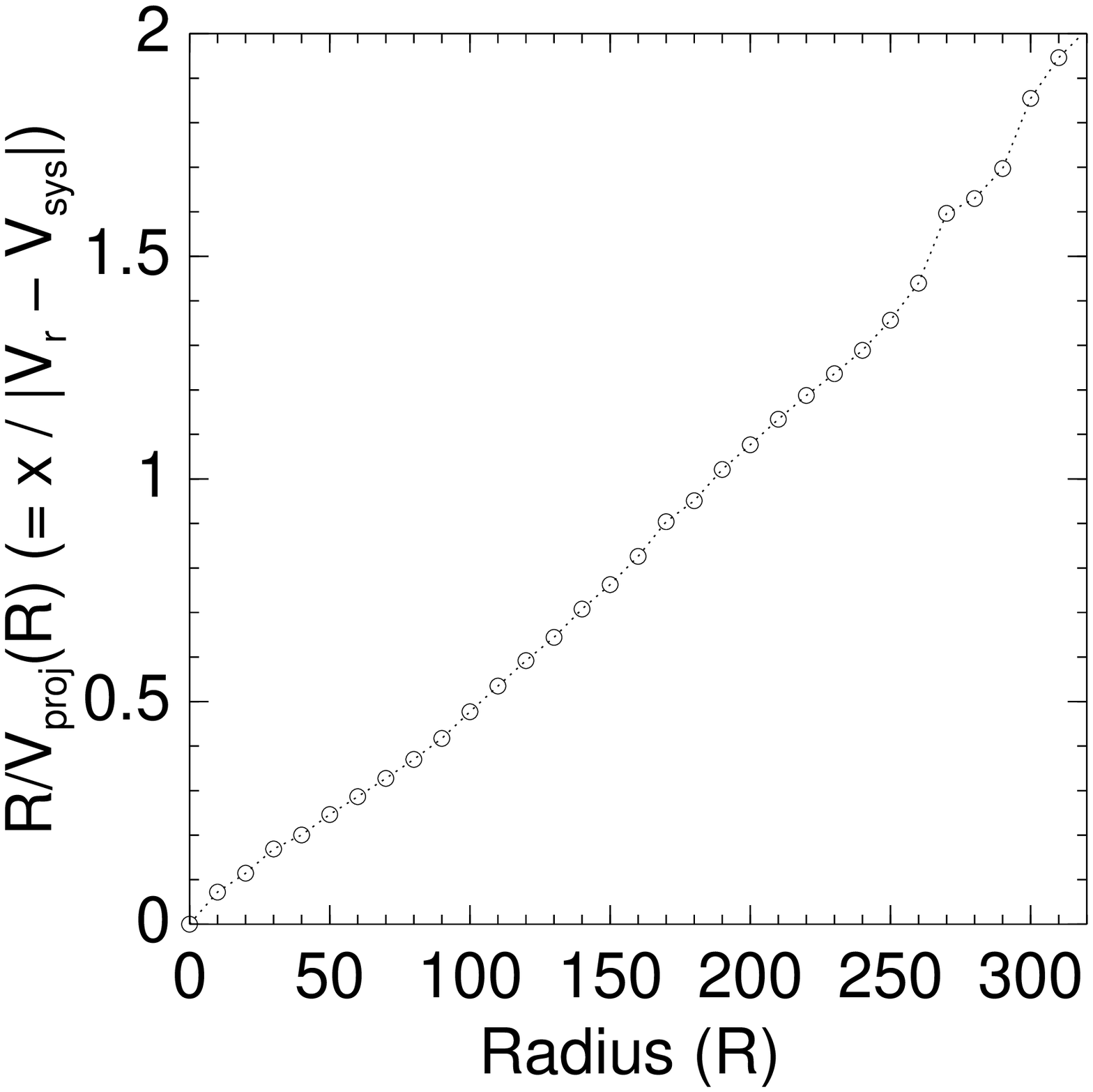}
\caption[Plot of R/V(R) vs. radius]{$R/V_{\rm proj}(R) (= x/|V_R - V_{\rm sys}|)$ against radius for NGC 4157. 
\label{getrad}}
\end{center}
\end{figure}

Finally, using all the parameters we obtained above,  we have derived inclinations ($i$) and the disk thicknesses as a function of radius for the galaxies using the equations given by    \cite{1996AJ....112..457O}: 
\begin{eqnarray}
i (x, V_R) &=& {\rm cos^{-1}}\frac{\Delta d (x, V_R)}{2\ y_{\rm off}(x, V_R)}, \\
{\rm FWHM^2} (x, V_R) &=& \frac{1}{{\rm sin^2} i}\{{\rm FWHM^2_{obs}} (x, V_R) - {\rm FWHM^2_{beam}} \nonumber\\
&-&\left[{\rm FWHM_p}(x, V_R) {\rm cos}\ i \right]^2\},
\label{olling}
\end{eqnarray}
where FWHM$_{\rm beam}$ is beam size of the  maps after convolving to the circular beam shown in Table \ref{obstable}. 
One double Gaussian fit to the butterfly shape of the data provides two different values of FWHM since the Gaussian widths from both sides are independent  unlike  the widths of the model. We use both FWHM values when the ratio of smaller to larger peak intensities is more than a factor of 0.6, while just one FWHM having the higher peak value is used when the peak ratio is less than 0.6.
We convert $x$ to a radius ($R$) using the projected rotation curve ($V_{\rm proj}$), which is used to obtain the velocity field model:
\begin{equation}
R = V_{\rm proj}(R) \frac{x}{|V_R - V_{\rm sys}|}.
\end{equation}
We plot $R/V_{\rm proj}(R) (= x/|V_R - V_{\rm sys}|)$ versus radius ($R$) in Figure \ref{getrad}, showing how we obtain a radius corresponding to each combination of ($x, V_R$).
From the inclination as a function of radius, we derived a weighted mean value of inclination and fixed the inclination to this value when deriving the FWHM values. 

\begin{figure}
\figurenum{A4}
\begin{center}
\begin{tabular}{c@{\hspace{0.1in}}c@{\hspace{0.1in}}c}
\includegraphics[width=0.45\textwidth]{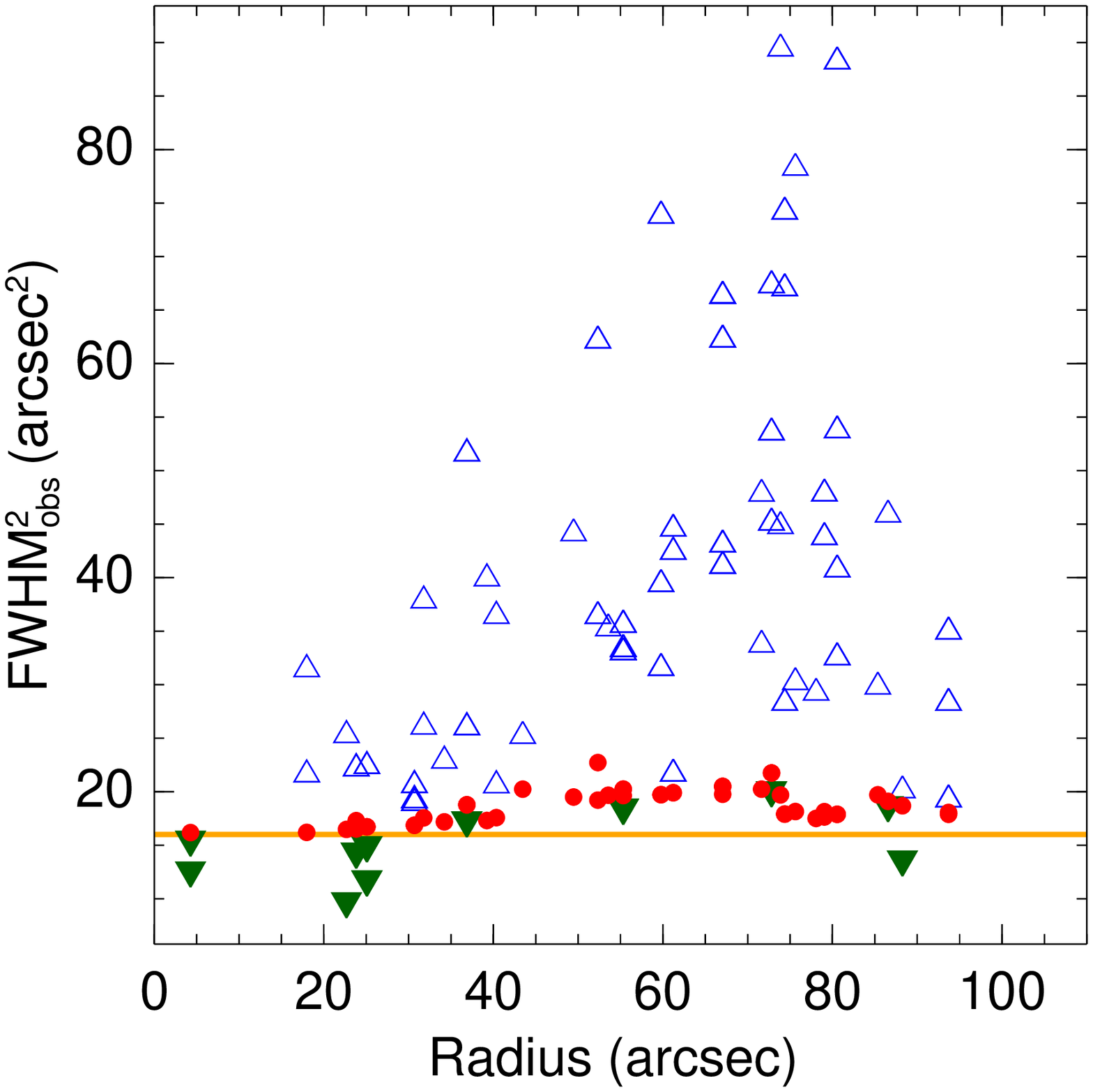}&
\includegraphics[width=0.45\textwidth]{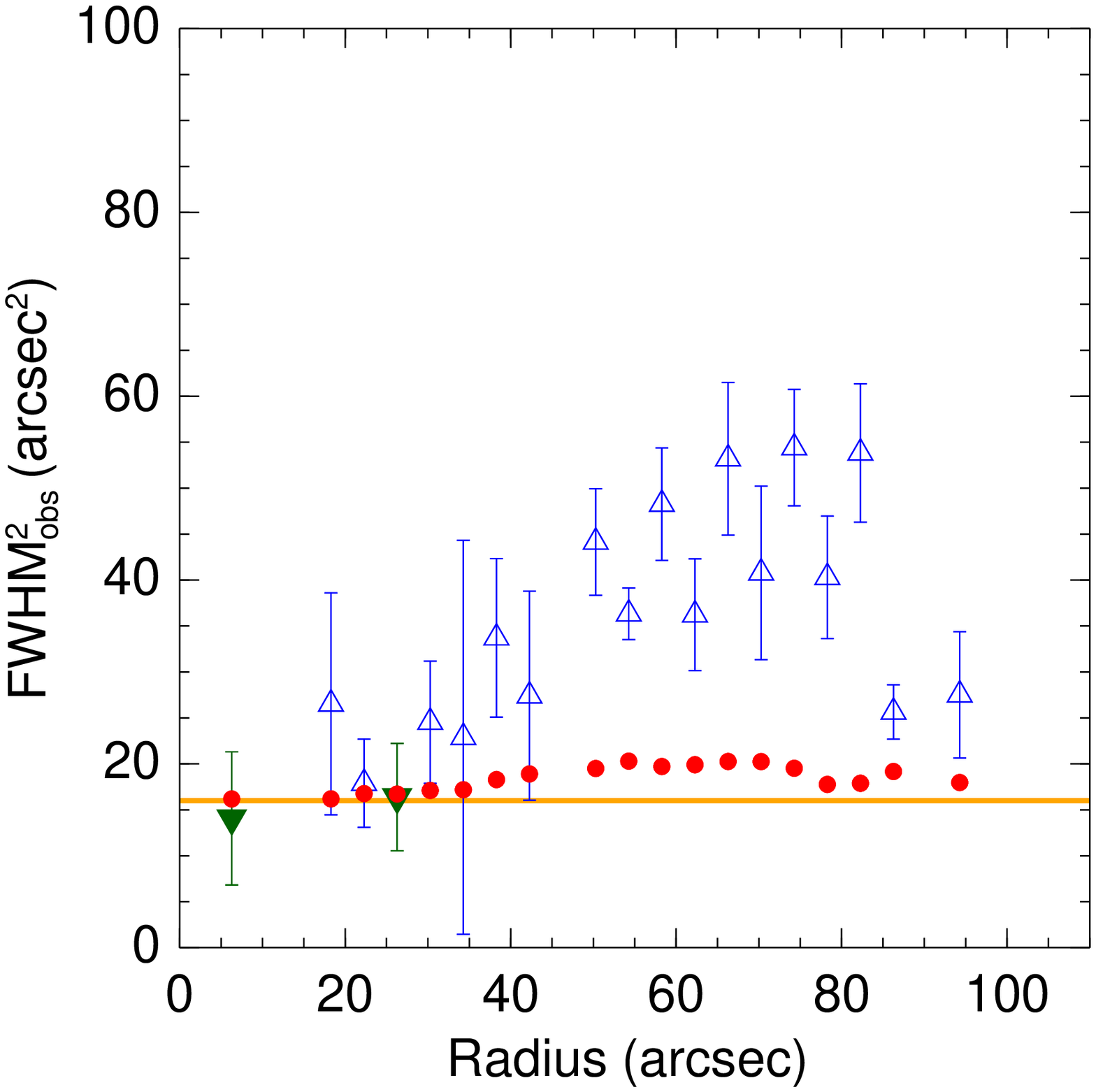}&
\end{tabular}
\caption[The observed FWHM values for CO in NGC 4157]{$Left$: The observed FWHM values (open blue and solid green triangles) and the sum of $\rm FWHM^2_{beam}$ and  $\rm FWHM^2_p cos^2{\it i}$ (solid red circle) as a function of radius for CO in NGC 4157. 
The inverted green triangles show the non-deconvolveable data since they are below the values of the solid circles.
$Right$: Averaged data points in 4\ac\ radial bins. The values are for valid double Gaussian fits. 
The inverted triangle points show the data, which are  non-deconvolveable. 
The horizontal line represents a value of $\rm FWHM^2_{beam}$. 
\label{fwhm4157}}
\end{center}
\end{figure}

As an example, the input needed to obtain FWHM thickness values are shown in Figure \ref{fwhm4157} (left panel) for the CO image of NGC 4157. The open triangles show the observed width ($\rm FWHM^2_{obs}$) and filled circles represent  the sum of the two broadening terms ($\rm FWHM^2_{beam} + FWHM^2_p cos^2{\it i}$) in Equation \ref{olling}. In the right panel,  all the points are averaged in a 4\ac\ radial bin and the averaged values are used to derive  the FWHM thickness following Equation \ref{olling}. The inverted triangle points show the data, which are  non-deconvolveable since they are below the values of the solid circles. 
The bin size is selected based on the angular resolution of each map: 4\ac\ for all the CO maps, 10\ac\ for the \HI\ map of NGC 4565, and 15\ac\ for the \HI\ maps of NGC 4157 and 5907.  The use of an average value in a bin provides a resolution similar to its map, as well as a better signal to noise ratio, hence we adopt this approach for our analysis.


\bibliographystyle{apj}
\bibliography{refer}

\end{document}